\documentclass[]{emulateapj}

\usepackage{longtable}
\usepackage{natbib}
\usepackage{graphicx}
\usepackage{graphics}

\begin{document}

\title{Spatially resolved stellar, dust and gas properties \\ of the post-interacting Whirlpool Galaxy system}

\author{\medskip 
Erin Mentuch Cooper\altaffilmark{1,2},
Christine D. Wilson\altaffilmark{1},
Kelly Foyle\altaffilmark{1},
George Bendo\altaffilmark{3},
Jin Koda\altaffilmark{4},
Marten Baes\altaffilmark{5}
M\'ed\'eric Boquien\altaffilmark{6},
Alessandro Boselli\altaffilmark{6},
Laure Ciesla\altaffilmark{6},
Asantha Cooray\altaffilmark{7},
Steve Eales\altaffilmark{8},
Maud Galametz\altaffilmark{9},
Vianney Lebouteiller\altaffilmark{10},
Tara Parkin\altaffilmark{1},
H\'el\`ene Roussel\altaffilmark{11}
Marc Sauvage\altaffilmark{10},
Luigi Spinoglio\altaffilmark{12}
Matthew W.L. Smith\altaffilmark{8}
}
\altaffiltext{1}{Department of Physics \& Astronomy, McMaster University, Hamilton, ON, L8S 4M1, Canada}

\altaffiltext{2}{Astronomy Department, University of Texas at Austin, Austin, TX 78712, USA}

\altaffiltext{3}{UK ALMA Regional center Node, Jodrell Bank center for Astrophysics, School of Physics \& Astronomy, University of Manchester, Oxford Road, Manchester M13 9PL, United Kingdom}

\altaffiltext{4}{Department of Physics \& Astronomy, SUNY Stony Brook, Stony Brook, NY 11794-3800, US}

\altaffiltext{5}{Sterrenkundig Observatorium, Universiteit Gent, Krijgslaan 281 S9, B-9000 Gent, Belgium}
\altaffiltext{6}{Laboratoire dÕAstrophysique de Marseille, UMR 6110 CNRS, 38 rue F. Joliot-Curie, F-13388 Marseille, France}

\altaffiltext{7}{University of California, Irvine, Department of Physics \& Astronomy, 4186 Frederick Reines Hall, Irvine, CA, USA}
\altaffiltext{8}{School of Physics \& Astronomy, Cardiff University, The Parade, Cardiff, CF24 3AA, United Kingdom}
\altaffiltext{9}{Institute of Astronomy, University of Cambridge, Madingley Road, Cambridge, CB3 0HA, UK}
\altaffiltext{10}{CEA/DSM/DAPNIA/Service dÕAstrophysique, UMR AIM, CE Saclay, 91191 Gif sur Yvette Cedex,
France}
\altaffiltext{11}{Institut dÕAstrophysique de Paris, UMR7095 CNRS, Universit\'e Pierre \& Marie Curie, 98 bis Boulevard Arago, 75014 Paris, France}
\altaffiltext{12}{ Istituto di  Fisica dello Spazio Interplanetario INAF, Via Fosso del Cavaliere 100, 00133 Roma, Italy}

\begin{abstract}


Using infrared imaging from the \textit{Herschel Space Observatory}, observed as part of the Very Nearby Galaxies Survey, we investigate the spatially resolved dust properties of the interacting Whirlpool galaxy system (NGC\,5194 and NGC\,5195),  on physical scales of $\sim1$\,kpc. Spectral energy distribution modelling of the new infrared images in combination with archival optical, near- through mid-infrared images confirms that both galaxies underwent a burst of star formation $\sim$370--480~Myr ago and provides spatially resolved maps of the stellar and dust mass surface densities. The resulting average dust-to-stellar mass ratios are comparable to other spiral and spheroidal galaxies studied with \textit{Herschel}, with NGC\,5194 at $\log(M_\mathrm{dust}/M_\star)= -2.5\pm0.2$ and NGC\,5195 at $\log(M_\mathrm{dust}/M_\star)= -3.5\pm0.3$. The dust-to-stellar mass ratio is constant across NGC\,5194 suggesting the stellar and dust components are coupled. In contrast, the mass ratio increases with radius in NGC\,5195 with decreasing stellar mass density.  Archival mass surface density maps of the neutral and molecular hydrogen gas are also folded into our analysis. The gas-to-dust mass ratio, $94\pm17$, is relatively constant across NGC\,5194, although there is some suggestion that it decreases radially but not significantly above our uncertainties.  Somewhat surprisingly, we find the dust in NGC\,5195 is heated by a strong interstellar radiation field, over 20 times that of the ISRF in the Milky Way, resulting in relatively high characteristic dust temperatures ($\sim30$\,K). This post-starburst galaxy contains a substantial amount of low-density molecular gas and displays a gas-to-dust ratio ($73\pm35$) similar to spiral galaxies. It is unclear why the dust in NGC\,5195 is heated to such high temperatures as there is no star formation in the galaxy and its active galactic nucleus is 5-10 times less luminous than the one in NGC\,5194, which exhibits only a modest enhancement in the amplitude of its ISRF.

\end{abstract}

\section{Introduction}

The fundamental building blocks of a galaxy -- its stars, gas and dust -- are expressed across its spectral energy distribution (SED). Hot gas ($>10^6$\,K) emits brightly at very short wavelengths in the X-Ray and ultraviolet (UV), while neutral and molecular gas are seen through emission lines primarily studied in the submm and radio. Young stellar populations dominate the UV and blue visible light of a galaxy and ionize gas leading to numerous emission lines in the UV, optical and infrared. Evolved stars peak at longer wavelengths in the red visible and near-infrared regime. At longer wavelengths and lower energies, mid-infrared light traces very small ($<$0.01\,\micron) dust grains and small organic molecules known as polycyclic aromatic hydrocarbons (PAHs) and far-infrared light traces re-processed stellar light emitted by dust grains. Multi-wavelength studies provide the synergy to compare these components and move toward building models that incorporate all the building blocks of a galaxy.

Standard stellar population techniques can tell us how much stellar mass is in a galaxy, its metallicity, the amount of dust attenuation and can even shed light on the history of star formation within a galaxy. Fundamentally, stellar emission models are built from the superposition of multiple simple stellar populations where assumptions regarding the IMF and stellar evolution lead to systematic but known uncertainties \citep{con09}. On the other hand, modelling the dust emission in a galaxy is not yet as evolved. Most dust emission models or templates \citep{dal02,gal11} measure the total dust mass and some \citep{des90,dra07,com11} the relative abundance of various dust emission components (eg. PAHs, very small grains and big grains). Fundamental to these models is the mechanism for dust heating, primarily provided by the interstellar radiation field (ISRF) produced by stars, although active galactic nuclei (AGN) can also play a role in nuclear regions. 

%
\begin{figure}[t]
\centering
\includegraphics[width=3.5in]{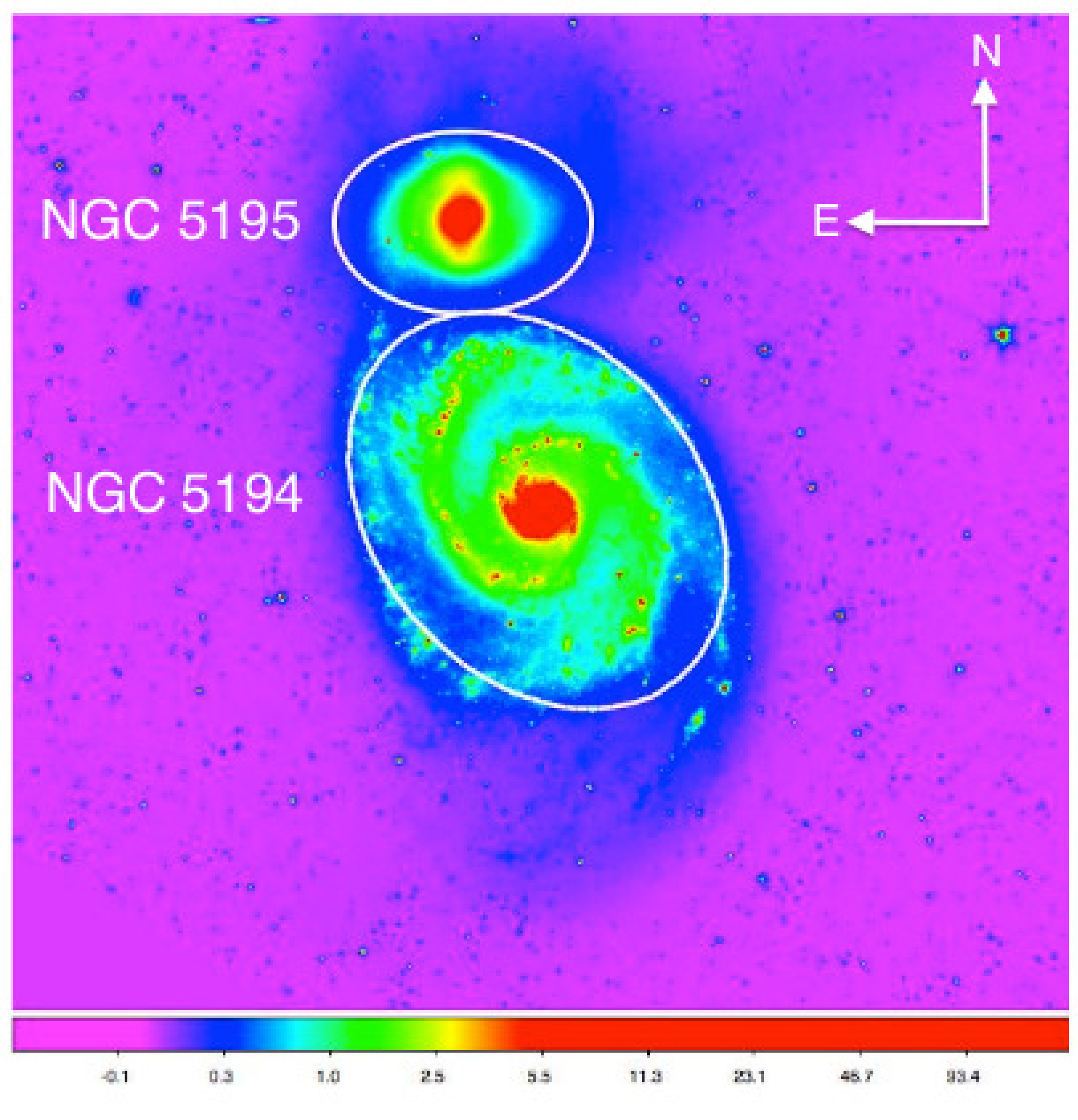}
\caption{Image of the Whirlpool galaxy system, NGC\,5194/5 (M51a/b) from the \textit{Spitzer Space Telescope's} IRAC camera at 3.6\,\micron~(in intensity units of MJy\,sr$^{-1}$). Measured apertures are shown for each galaxy and were defined by a constant stellar mass surface density of 50\,M$_\sun$\,pc$^{-2}$.}
\label{fig:app}
\end{figure}

Nearby galaxies offer unique laboratories to explore how dust emission is tied to the underlying stellar populations in the context of standard SED modelling techniques. The post-interacting Whirlpool galaxy system was observed as part of the Very Nearby Galaxies Survey (VNGS; PI Christine Wilson), a guaranteed-time photometric and spectroscopic survey of a diverse range of 13 nearby galaxies with \textit{Herschel's} Photodetector Array Camera and Spectrometer (PACS; \citealt{pog10}) and Spectral and Photometric Imaging Receiver (SPIRE; \citealt{gri10}) instruments.  

This system consists of the diverse galaxy pair of NGC\,5194/M51a, a spiral SABbc type galaxy, and NGC\,5195/M51b, a dusty spheroid galaxy whose peculiar dust lanes have given it an SB0p classification \citep{dev95}. An encounter between the two galaxies 300--500\,Myr ago is inferred from both kinematic \citep{sal00} and hydrodynamic modelling \citep{dob10} of the M51 system. In agreement, a recently constructed color-magnitude diagram of its individual stars, measured using \textit{Hubble Space Telescope} archival images, suggests a burst of star formation occurred 390--450\,Myr ago in NGC 5195 \citep{tik09}. Star formation also peaked around 380-450\,Myr in the spiral NGC\,5194 and has continued until the present day. While star formation is ubiquitous and ongoing across NGC\,5194/M51a as shown by its $\sim9\,600$ HII regions \citep{lee11} and optical colors \citep{tik09}, the spheroid neighbor NGC\,5195 shows no evidence of recent star formation through a deficiency of ionized gas \citep{thr91}. This is further supported by evidence for dynamically stable molecular gas, which is unable to form stars.  \citet{koh02} measured both CO(1-0) and HCN(1-0) in the galaxy and found an HCN-to-CO intensity ratio 5-15 times smaller than that seen in starburst regions \citep{hel93,koh02}  and 5 times smaller than that usually found in spiral disks \citep{hel95,koh99}, indicative of a lack of dense molecular cores.

\begin{figure*}[t]
\centering
\includegraphics[angle=-90,width=4.5in]{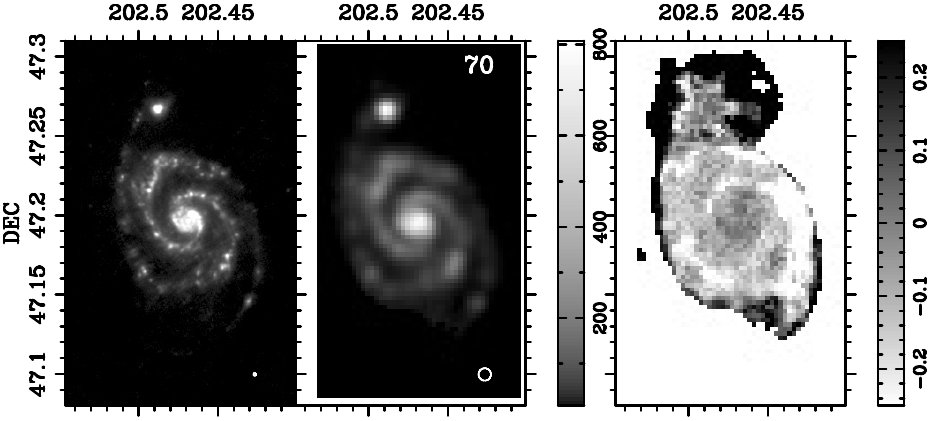}
\includegraphics[angle=-90,width=4.5in]{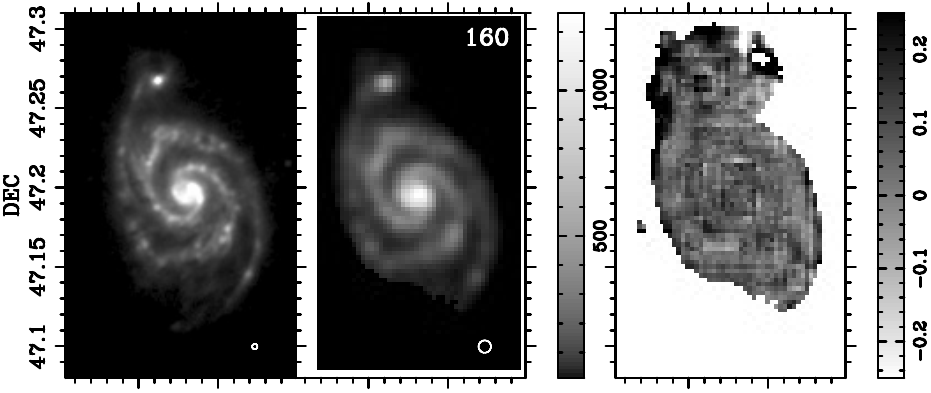}
\includegraphics[angle=-90,width=4.5in]{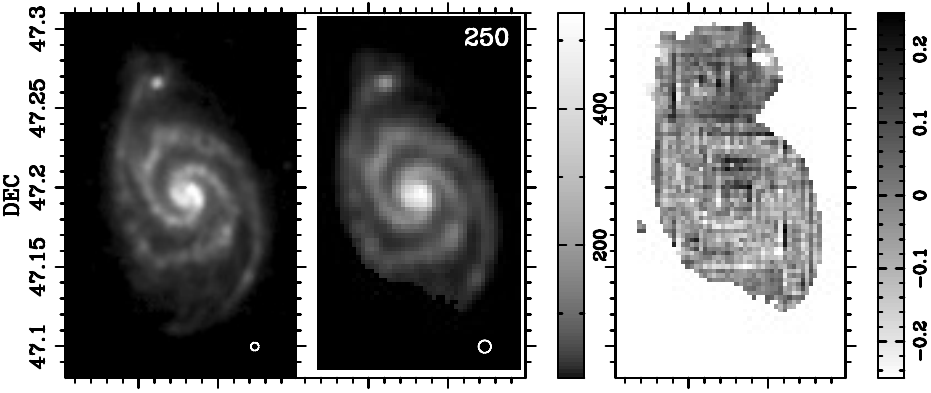}
\includegraphics[angle=-90,width=4.5in]{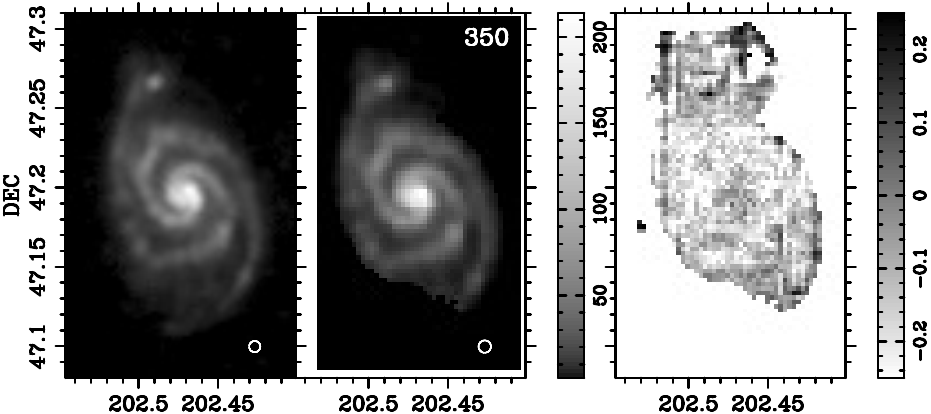}
\caption{Far-infrared images of NGC\,5194/5 from \textit{Herschel}/PACS and SPIRE as part of the VNGS survey from top to bottom at 70\,\micron, 160\,\micron, 250\,\micron~and 350\,\micron.   The left panel is the original image, while the middle panel shows the image after matching it to a common resolution of 28\arcsec~and platescale of 10\arcsec\,pix$^{-1}$. Both images are shown in MJy~sr$^{-1}$ and the beam area for each image is shown in the bottom-right. The right shows the residual image between the models and observations, normalized by the observations: ($I_\mathrm{\lambda,obs}-I_\mathrm{\lambda,model})/I_\mathrm{\lambda,obs}$. The best-fit model images from \citet{dra07} match the images at 160\,\micron~and 250\,\micron~to within $\sim$10\%, but underestimate emission in some regions of the 70\,\micron~ (top) and 350\,\micron~(bottom) images by over 20\%.}
\label{fig:montageHerschel}
\end{figure*}

\begin{table*}
\caption{Structural Properties}
\begin{center}
\begin{tabular}{c|c|c|c|c|c|c}
Galaxy & RA (J2000) & DEC (J2000) & 2$a$  & 2$b$ & P.A. & Aperture Radius\\ 
& (deg) & (deg) & (arcsec) & (arcsec) & (deg) & (arcsec) \\
\hline
NGC\,5194 & 202.47065 & +47.19517 & 383 & 286 & 50 & 295 \\
NGC\,5195 & 202.49726 & +47.26531 & 285  & 225 & 90 &  115 \\
\end{tabular}
\end{center}
\label{tab:param}
\end{table*}%

This paper takes advantage of a large multi-wavelength dataset encompassing observations of the stars, dust and gas at a common spatial resolution to try to decode the star formation history of the galaxy and how it has impacted the dust and gas properties. Our ultimate goal is to present a story of these components in the context of the interaction and star formation history of the system. Our observations and the steps involved to match the spatial and surface brightness distributions are given in \S\ref{s:data}. We apply standard SED fitting techniques to the dust and stars separately but do so on a pixel level in order to map any gradients and spatial dependencies of the parameters. The models, our fitting procedure and output parameter maps are all presented in \S\ref{s:methods}. We provide dust-to-stellar and gas-to-dust mass maps and 1D radial plots in \S\ref{s:discussion} and discuss how the observed dust properties relate to the stellar and gas properties of the system.

\section{Observations and image processing}\label{s:data}

Our multi-wavelength analysis includes twenty images of the Whirlpool system, and thus exploits the great efforts of a number of large legacy surveys. In this section we provide a summary of the archival observations and describe our new \textit{Herschel} observations from the \textit{Very Nearby Galaxies Survey} (VNGS). A number of steps were taken to ensure each pixel represents a common measure of surface brightness. We have opted to exclude the 500\,\micron~image as a compromise for higher spatial resolution. \citet{bos12} demonstrate that for late-type spiral galaxies the SPIRE FIR colours are tightly correlated and the 500\,\micron~band is not necessary to constrain a dust model. In addition, because our analysis hinges on comparing a large number of observations (4000 pixels in 20 images) to large libraries of stellar and dust models, we require satisfactory signal-to-noise levels in each pixel to determine parameter likelihoods. In this section we describe image processing steps and discuss our efforts to accurately quantify uncertainties in post-processed images.

\subsection{Adopted Galaxy Parameters}

For our analysis we use a common distance of $9.9\pm0.7$\,Mpc \citep{tik09}, although a wide range of distances (anywhere from 6-10\,Mpc) can be found in the literature (e.g. 8.4\,Mpc from \citealt{fel97} is often used). This latest distance is from Hertsprung-Russell diagram matching using individual red giant stars seen in \textit{HST} archival images. We also measured our own aperture and galactic center for each galaxy based on constant stellar mass density rather than surface brightness. Our apertures, shown in Figure~\ref{fig:app}, were defined at a constant stellar mass of 50\,M$_\sun$\,pc$^{-2}$ and ensured that neither galaxy's aperture overlapped. When total mass estimates are measured for each galaxy, it is the sum of pixels within this aperture. No aperture corrections are applied extending to larger galactic radii because beyond our chosen apertures, the galaxies begin to blend. For this reason, mass surface density ratios are more relevant to our analysis. We also use the apertures to define an effective radius, 
$r_\mathrm{eff}$, in order to measure 1D radial profiles of mass densities and other derived SED properties. Our effective radius is defined as

\begin{equation}
\left(\frac{r_\mathrm{eff}}{r_\mathrm{app}}\right)^2 = \left(\frac{x}{a}\right)^2 + \left(\frac{y}{b}\right)^2
\end{equation}

where $a$ is the semi-major axis, $b$ is the semi-minor axis, and $r_\mathrm{app}^2 = a^2 + b^2$. The position angle, P.A., is measured east from north. These parameters are given for reference in Table~\ref{tab:param} and shown in Figure~\ref{fig:app} imposed on a \textit{Spitzer}/IRAC 3.6\,\micron~image of the galaxies.


\subsection{Far-infrared data}

\begin{figure*}[t]
\centering
\includegraphics[angle=-90,width=6in]{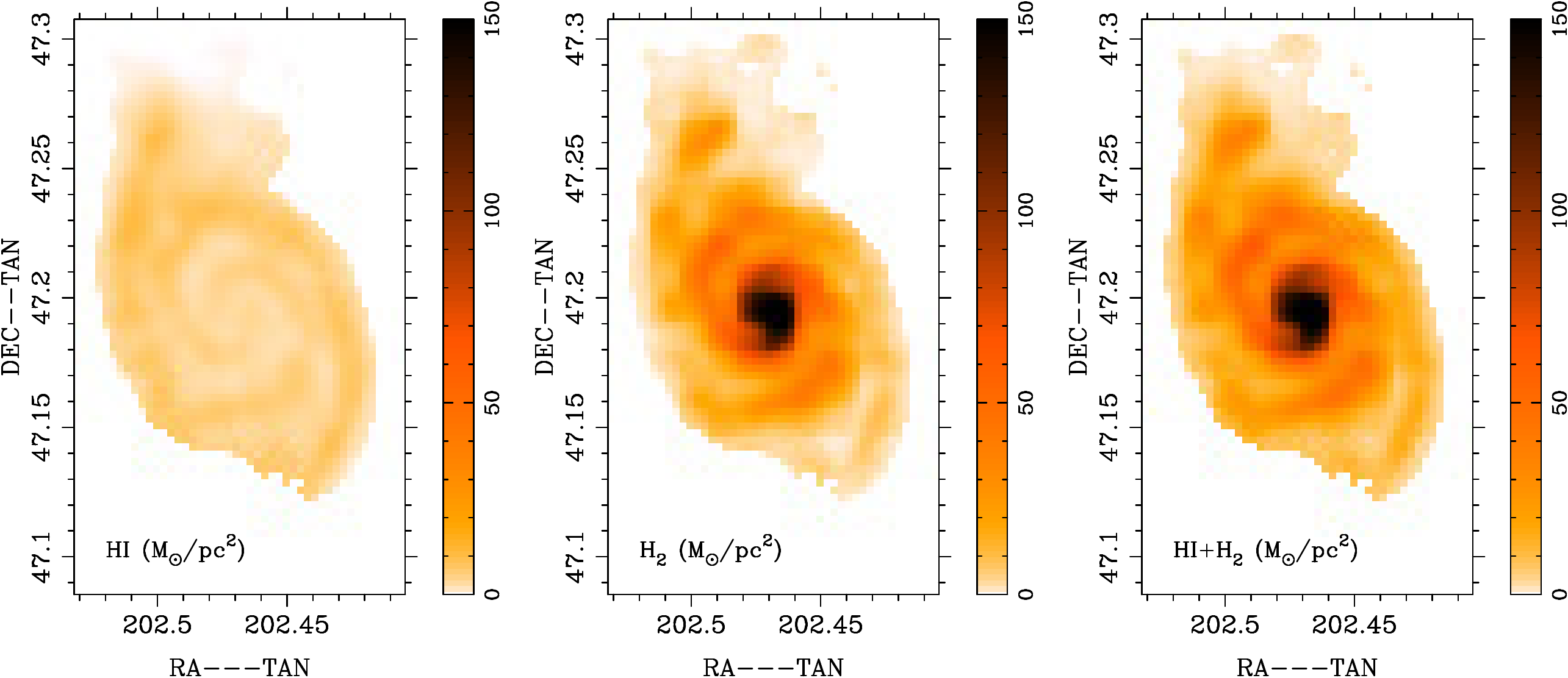}
\caption{\textit{Left:} HI mass surface density image of NGC\,5194/5 from the THINGS survey \citep{wal08} convolved to match the PSF (FHWM=28\arcsec) of our other observations. \textit{Middle:} the CO(1-0) image from the Nobeyama 45\,m telescope \citep{kod11} converted to an H$_2$ mass surface density image. \textit{Right:} Total gas mass surface density image is created by summing the two left images. All images are in units of M$_\sun$~pc$^{-2}$ and are shown on the same intensity scale. Molecular gas dominates the gas emission in the central area of the galaxy and along the spiral arms. To convert to total gas mass surface densities, a 1.36 scaling is required to account for helium.}
\label{fig:gasimages}
\end{figure*}

The VNGS observed the Whirlpool system with \textit{Herschel}/PACS at 70\,$\mu$m and 160 $\mu$m and \textit{Herschel}/SPIRE at 250\,$\mu$m and 350\,$\mu$m. Images at each wavelength at their native resolution are shown on the left column of Figure~\ref{fig:montageHerschel}. More details on the observational setup can be found in \citet{ben12}. 

The PACS images were processed using both \textsc{HIPE} v5 and \textsc{SCANAMORPHOS} v8\footnote{http://www2.iap.fr/ users/roussel/herschel/}. To account for changes in calibration from HIPE v5 to HIPE v6, the 70 and 160\,\micron~images were divided by factors of 1.119 and 1.174. The PACS images were converted to Jy~sr$^{-1}$ from Jy\,pix$^{-1}$ with platescales of 2\arcsec~and 4\arcsec~ for the 70\,$\mu$m and 160\,\micron~images respectively. 

The SPIRE images were processed using \textsc{HIPE} and \textsc{BriGAde} as described in \citet{smi12}. The final SPIRE maps were created using the na\"ive mapper provided in the standard pipeline with pixel sizes of 6\arcsec and 8\arcsec at 250\,$\mu$m and 350\,$\mu$m. The FWHM of the SPIRE beams for this pixel scale are 18.2\arcsec~ and 24.5\arcsec~ at 250\,$\mu$m and 350\,$\mu$m~\citep{swi10}. In addition, the 350\,\micron~data are multiplied by 1.0067 to update our flux densities to the latest v7 calibration product. The SPIRE images required conversion from Jy~beam$^{-1}$ to Jy~sr$^{-1}$ where beam is the PSF area size of the instrument at the observed wavelengths taken for this analysis as 423 and 751 arcsec$^2$~beam$^{-1}$ for the 250\,$\mu$m and 350\,$\mu$m images respectively. SPIRE images are initially in point source relative spectral response function (RSRF) weighted intensities. To convert to extended source RSRF-weighted intensities we divide each image by the $K_{4p}$ terms given in the SPIRE handbook of (1.0113, 1.0087) for the 250\,$\mu$m and 350\,$\mu$m images. The images remain in RSRF-weighted intensities (ie. not converted to monochromatic intensities or color-corrected). Model band fluxes are computed by integrating the model spectra with each filter's spectral response function and thus account for the filter's effective wavelength and throughput, as well as the galaxy's varying spectral slope.

The sky background in each image is measured using 10 pixel wide boxes surrounding the system. For all of our \textit{Herschel}/PACS and \textit{Herschel}/SPIRE images, we subtracted the median value of the sky region pixels as we found values that were on order 1-2\,$\sigma$ higher (in the case of PACS images) or lower (in the case of our SPIRE images) than the standard deviations in the background pixels. All images are matched to a common resolution using kernels created from the PSFs from \citet{ani11}. They measured azimuthally averaged empirical PSFs for the \textit{Herschel} instruments. The images were all convolved to a matched gaussian PSF (FWHM=28\arcsec) slightly higher than the SPIRE 350\,\micron~image to ensure the power in the non-gaussian beam of SPIRE is conserved to the assumed gaussian beam \citep{ani11}.  All images are remapped to a common platescale of 10\arcsec\,pix$^{-1}$ using the WCS tools program \textsc{remap}\footnote{http://http://tdc-www.harvard.edu/wcstools/}. The resulting processed images are shown in the middle panel of Figure~\ref{fig:montageHerschel}.

\subsection{Optical and near/mid-infrared data}

Our optical ($B\,V\,R\,I\,H\alpha$) and mid-infrared (IRAC 3.6, 4.5, 5.6, 8.0 and MIPS 24\,$\mu$m) observations come from the \textit{Spitzer Infrared Nearby Galaxies Survey} (SINGS; \citealt{ken03}). Near-infrared $JHK$ imaging is from the \textit{2MASS Large Galaxy Atlas survey} \citep{jar03}. All images are converted to units of Jy~sr$^{-1}$ and we account for loss of light due to foreground dust attenuation from the Milky Way using the extinction values provided by NED\footnote{http://ned.ipac.caltech.edu/} using maps from \citet{sch98} for all bands up to and including the IRAC\,3.6\,$\mu$m image. 

We replace emission from foreground stars (in 10\arcsec~diameter apertures) with the median value of emission of the pixels contained in its surrounding annulus (4\arcsec~wide). The stars are identified from the 2MASS Point Source catalogue. In order to ensure we do not remove any HII regions belonging to the galaxy we only remove point sources with mid-infrared to optical light ratios comparable to stars, taking the criterium that the 8\,\micron~surface brightness cannot be greater than 1.5 times the $R$-band surface brightness. If the ratio is higher, I$_8$/I$_R>1.5$, the point source is an HII region (identified by its bright PAH emission at 8\,\micron) within the galaxy and is not removed. 

Also, we correct for any constant in the sky background in each image by placing 10 boxes, each 10 pixels wide throughout the sky in each image. If the median in this background value is larger than the standard deviation across the boxes, we subtracted this constant from the image. The narrowband H$\alpha$ image (not continuum subtracted) was the only image which required this step. Finally, we convolve the images with appropriate kernels to our chosen common resolution of 28\arcsec~assuming gaussian PSFs for the optical and near-infrared images. The FWHMs of the PSFs for the optical images are 2.0\arcsec, 1.9\arcsec, 1.6\arcsec, 1.9\arcsec, and 1.8\arcsec~for the $B, V, H\alpha, R, $ and $I$ bands respectively. For the NIR images, the FWHM of the PSFs are 3.3\arcsec, 3.2\arcsec, and 3.5\arcsec~for the $J$, $H$, and $K$ bands.

The \textit{Spitzer} images are convolved to a 28\arcsec~gaussian using kernels from \citet{ani11}. The FWHMs of the \textit{Spitzer} images are 1.90\arcsec, 1.81\arcsec, 2.11\arcsec, 2.82\arcsec~and 6.43\arcsec~ for the IRAC 3.6\,\micron, 4.5\,\micron, 5.6\,\micron, 8.0\,\micron~and MIPS 24\,\micron~images respectively. All images are remapped to a common platescale of 10\arcsec\,pix$^{-1}$ using the WCS tools program \textrm{remap}\footnote{http://http://tdc-www.harvard.edu/wcstools/}. For reference, we have included images of the galaxy in each band, before and after image processing in the appendix of this paper.

\subsection{Neutral and molecular gas}

Imaging of the neutral gas is provided by the HI Nearby Galaxy Survey (THINGS; \citealt{wal08}). Already in units of M$_\sun$\,pc$^{-2}$, we match plate scales and convolve the beam assumed to be a gaussian of FWHM=6\arcsec~to a 28\arcsec~beam to match our other observations. Molecular gas observations at CO(1-0) from the Nobeyama 45\,m telescope \citep{kod11} are also included in the analysis.  The CO(1-0) image, originally in antenna temperature units of K~km\,s$^{-1}$, is divided by 0.4 to convert to main beam temperature units. We then use a constant conversion factor of $X_{CO}=2.0\times10^{20}$~mol~cm$^{-2}$ (K~km~s$^{-1}$)$^{-1}$ \citep{str88} which corresponds to 3.2 M$_\sun$~pc$^{-2}$ (K~km~s$^{-1}$)$^{-1}$ \citep{wil09} to convert the CO intensities to an H$_2$ gas mass surface density. This image is then convolved from a 22\arcsec~beam to a 28\arcsec~beam to match all of other observations as well as changing the platescale to 10\arcsec\,pix$^{-1}$. Both images of the gas mass surface densities are shown in Figure~\ref{fig:gasimages}. The sum of both the neutral and molecular gas is shown on the right. All images are multiplied by 1.36 to account for helium and thus represent the total, not just hydrogen, gas in the galaxy. We use the total gas mass surface density in our comparisons to the stellar and dust mass surface densities.

\begin{figure*}[t]
\begin{center}
\includegraphics[clip=true,angle=-90,width=6in]{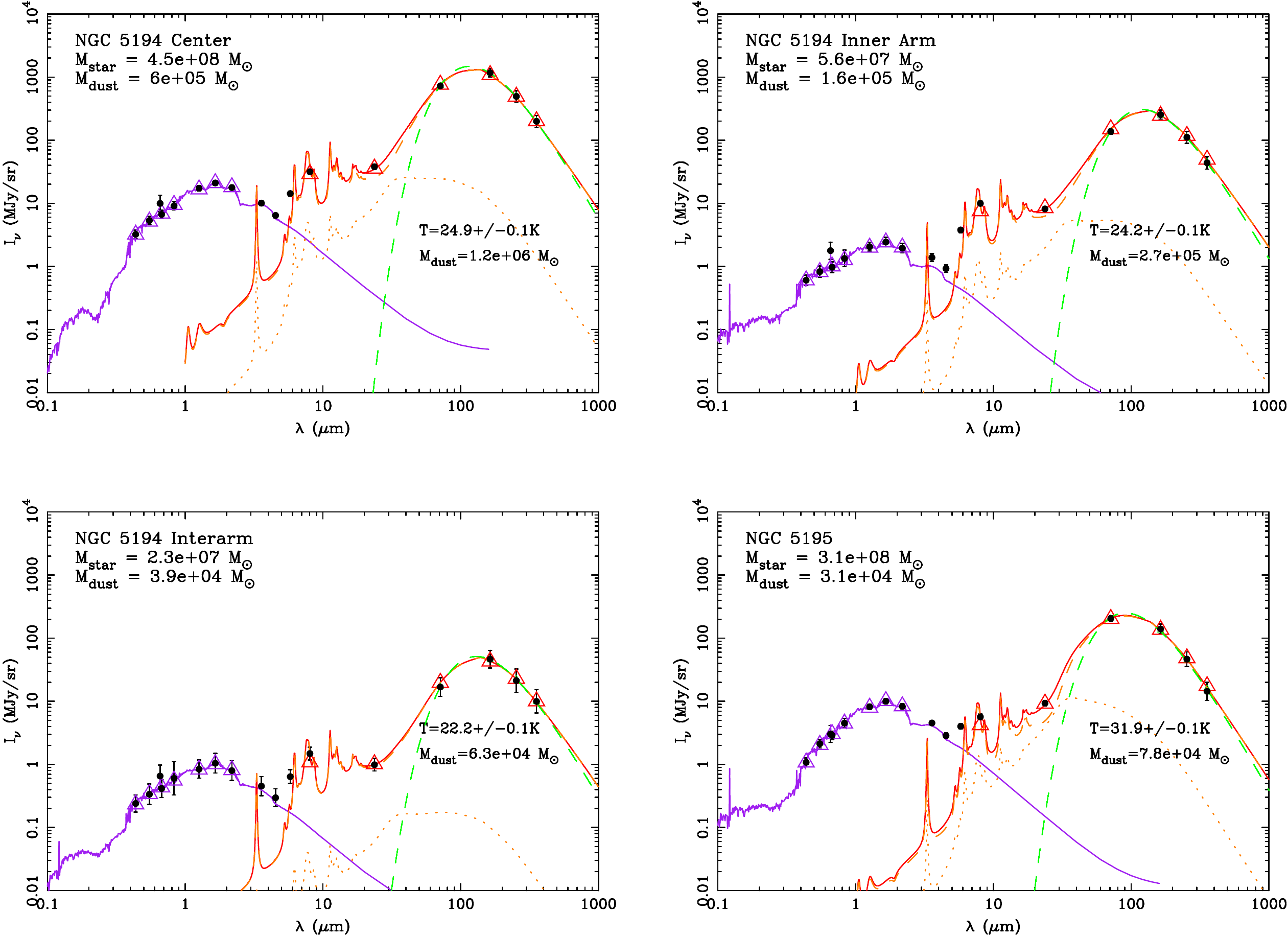}
\caption{We plot four example SED fits for single representative pixels from the nuclear region (top-left), inner arm (top-right) and interarm (bottom-left) regions of NGC\,5194 and in the bottom-right, a pixel from the center of the post-starburst NGC\,5195. Model band fluxes of each best-fit SED are shown in open triangles. The purple curve is the best fit stellar emission model (PEGASE.2, \citealt{fio97}) to the $B$-$K$ images (H$\alpha$ excluded) and the red curve is the best fit dust emission model from DL07 to the 5.6-350\,\micron~images. We separate the PDR component (dotted orange line) from the dominant ISM component (dashed line) for each DL07 model. High PDR fractions are found on the spiral arm of NGC\,5194 (top-right) and in the central region of NGC\,5195 (bottom-right). For comparison, we also show the best fitting modified greybody (dashed green curve) for both examples and give their characteristic temperatures and dust mass. Note that the dust and stellar SEDs have not been added together and neither SED was constrained by the IRAC images at 3.6\,\micron~ and 4.5\,\micron.}
\label{fig:SED}
\end{center}
\end{figure*}

\subsection{Uncertainties and qualifying signal-to-noise criteria}\label{s:errors}

We generally considered two types of uncertainties in each pixel value of each image: the error due to the uncertainty in the sky background and also the calibration uncertainty. The uncertainty in the sky background is measured in each image after background subtraction and convolving to lower spatial resolution by selecting ten 100\,\arcsec~wide boxes throughout the image. The standard deviation in the median value of the pixels in each box leads to our uncertainty in the sky background. Calibration uncertainty is taken from the various instrumental handbooks or survey description papers. Calibration uncertainties in the optical data are assumed to be 5\%, except for a 10\% uncertainty in the narrowband H$\alpha$ image. Both 2MASS and \textit{Spitzer} images have calibration uncertainties on the order of 3\%. Uncertainties for \textit{Herschel} observations are 3\% and 5\% at 70\,$\mu$m and 160\,$\mu$m and 7\% at 250\,$\mu$m and 350\,$\mu$m. In addition, error maps of the \textit{Herschel} images were also included in the analysis. These images were processed to a common spatial scale as the images above by taking the square root of the convolution of the square of the error image with the square of the kernel. Furthermore, the error images were matched to the 10\arcsec/pix platescale. Uncertainties in each pixel were added in quadrature to produce a combined error map at each bandpass.

We want to include as many pixels in our analysis as possible and at the same time exclude regions in which the majority of pixels have low signal-to-noise (S/N) levels. A fit to the SED of a pixel was performed if at least 5 bands had S/N$>3$ in the optical/near-infrared data from $B$ through $K$-band. We also required S/N$>3$ in at least 3 bands for the infrared data from 5.6\,$\mu$m to 350\,$\mu$m. In practice this selection was dominated by the relatively less sensitive infrared data; every pixel which satisfied the infrared criteria had S/N$>3$ in all optical bands.

\begin{figure*}[tb]
\begin{center}
\includegraphics[angle=-90,width=5.5in]{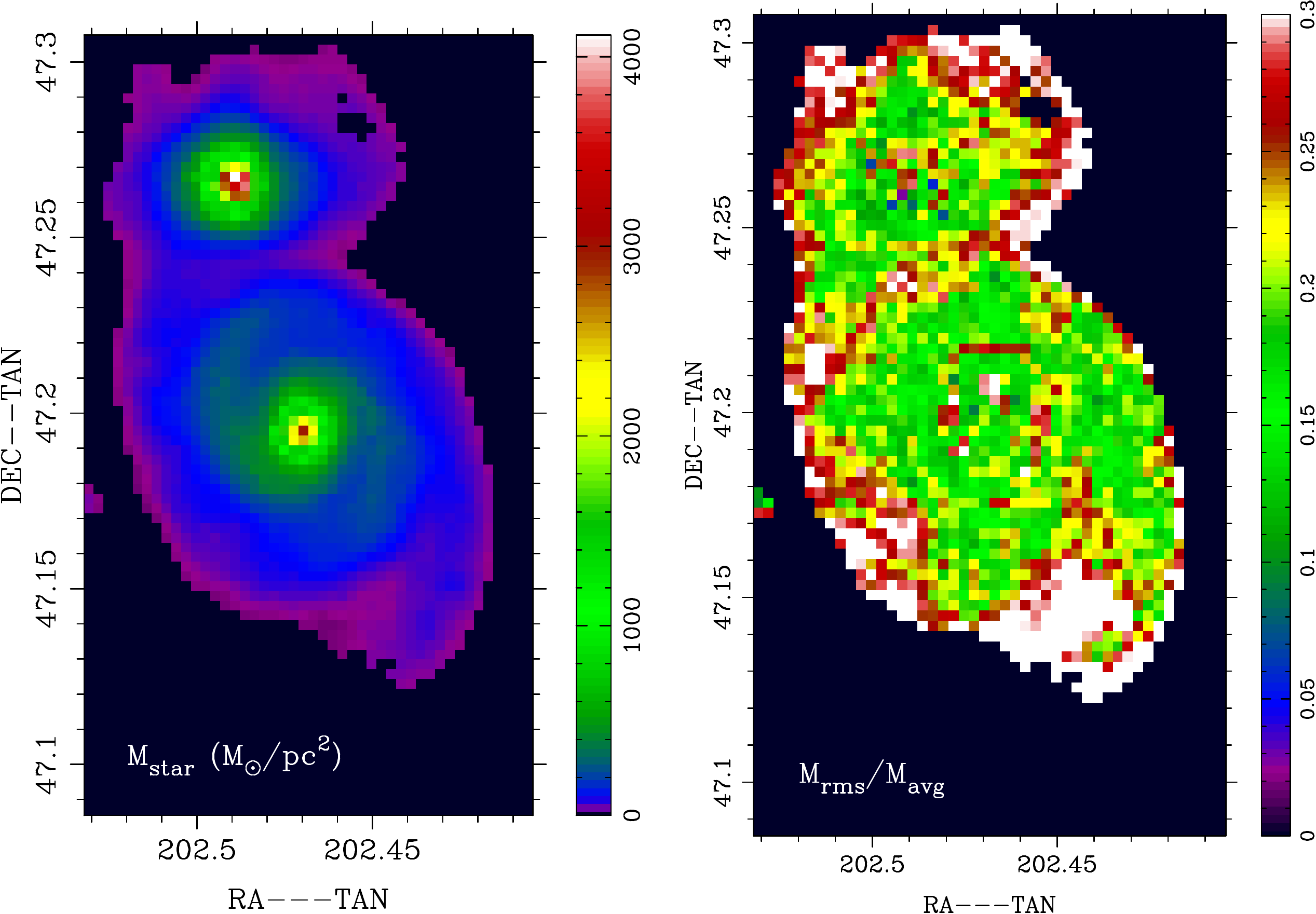}
\caption{Stellar mass surface density in M$_\sun$\,pc$^{-2}$ is shown in the left panel and the normalized uncertainty in the pixel mass determination is shown in the right panel as M$_\mathrm{rms}$/M$_\mathrm{avg}$.  }
\label{fig:stellarmass}
\end{center}
\end{figure*}

\section{SED libraries and fitting results}\label{s:methods}

Although models exist (eg. \citealt{dev99}, \citealt{dac08}, \citealt{nol09}, \citealt{pop11}) which describe both stellar and dust emission together, they tend to be either computationally expensive (requiring many fitted parameters and slow codes computing the uncertainties in these parameters) or lack the most up-to-date SEDs. Unlike today's common stellar population synthesis (SPS) codes -- PEGASE.2 \citep{fio97}, GALEXEV \citep{bru03}) and FSPS \citep{con09} -- these self-consistent stellar and dust emission codes are not extensively tested and their use is not widespread. For this reason, we opt to fit the stellar SED and dust SED components separately. This approach is similar to that of \cite{nol09} but by considering both components separately we can fit the SED at a much faster speed; this is primarily out of necessity since we are dealing with a very large dataset (20 images with 4000 pixels each). The downfalls of this approach are first that energy is not necessarily conserved from the optical to IR (such that the re-processed infrared emission is linked to the ionizing radiation provided by the stellar populations), something that both MAGPHYS \citep{dac08} and CIGALE \citep{nol09} do achieve and second the near- to mid-infrared regime from $\sim1-5$\,\micron~where both dust and stellar emission is significant cannot be used to constrain the fit. However, because both pre-main-sequence and post-main-sequence stellar emission is not well calibrated in the SPS models at these wavelengths, nor is PAH emission, it is not ideal to fit this regime in any case.

In our method, we first fit our $B$ through $K$ band photometry to stellar emission models using standard SPS codes as we describe in \S\ref{s:SPS}. Using the predicted stellar emission from these fits, we are able to subtract stellar emission from the mid-infrared images at 5.6\,$\mu$m and 8.0\,$\mu$m. As we will describe in \S\ref{s:dustmodels}, we then fit the 5.6 through 350\,$\mu$m images to the mid- through far-infrared dust emission models from \citet{dra07} (hereafter DL07), as well as fitting a single temperature modified blackbody model.


\subsection{Stellar SEDs}\label{s:SPS}

A number of codes exist for a user to generate a model galaxy SED. The choice of parameters and assumptions that are entered into the codes can lead to systematic differences in the output parameters. Stellar masses are most robust and can be predicted to within 0.2 dex \citep{muz09,con10a} and often to higher precision if near-infrared photometric bands are included, although masses will vary systematically depending on the choice of initial mass function (IMF). Other parameters tend to have larger dispersions across different codes \citep{muz09}. If uncertainties in the inputs (such as evolutionary isochrones, stellar libraries, binary fraction) of the code are considered, then masses are only good to within $\sim$0.3\,dex \citep{con10a}. For a more intensive discussion on systematic differences and known uncertainties of SPS models, the reader is referred to papers by \citet{muz09} and \citet{con09,con10a}. 

In this study we take a more empirical approach and try to compare the best fit model SED parameters to information from other observations and independent studies. For example, we have some prior knowledge regarding the star forming history of the system based on isochrones (which have some of the same uncertainties as those found in SPS codes) fit to individuals stars, most recently from \citet{tik09}. They indicate an enhancement in star formation around 400\,Myr ago in both galaxies. Independently, previous dynamical studies of the system provide a consistent temporal view of the system. The motions of gas and stars in the galaxies suggest at least one encounter between the two galaxies 300-500\,Myr ago from two independent studies from \citet{sal00} and \citet{dob10}. This prior knowledge of the star formation history could be used to constrain our fitting, but instead we have opted to keep the star formation history (SFH), stellar population ages and dust attenuation as free parameters to see how they compare to these independent studies.

\begin{figure*}[t]
\begin{center}
\includegraphics[angle=-90,width=2.3in]{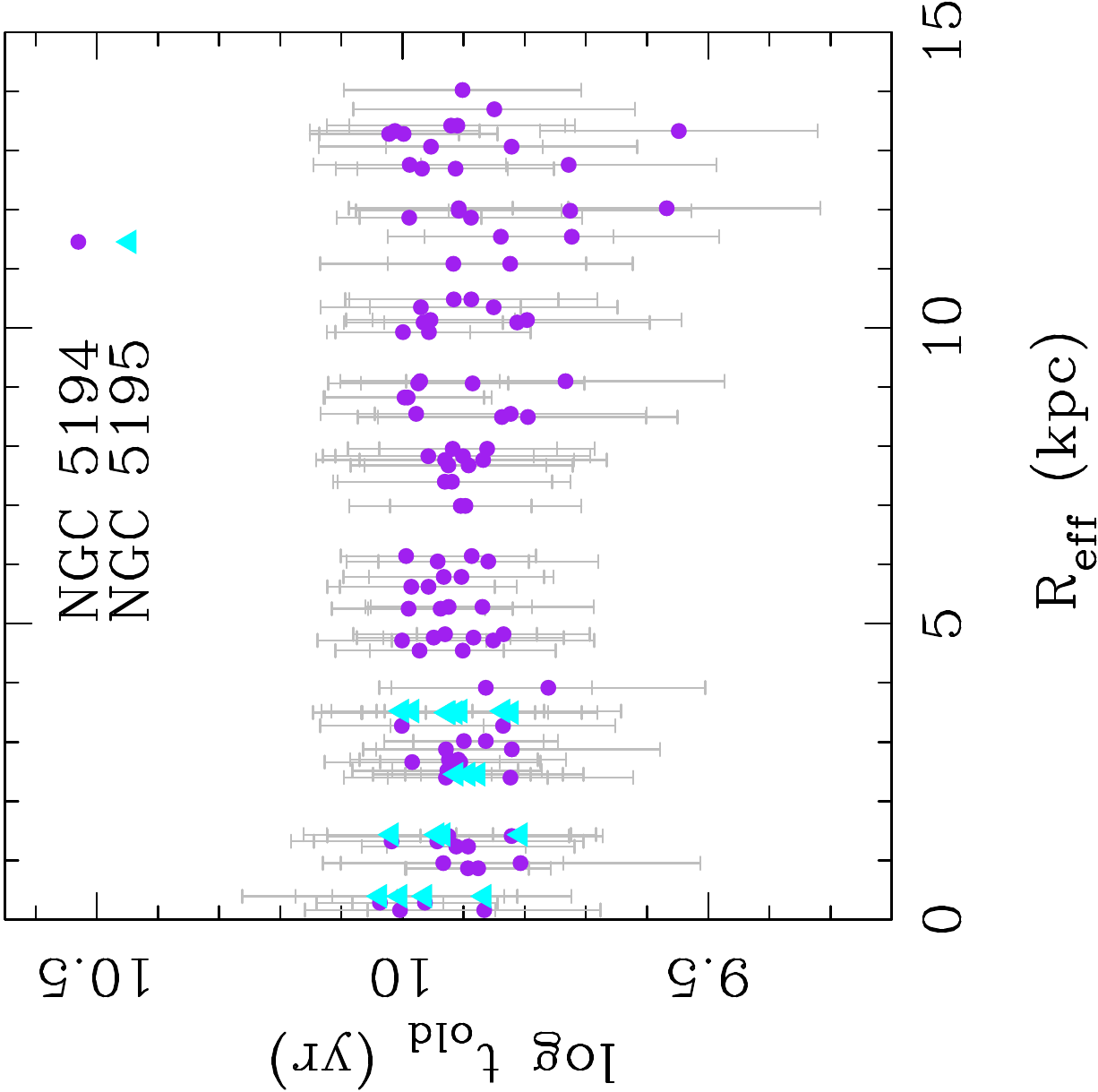}
\includegraphics[angle=-90,width=2.3in]{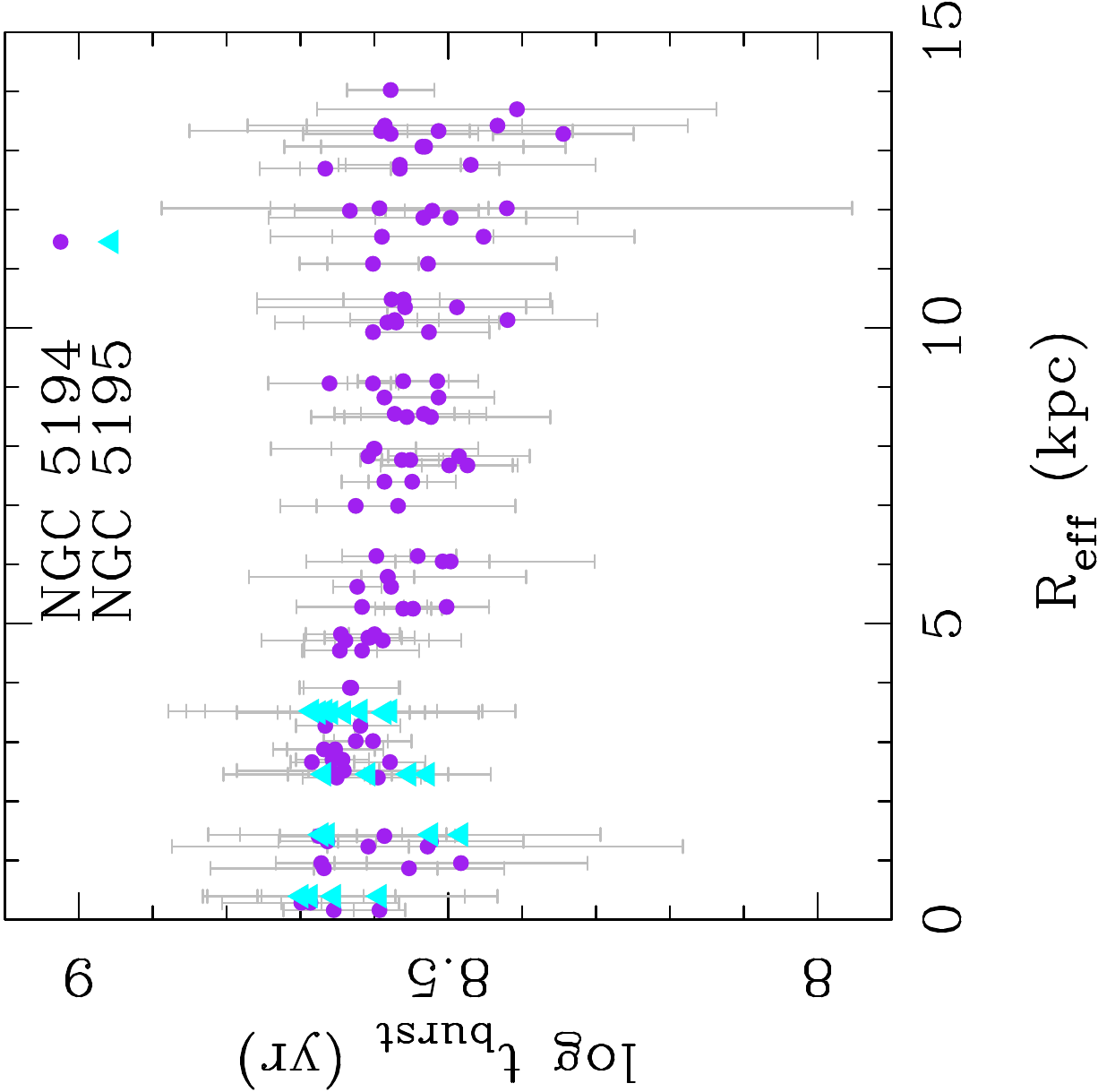}
\includegraphics[angle=-90,width=2.3in]{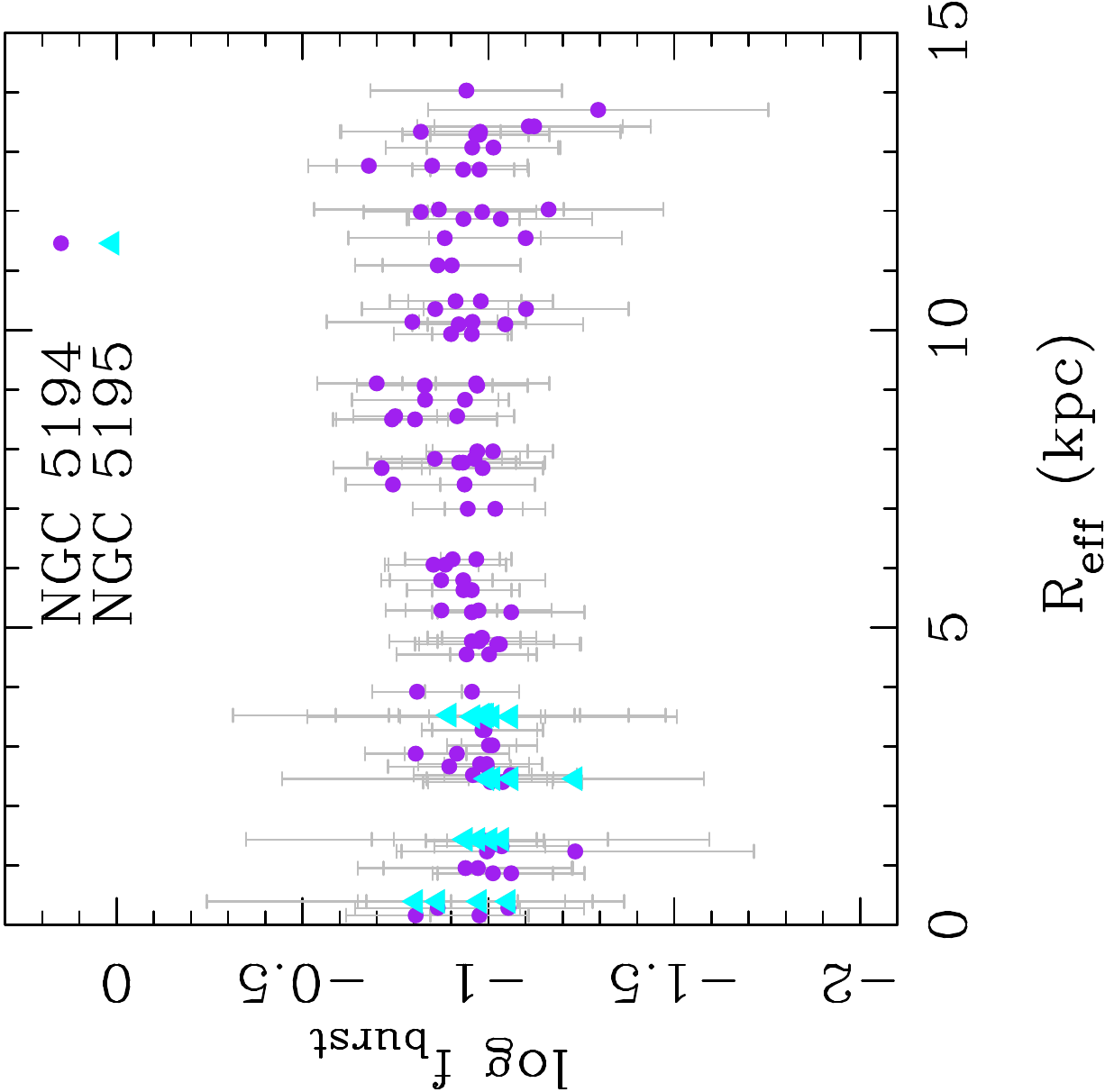}

\caption{Radial profiles of SED derived star formation history parameters for NGC\,5194 in purple dots and NGC\,5195 in cyan triangles. \textit{Left:} The age of the old stellar population is $\log t_{\star} = 9.9 \pm 0.2$. \textit{Middle:} Age of the burst component. A secondary star formation burst period happened 300-600\,Myr ago. \textit{Right:} The mass fraction of the secondary burst component relative to the older primary stellar population is $\sim5-15$\% throughout the galaxy. Note that the deviation from pixel to pixel tends to be less than the uncertainty in the parameters indicating that at the spatial scale of our analysis (28\arcsec), the stars are well mixed.  }
\label{fig:SEDparams}
\end{center}
\end{figure*}

%
Our stellar libraries are generated by the extensively tested \textsf{PEGASE.2} \citep{fio97} code and are an extension of work done in \citet{gla04} and \citet{men09} and include four metallicities: Z$=\{0.004,0.008,0.02,0.05\}$. We assume no evolution in the metallicity and no gas infall, although this is likely not the best explanation for NGC\,5195 in particular since it appears to be accreting material from NGC\,5194. Our adopted IMF is that of \citet{kro01}. The SFH is modelled as a sum of a continuous star forming population described by a tau exponential and an additional burst of star formation whose amplitude and age are both free parameters. The SFH of the continuous population is modelled as an exponentially decreasing function of time ($\exp^{-t/\tau}$), parameterized by an e-folding timescale of $\tau=\{100,500,1000,5000,500000\}$\,Myr where 100\,Myr is a short burst of star formation or roughly a simple stellar population and the largest timescale of 500 Gyr represents a constant star formation history. The burst component is modelled as a tau-exponential with $\tau=100$\,Myr. The time since the star formation burst is allowed to be an additional parameter and was given the largest range and sampling at steps of 50\,Myr from 50 to 1000\,Myr and then steps of 250\,Myr from 1\,Gyr to 5\,Gyr and then steps of 500\,Myr from 5\,Gyr to 13\,Gyr. The age of the galaxy could be anywhere between 1 Gyr and 13\,Gyr with fairly high temporal sampling as well. The mass of the burst compared to the mass of the continuous star forming population (CSP) ranged as log\,($M_\mathrm{burst}/M_\mathrm{CSP})=\{-3,-2,-1,-0.75,-0.5,-0.25,-0.1,0,0.1\}$. Thus, the burst SSP could be anywhere from 0.001 to 1.25 times the mass of the primary CSP.



As we did not include UV observations in this analysis, the nuances of the 2175\,\AA~bump did not have to be considered in the choice of a dust attenuation law. We opted to use the Milky Way dust attenuation curve from \citet{pei92} and only considered a screen type geometry of the dust. This law, compared with those of \citet{cal94} and \citet{cha00} for example, more accurately describes the dust attenuation of stars in the Milky Way \citep{ind05} observed in the near- and mid-infrared \citep{men10}. Curves that represent starburst galaxies \citep{cal94,cha00} incorrectly model the dust attenuation because they have not accounted for non-stellar emission that is quite large in star forming regions \citep{men10} which becomes significant at wavelengths beyond $\sim$1\,$\mu$m \citep{men09}. The amplitude of the dust attenuation was an additional free parameter ranging from $A_{V}=0-2$\,mag in steps of 0.25\,mag. 

We place the modelled galaxies at a distance of 9.9\,Mpc to match the system's distance and integrate the synthetic SEDs across each observed band's filter response curve to get theoretical band fluxes. A least-squares comparison between observed and modelled band fluxes in each pixel selects the best fit model from our suite of stellar models to our $BVIRJHK$ maps. The stellar mass in each pixel is inferred from the least-squares normalization of the model to observed band fluxes.

Following standard Monte Carlo error simulations, our errors are derived for individual parameters by running our fitting routine one hundred times, each time allowing the pixel values in each band to sample the gaussian probability distribution of each measurement. The uncertainty, or 1\,$\sigma$ spread in the gaussian distribution, accounts for calibration uncertainty, sky background limitations and other uncertainties as described in \S\ref{s:errors}. The mean and standard deviation of the best fit parameters from all trials lead to our best estimate parameter value and its uncertainty in each pixel. 

Figure~\ref{fig:SED} shows four examples of fits to individual pixels in different regions of NGC\,5194/5. The purple curve shows the best-fit SED to the $B$-$K$ pixels. The large decrease in amplitude from the nuclear region of NGC\,5194 (top-left) to the arm regions (top-right and bottom-left) is due to a decrease in stellar mass density. Overall we find that the models fit the observations to better than 10\% and are consistent with the observations within our estimated uncertainties. In an appendix, we show quality of fit images for the optical, near- and mid-infrared images. Consistently we have very good fits of the stellar emission, with $\chi^2_\mathrm{reduced}\sim1$ for most of NGC\,5195 and  $\chi^2_\mathrm{reduced}\sim0.5-1$ for NGC\,5194. Only the H$\alpha$ image (which we don't use to constrain the SPS model) is slightly mismatched. We have simplified the SFH in the model to consist of just two components. But we know that for NGC\,5194, it consists of at least two continually star forming episodes: one old stellar population and then one that has been forming stars since at least its encounter with NGC\,5195 and likely before given its gas fraction and bulge-disk ratio. But our simplified SFH only consists of one continually star forming population and then a burst population. Because the current star formation is not being accounted for properly, neither is its associated emission in the H$\alpha$ narrowband image. 

Maps of the stellar mass surface density and fractional uncertainties (M$_\star,\mathrm{rms}$/M$_\star,\mathrm{avg}$) are shown in Figure~\ref{fig:stellarmass}. On average, stellar masses in each pixel can be found to a precision of 0.1\,dex based on our Monte Carlo simulations. The distribution of stellar mass radially decreases from the nuclear regions of both galaxies and there is a slight enhancement of stellar mass along the spiral arms in NGC\,5194. 
\begin{figure*}[t]
\centering
\includegraphics[angle=-90,width=2.2in]{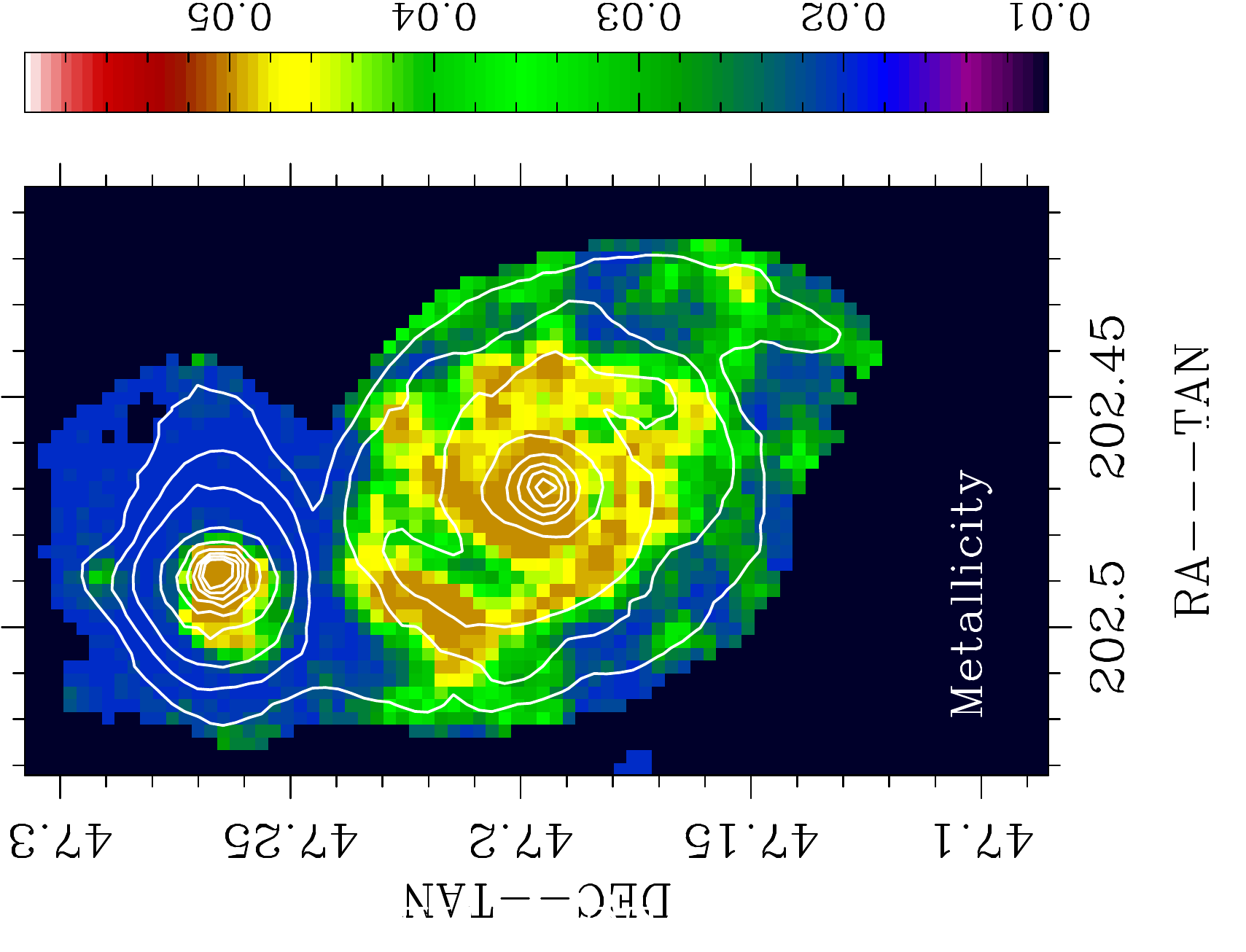}
\includegraphics[angle=-90,width=2.7in]{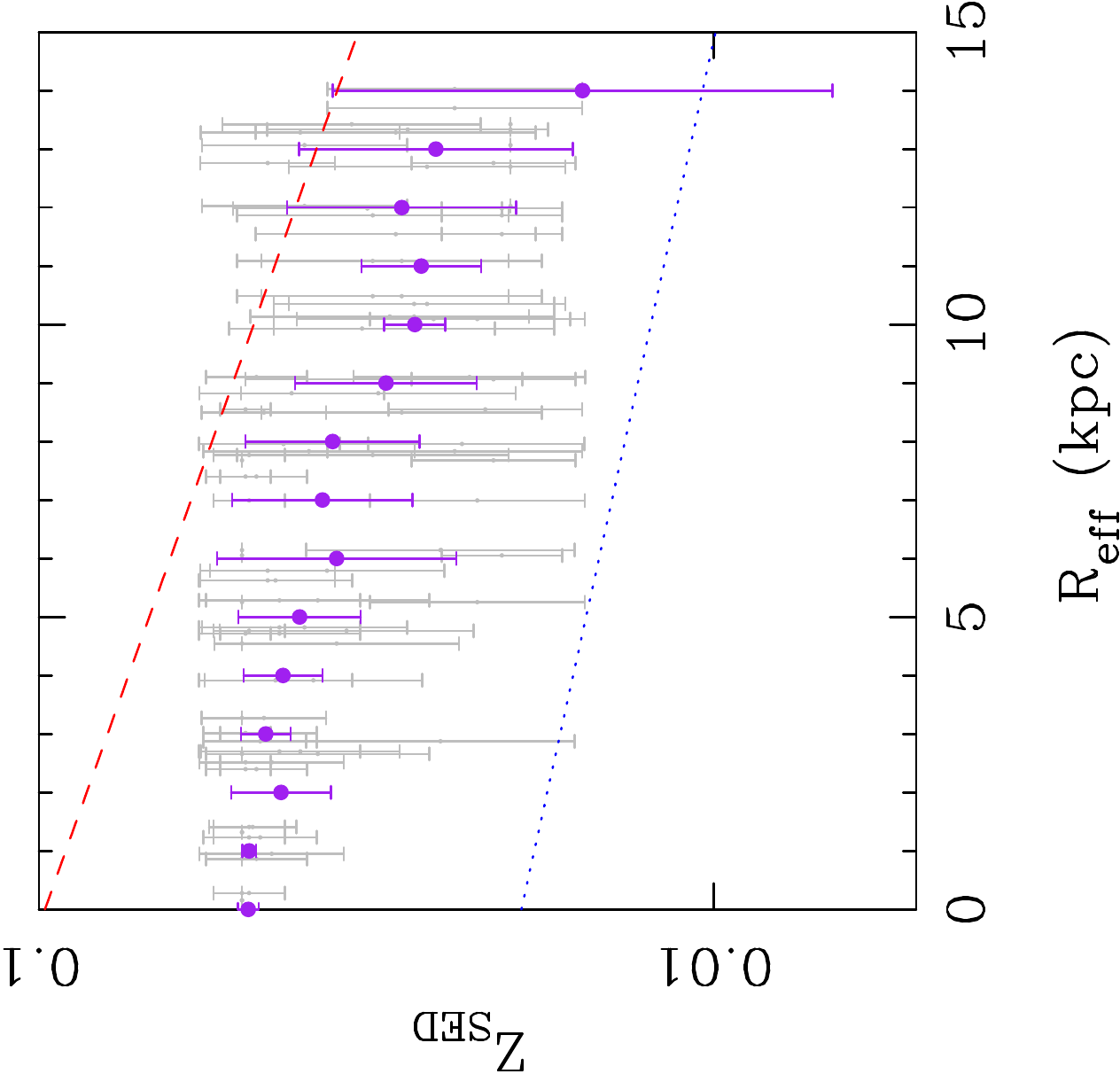}
\caption{The left panel displays a map of metallicity across the NGC\,5194/5195 system showing increased metallicity in the centers of both galaxies, relative to larger radial distances. The stellar mass surface density of the system is shown in contours. Uncertainties on this parameter tend to be quite high as can be better seen in the 1D radial plot of metallicity for NGC\,5194 shown on the right. Grey points are for individual 30\arcsec~pixels and the purple points are the pixels binned radially The metallicity tends to decrease with radial distance consistent with results from \citet{mou10}. Their abundance measurements are calibrated using two methods. Here we show best-fit lines to the radial abundance profile derived using two different calibration methods:  the \citet{kob04} calibration method in dashed red and the \citet{pil05} calibration method in dotted blue. }
\label{fig:metals}
\end{figure*}

\medskip

\noindent \textit{Star formation history parameters:} In Figure~\ref{fig:SEDparams} we give radial plots of parameters describing the star formation history (SFH) of the system. The profiles go out to a radial distance of 295\arcsec~for NGC\,5194 and 115\arcsec~for NGC\,5195. Beyond these radii the two systems significantly overlap. The left panel shows the age of the old stellar population. There is no significant radial dependence and both systems have similar old stellar population ages of $\log t_{\star} [yr] = 9.9\pm0.2$. Thus, the age is not well constrained for the system, ranging from $\sim$7--10\,Gyr. Somewhat reassuringly, the SED fits reveal, as we show in the middle panel of Figure~\ref{fig:SEDparams}, that there was an increased episode of star formation across the system. For the inner radius of 295\arcsec~of NGC\,5194 we find a burst occurred $3.8\pm0.9\times10^8$\,yrs ago and for the inner 115\arcsec~of NGC\,5195 the average time since the burst is $4.2\pm0.8\times10^8$\,yrs ago, consistent with the dynamical models of \citet{sal00} and \citet{dob10}, and the color-magnitude diagrams of individual red giant branch stars seen by \textit{HST} \citep{tik09}. Our analysis suggests the fraction of mass, relative to the primary population, formed in the burst in NGC\,5194 is $0.12\pm0.05$ and in NGC\,5195 is $0.10\pm0.05$. A radial profile of the burst fraction for both galaxies is shown in the right panel of Figure~\ref{fig:SEDparams} and again, no spatial dependence is found. 

\medskip

\noindent \textit{Metallicity:} While the SFH parameters are generally well mixed in the galaxy at the spatial resolution of this work ($\sim1$\,kpc), the metallicity of the stars shows a trend of decreasing metallicity values with radius. In Figure~\ref{fig:metals}, we show a map of the metallicity, where the metallicity transitions from being super solar in the nuclear regions of both galaxies (topping out at Z=0.05, our maximum fitted metallicity value), to being near solar (Z=0.02) on the outer galaxy edges. The uncertainty from our Monte Carlo simulations is quite large because of the poor sampling of this parameter available by PEGASE.2. 

The right plot shows the 1D radial profile of the metallicity for NGC\,5194. Our results are consistent, but not as precise, with radial metallicity profiles from \citet{bre04} and \citet{mou10} who also find metallicity decreases with radius in NGC\,5194. Most late-type galaxies exhibit decreasing metallicity gradients \citep{mou10}, whose slopes depend on the calibration method to convert line strengths to abundances and differ from galaxy to galaxy. Both \citet{bre04} and \citet{mou10} measured the metallicity radial profile in NGC\,5194 using abundances derived from optical spectroscopy of HII regions. We convert the best-fit abundance profiles from \citet{mou10} to a metallicity radial profile (assuming $Z_\sun=0.02$) using the value for solar metallicity ($12+\log(O/H)_\sun = 8.69\pm0.05$) from \citet{asp09}. \citet{mou10} use two calibration methods to convert line strength to chemical abundances. We plot both calibration methods in Figure~\ref{fig:metals}. The red dashed line has been converted following the \citet{kob04} calibration method and the blue dotted line following the \citet{pil05} calibration method. Our values lie somewhere in the middle, but the same trend in decreasing metallicity is seen. Systematic offsets result from our assumed solar metallicity as well as the calibration method. The reader is referred to \citet{mou10} for more precise metallicity results. 

\medskip

\begin{figure}[t]
\centering
\includegraphics[angle=-90,width=2.7in]{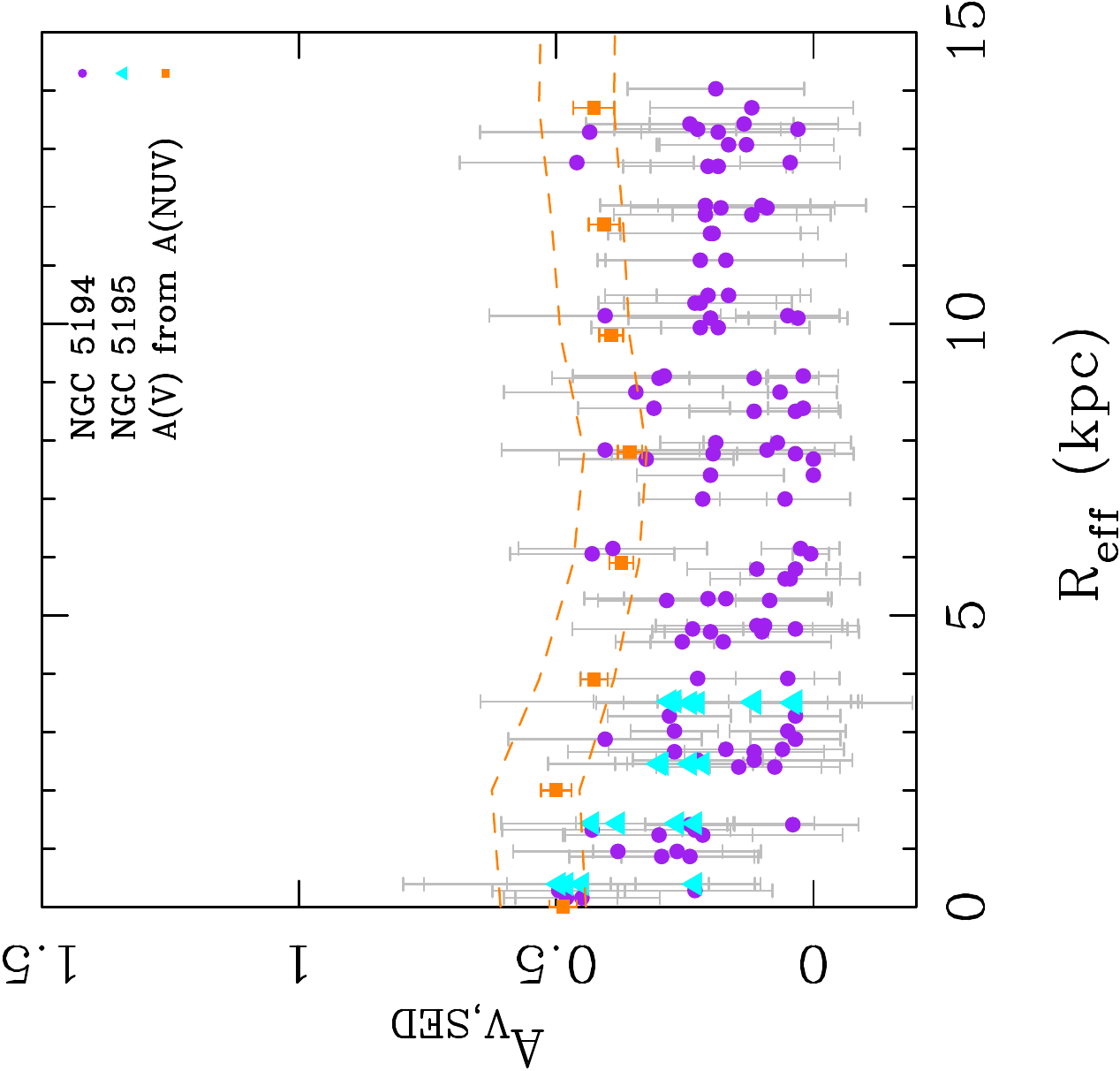}
\caption{Radial profiles of visible dust attenuation, A$_\mathrm{v}$ in units of mags shown for NGC\,5194 (purple dots) and NGC\,5195 (cyan trianges). A$_\mathrm{v}$ is a fitted parameter in our SED stellar population fitting analysis. For comparison, the orange squares show the measured visual extinction from \citealt{mun09} inferred from their measured near-UV attenuation. The dashed orange line show the systematic uncertainty due to the 2175\,\AA~bump. }
\label{fig:dustex}
\end{figure}

\noindent \textit{Dust attenuation:} SPS models also include a parameter that quantifies the intrinsic dust attenuation. Figure~\ref{fig:dustex} shows a radial plot of the dust attenuation for each galaxy. There is enhanced dust attenuation in the central regions of both galaxies which decreases with radius from 0.5 mag of visual dust attenuation in the center down to about 0.2 mag.  For comparison we plot the radial profile of dust extinction for NGC\,5194 measured in the near-UV by \citet{mun09}. We convert from A(NUV) to A(V) using a factor of 3 based on the dust extinction models of \citet{pei92} and \citet{dra03}  but caution that this attenuation ratio is sensitive to the UV bump at 2175\,\AA~and can range anywhere from $\sim 2.4$ to $3.3$. The dashed orange lines on Figure~\ref{fig:dustex} demonstrate the systematic uncertainty due to this conversion factor. Error bars show the precision in the measurement from \citet{mun09} which is more precise than ours and generally higher. Dust attenuation is modestly underestimated likely because of exclusion of UV and $U$-band data in our analysis.

\subsection{Dust SEDs}\label{s:dustmodels}

We use the modelled stellar maps of the galaxy to subtract the contribution of stellar emission from our mid-infrared images of the IRAC 5.6\,$\mu$m  and 8.0\,$\mu$m bands. These resulting images, which probe emission related to polycyclic aromatic hydrocarbons (PAHs), and our other infrared (24-350\,$\mu$m) images are fit using the two-component emission model put forth by \cite{dra07} (hereafter DL07) as well as a conventional single temperature modified blackbody component.

\subsubsection{Draine and Li 2007 dust models}

\begin{figure*}[tb]
\begin{center}
\includegraphics[angle=-90,width=5.5in]{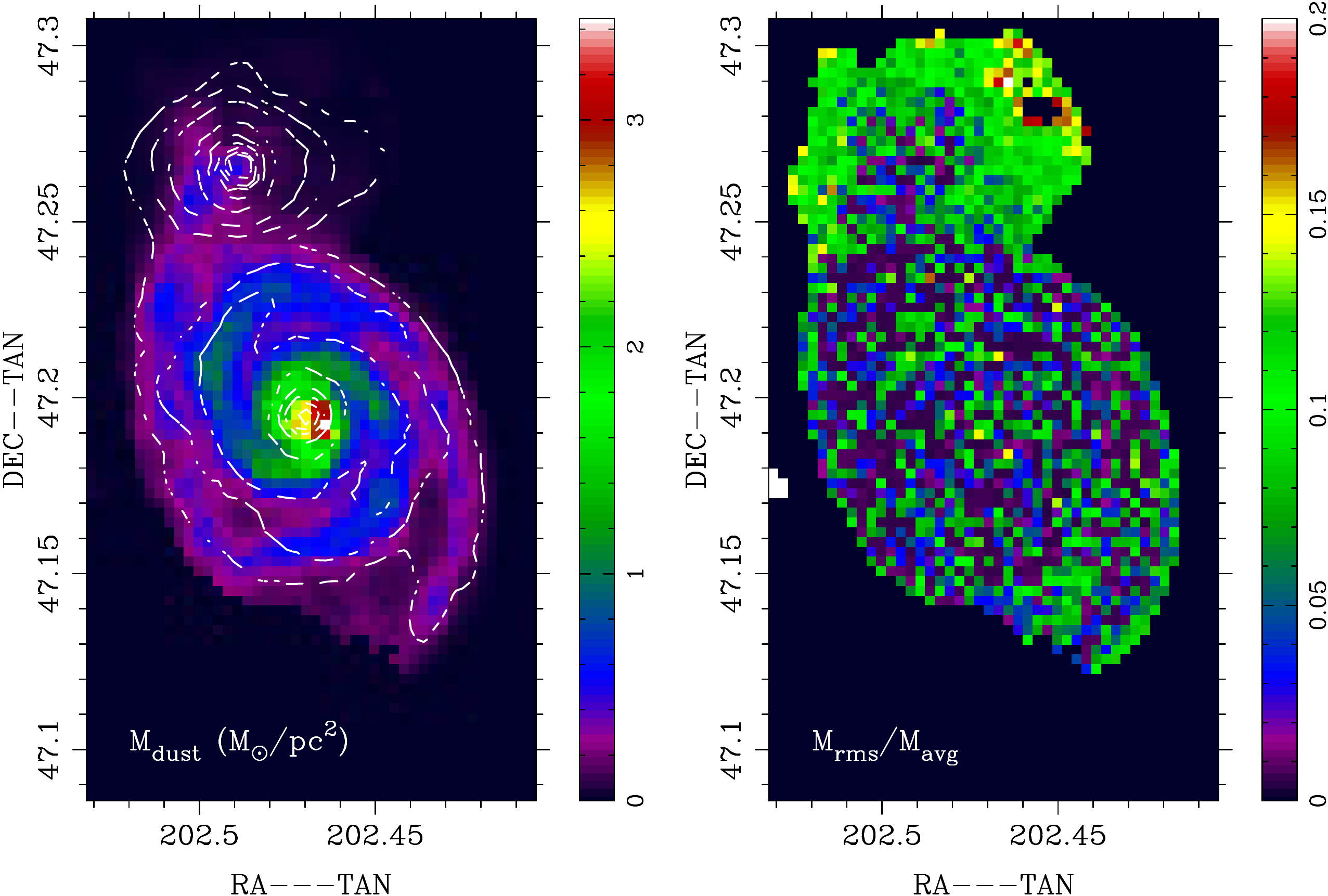}
\caption{Maps of the dust mass surface density in M$_\sun$\,pc$^{-2}$ (left) with the stellar mass surface density from Figure~\ref{fig:stellarmass} shown in contours at levels of 50, 100, 200, 500, 1000, 1500, 2000, 2500 and 3000 M$_\sun$\,pc$^{-2}$. The normalized uncertainty determined from our Monte Carlo simulations in the pixel mass determination is shown in the right panel as M$_\mathrm{rms}$/M$_\mathrm{avg}$.  }
\label{fig:dustmass}
\end{center}
\end{figure*}

DL07 model the mid- through far-infrared emission due to PAHs, very small dust grains and larger dust grains expressed as a function of the underlying interstellar radiation field (ISRF). The dust is assumed to consist of a mixture of carbonaceous grains and amorphous silicate grains, with size distributions that are consistent with the observed wavelength-dependent dust attenuation in the local Milky Way \citep{wei01}, and allows for varying PAH abundances. The infrared emission is a function of the illuminating radiation field which in this model is assumed to come from two components. Most of the emission comes from diffuse dust heated by the ISRF. The bulk of the dust mass in a galaxy is in this component. A second component represents emission from photodissociation regions (PDRs) where massive star formation creates much stronger radiation fields, leading to relatively higher emission at shorter infrared wavelengths. The diffuse ISRF is scaled to match the spectrum of the ISRF of the MW. The amplitude of the diffuse ISRF is a fitted parameter, U$_\mathrm{min}$, normalized to the amplitude of the ISRF of the MW. The radiation field of the PDR component consists of a range of energies from U$_\mathrm{min}$ to U$_\mathrm{max}$, where U$_\mathrm{max}$ can be an additional fitted parameter and U$_\mathrm{min}$ is the ISRF of the diffuse component. We opt to fix U$_\mathrm{max}=10^6$\,U$_\mathrm{MW}$ as \citet{dra07b} showed that it was not a highly sensitive parameter and that most galaxies from the SINGS survey were fit well by this fixed value parameter. 

\begin{figure*}[t]
\begin{center}
\includegraphics[angle=-90,width=2.3in]{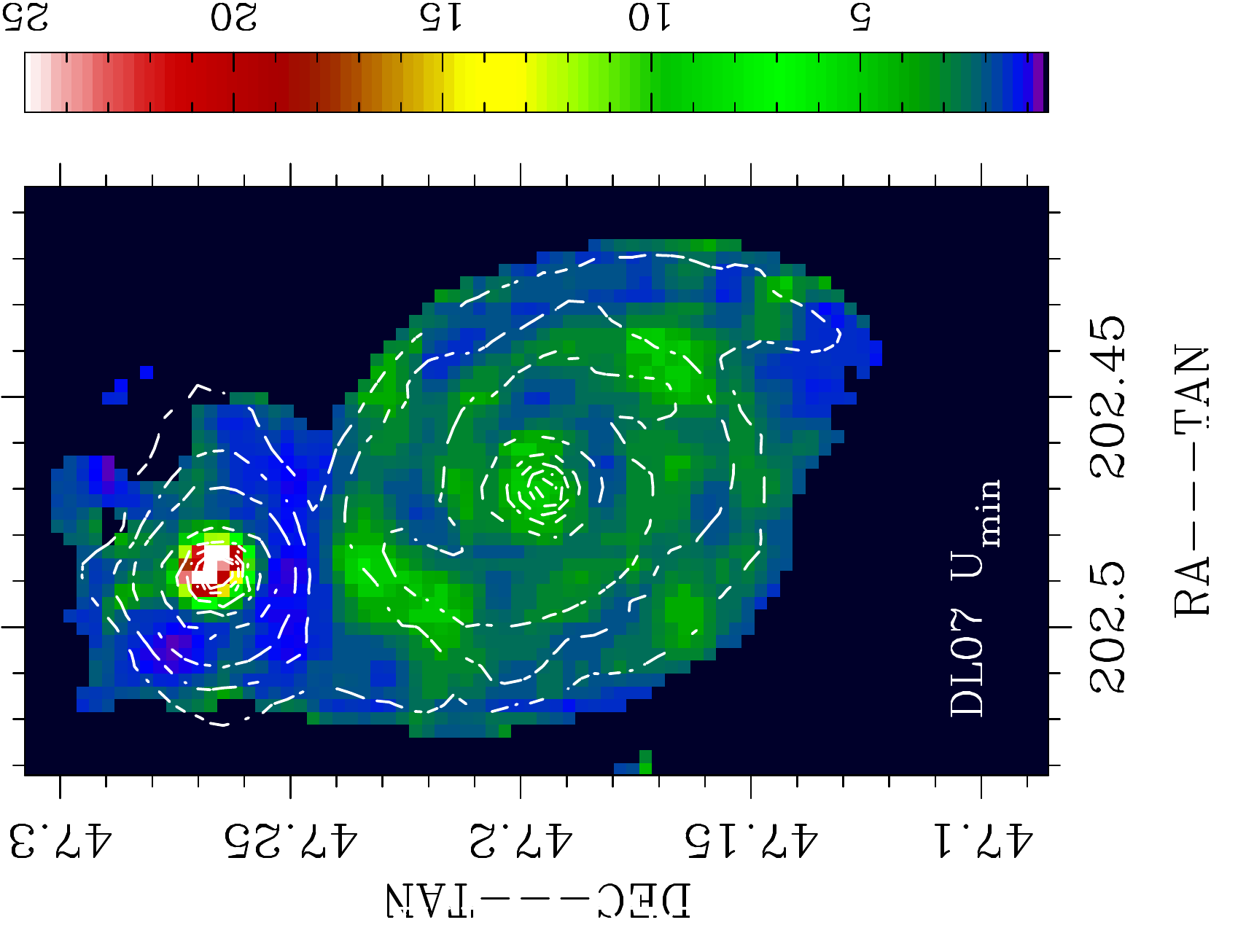}
\includegraphics[angle=-90,width=2.3in]{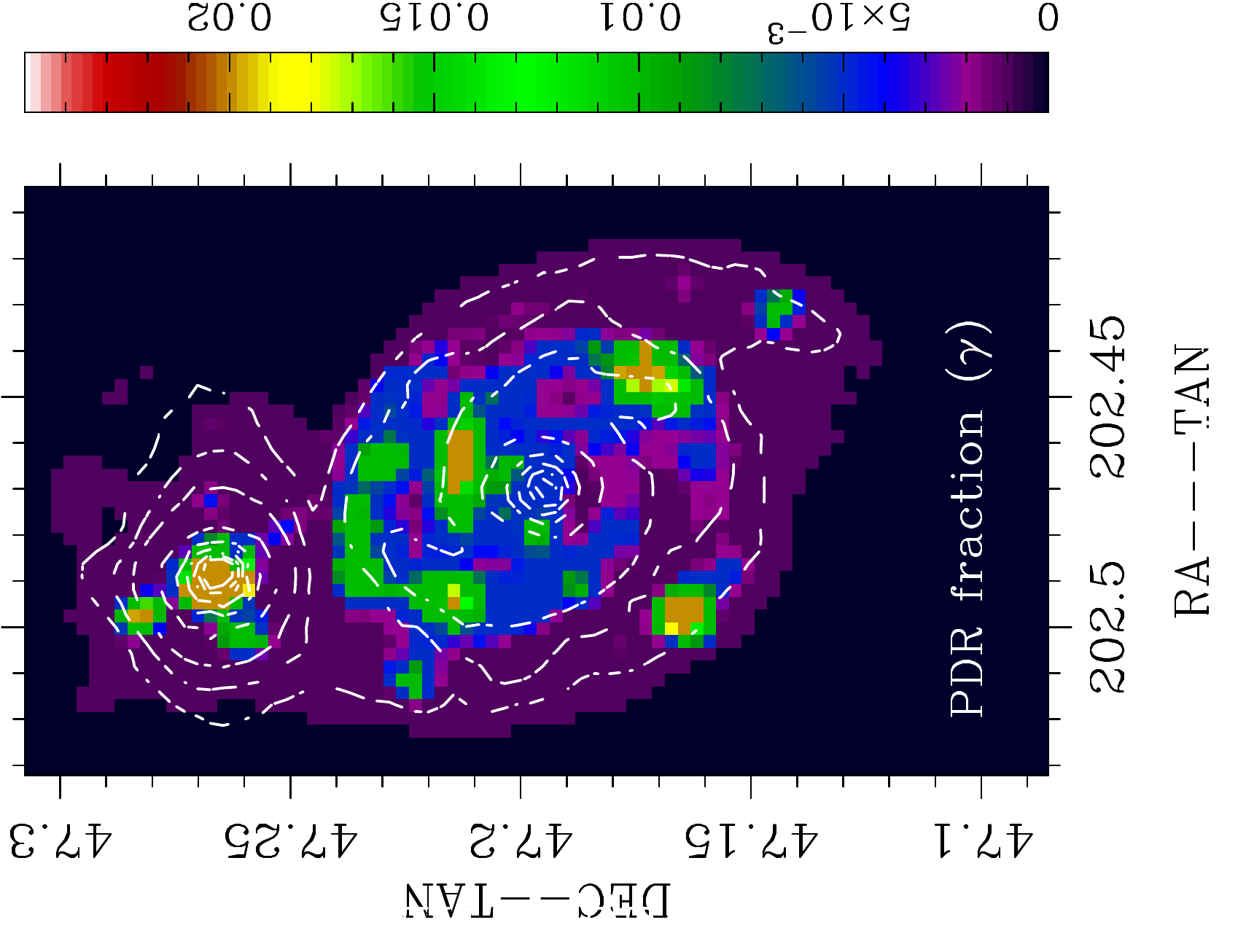}
\includegraphics[angle=-90,width=2.3in]{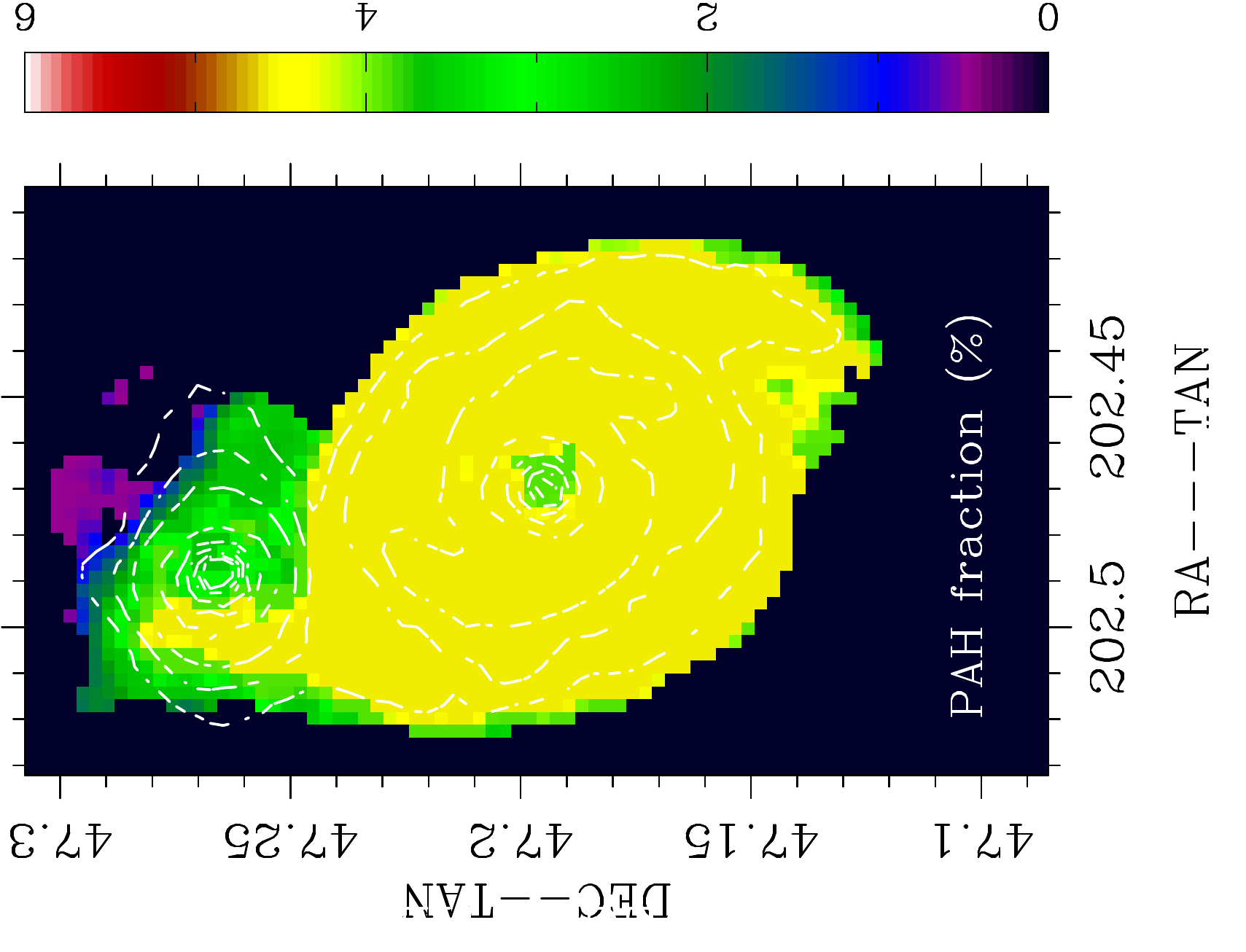}
\caption{\textit{Left:} Dust parameter map of the amplitude of the ISRF of the diffuse dust component, U$_\mathrm{min}$. Stellar mass density is shown in dashed contours in all three panels at levels of 50, 100, 200, 500, 1000, 1500, 2000, 2500 and 3000 M$_\sun$\,pc$^{-2}$. The ISRF is enhanced along the spiral arms by a factor of $\sim7-15$ in NGC\,5194. Surprisingly, U$_\mathrm{min}$ is highest in the post-starburst galaxy, NGC\,5195, and is up to 25 times the amplitude of the ISRF of the Milky Way. \textit{Middle:} Parameter map of the fractional contribution of emission from PDR regions to the total ISRF. The PDR fraction is highest in NGC\,5195 even though there is no star formation. \textit{Right:} Mass fractional contribution of PAH molecules to the total dust mass. Over most of the system, the abundance is 4.58\%, the highest allowed in the models. Both nuclear regions show small depressions where both harbor low-luminosity AGN. }
\label{fig:dustparam}
\end{center}
\end{figure*}

Our model library consists of MW type dust models from DL07 and varies in U$_\mathrm{min}$ from $0.1-25\times$U$_\mathrm{MW}$. The relative contribution of the secondary PDR component is quantified by the parameter, $\gamma$, and ranges from 0.001 to 1 times the primary ISM component. An additional parameter quantifies the mass fractional contribution of PAH emission, $q_\mathrm{PAH}$,  in the mid-infrared region of the SED with following range of values: q$_\mathrm{PAH}=\{0.47,1.12,1.77,2.5,3.19,3.9,4.58\}$\%. Finally, we convert the SEDs from units of per hydrogen nucleon to units per dust mass assuming M$_\mathrm{dust}$/M$_\mathrm{H} = 0.01$.

As with the stellar SEDs, model band fluxes are generated by integrating the PACS and SPIRE filter response functions over the model SED. The SPIRE filter curves were multiplied by a normalized $\lambda^2$ function to convert the response curves from those defined for point sources to those of extended source emission. A least-squares fit is done between the observed bands from 5.6\,$\mu$m to 350\,$\mu$m to obtain the best fit parameters. Again, average best fit parameter values and uncertainties are measured from 100 Monte Carlo simulations of the probability distribution of the photometry. The dust mass in each pixel is inferred from the least-squares normalization of the model to observed band fluxes. 

Overall, the fits reproduce the 5.6-350\,\micron~emission fairly well, leading to $\chi^2_\mathrm{reduced}=1-2$ over much of NGC\,5195 and the central 200\arcsec~aperture of NGC\,5194. Outside of r=100\arcsec, the quality of fits degrades, resulting in  $\chi^2_\mathrm{reduced}>10$. This can be seen in the disagreement between models and observations at 70\,\micron~and 350\,\micron~in Figure~\ref{fig:montageHerschel}. The right panel shows the quality of fit of the model, defined by the following equation:

\begin{equation}\label{eq:res}
\frac{I_\mathrm{\lambda,obs}-I_\mathrm{\lambda,model}}{I_\mathrm{\lambda,obs}},
\end{equation}

In general we find that the models underestimate the emission at 70 and 350\,\micron~by as much as 20\% but at all other wavelengths the photometry and models match to better than 5\% over most of the galaxy. 

The dust mass surface density map is shown in Figure~\ref{fig:dustmass} along with the fractional uncertainty in deriving the parameter. The dust mass is fairly well constrained by the models with the parameter having less than 10\% uncertainty over most of the galaxy. The dust mass is better constrained by the dust models than the stellar mass is by SPS models because of the smaller variation in the mass-to-light ratio of dust emission. Unlike the stellar mass map (Fig.~\ref{fig:stellarmass}), there is more spatial structure in the dust mass which is particularly enhanced along the spiral arms, similar to the mass distribution seen in the molecular gas in Figure~\ref{fig:gasimages}. NGC\,5195 does not show a radially decreasing profile as seen in its stellar mass map but still does contain a fair amount of dust in its vicinity and there is some indication of central concentration. If we consider the higher spatial resolution images from 70\,\micron-350\,\micron~in Figure~\ref{fig:montageHerschel}, we observe that the thermal dust emission is very peaked, which could suggest that the dust is concentrated. However, the main reason for this is not that the dust mass is higher, but rather the illuminating radiation is much more intense and concentrated, leading to the peaked infrared emission seen in the infrared images.

The left panel in Figure~\ref{fig:dustparam} shows the U$_\mathrm{min}$ parameter map of the amplitude of the diffuse ISRF component. The ISRF in NGC\,5195 is extremely high, up to 25 times that of the ISRF in the Milky Way. Over most of NGC\,5194 the ISRF is very close to the Milky Way value ($\sim$2-3\,U$_\mathrm{MW}$), but increases by a factor of 7-15\,U$_\mathrm{MW}$ along the spiral arms. The nuclear region is also higher by a factor of 10\,U$_\mathrm{MW}$.  Our formal fitting uncertainties in this map and the one in the middle of the PDR parameter are quite small (about 5\%) indicating the infrared data was quite sensitive to both parameters. 

The middle panel shows the fractional contribution of radiation associated with PDR regions, that is regions with intensities as high as 10$^6$ times that of the MW. The radiation coming from PDR regions in NGC\,5194 is concentrated along the spiral arms but not the nuclear region. About 2\% of the emission in NGC\,5195 can be associated with this type of emission, even though there is no massive star formation providing the radiation. The bright emission seen in the infrared images comes from intense radiation fields, not enhanced dust mass. The high dust heating found in NGC\,5195 provides strong support for the dust to be located within the galaxy rather than in front of it. But the question then remains as to what is heating the dust? The stellar density is higher in NGC\,5195, but not that much higher than in the bulge of NGC\,5194, so there is another mechanism at play. We leave this for later discussion in \S\ref{s:heating} and \S\ref{s:quench}.

Finally, our SED fits constrain the PAH emission and is shown in the right panel of Figure~\ref{fig:dustparam}. The PAH fraction is the maximum value allowed by the models of 4.58\% (and thus could feasibly be higher) across most of the system. The fraction remains high along the spiral arms  of NGC\,5194 extending out to NGC\,5195 but then decreases across NGC\,5195 to about half the value ($q_{PAH}=3.35\pm1.07$) where a compact radio source  \citep{van88} attributed to an AGN exists. There is also a small depression in the abundance in the center of NGC\,5194, which is a low-luminosity Seyfert 2 nucleus \citep{ter01}.

\subsubsection{Modified black body models}

\begin{figure}[t]
\begin{center}
\includegraphics[angle=-90,width=1.6in]{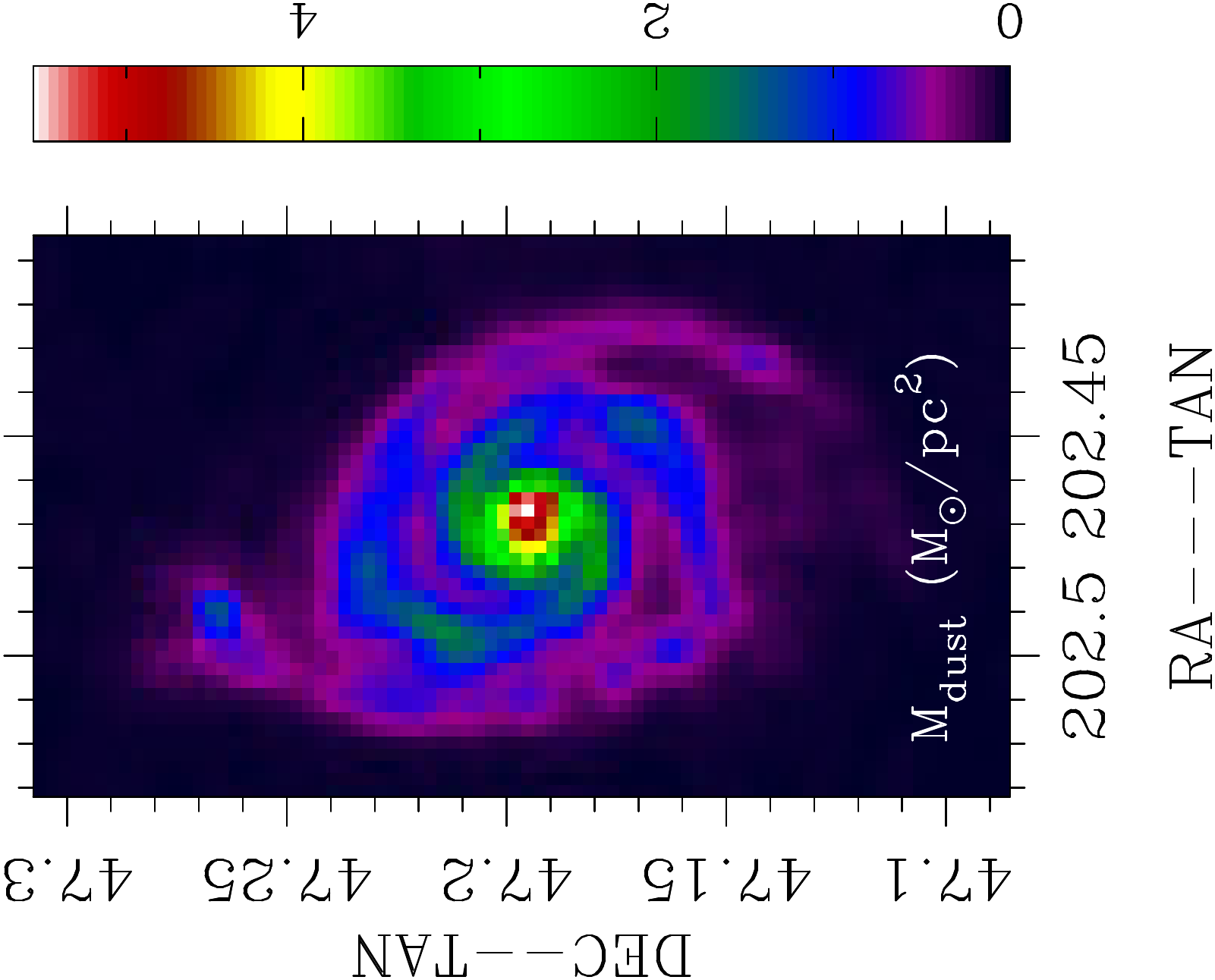}
\includegraphics[angle=-90,width=1.6in]{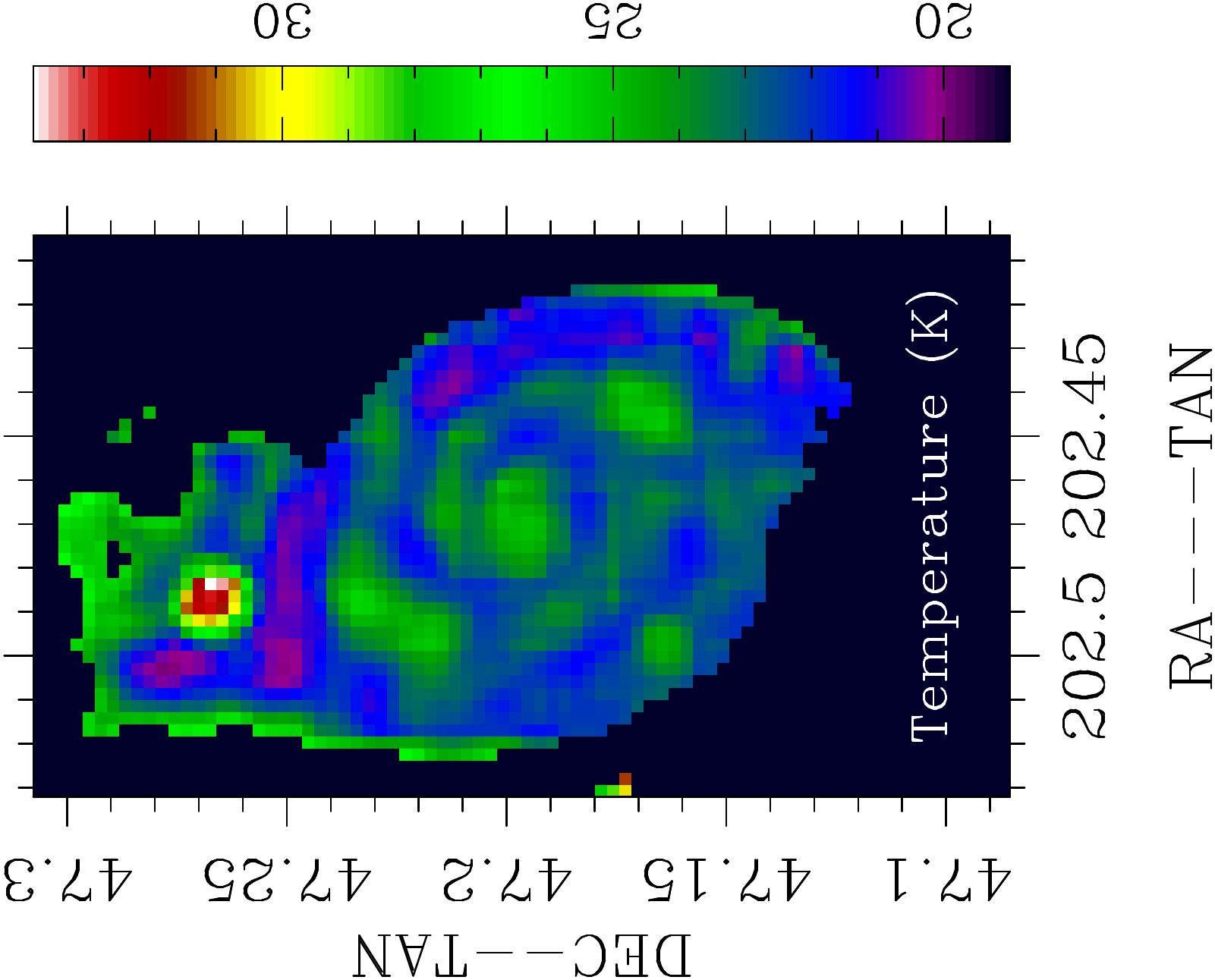}
\caption{Dust mass surface density (left) and temperature (right) from single temperature modified blackbody models. Just as with the ISRF parameter, U$_\mathrm{min}$ from DL07, the dust temperature from modified black body fits is largest in the post-starburst galaxy NGC\,5195. Temperatures are slightly higher in the nuclear region and along the spiral arms. }
\label{fig:BBdustmass}
\end{center}
\end{figure}

A modified black body model was also fit to the 70 though 350\,$\mu$m images as:

\begin{equation}
S_\nu = \frac{ \kappa_{\nu} M_{d} B(\nu,T_{d})}{D^2}
\end{equation}

\noindent where $M_{d}$ is the dust mass with a temperature of $T_{d}$ at a distance, $D$. The emission is expressed as the Planck function, $B(\nu,T_{d})$, modified by $\kappa_\nu \propto \nu^\beta$.  \citet{bos12} indicate that for at least late-type, metal-rich galaxies a value of $\beta=2$ appropriately models the SPIRE infrared colors, but caution that $\beta=1.5$ appears to be more appropriate for less metal-rich, lower surface brightness late-type galaxies in the \textit{Herschel Reference Survey} (HRS; \citealt{bos10}). As a test, we verify that a $\beta=2$ value is sufficient. We performed similar modified blackbody fits allowing $\beta$ to vary from 1 to 3, but this time fit at lower resolution and include our 500\,\micron~SPIRE image. We found that $\beta=2.0\pm0.6$ over the whole system.

We normalize for the dust mass at 250\,\micron~with $\kappa_{(\lambda = 250\,\micron)} = 0.398~\mathrm{m^2\,kg^{-1}}$, using the opacity function for the Milky Way from \citet{dra03}, updated from their dust model presented in \citet{wei01} and consistent with the opacity model from the DL07 models. We assumed a single temperature model, although it should be noted that recent results from \textit{Herschel} (e.g. \citealt{ben10,ben12,boq11}) suggest such a simplistic model is not appropriate since the dust emitting at $<160$\,\micron~is not the same thermal component as the dust emitting at $>250$\,\micron~in nearby spiral galaxies.

As above, band fluxes were generated by integrating the model SED over the filter response curves appropriate for extended source emission. The normalization between the model and observed band fluxes gave the dust mass. Even though our models were normalized at 250\,$\mu$m, the fits themselves were not particularly biased towards this band since our least-squared normalization yields the dust mass not a single band flux. The infrared observations were sufficiently modelled by the single temperature model resulting in $\chi^2_\mathrm{reduced}\sim1$ over the entire system, even though other galaxies are not so easily described. In fact, single temperature models fit all the images from 70-350\,\micron~as well as the DL07 models if fit to the same band fluxes.

\begin{figure}[t]
\begin{center}
\includegraphics[angle=-90,width=1.6in]{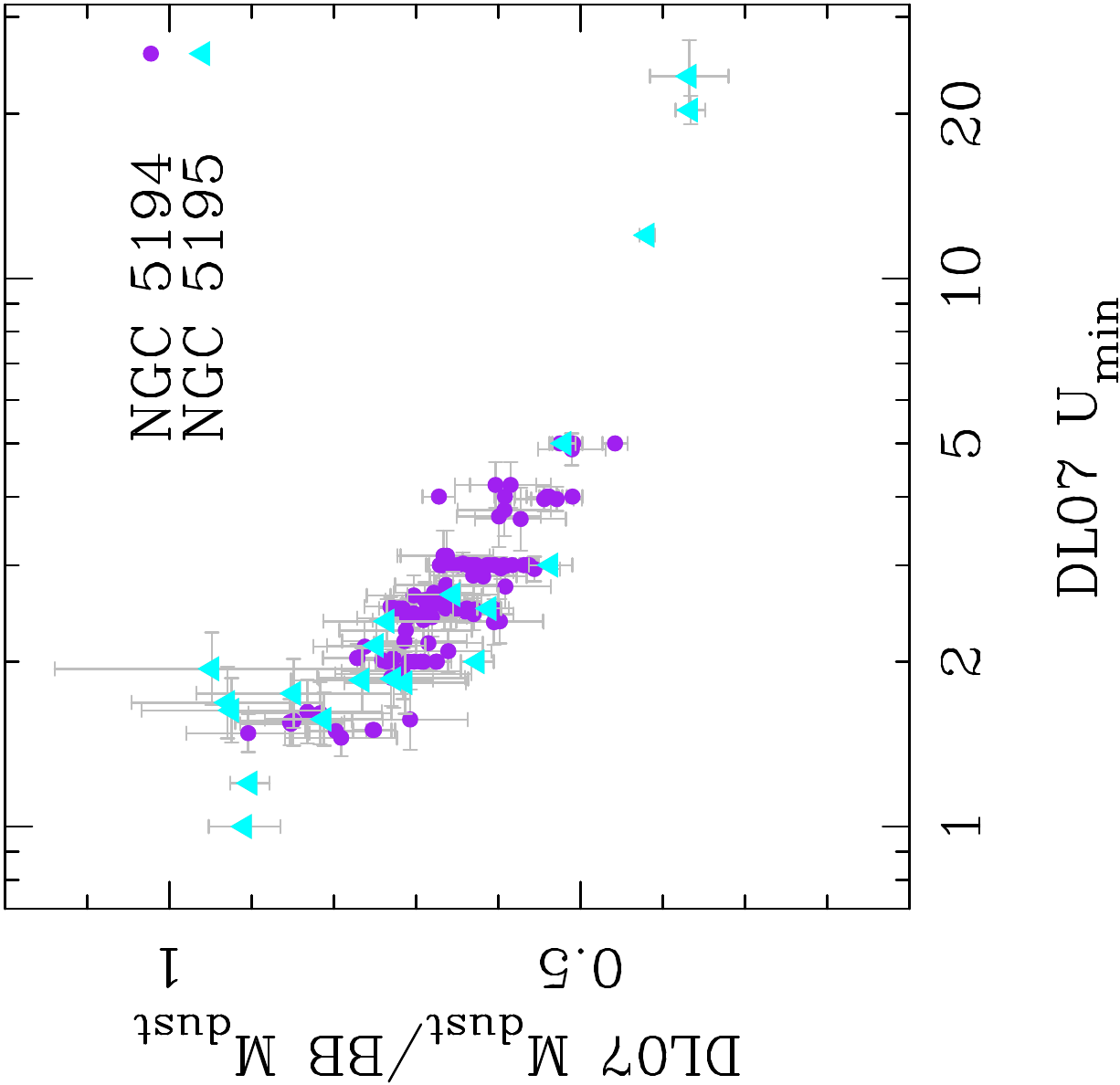}
\includegraphics[angle=-90,width=1.6in]{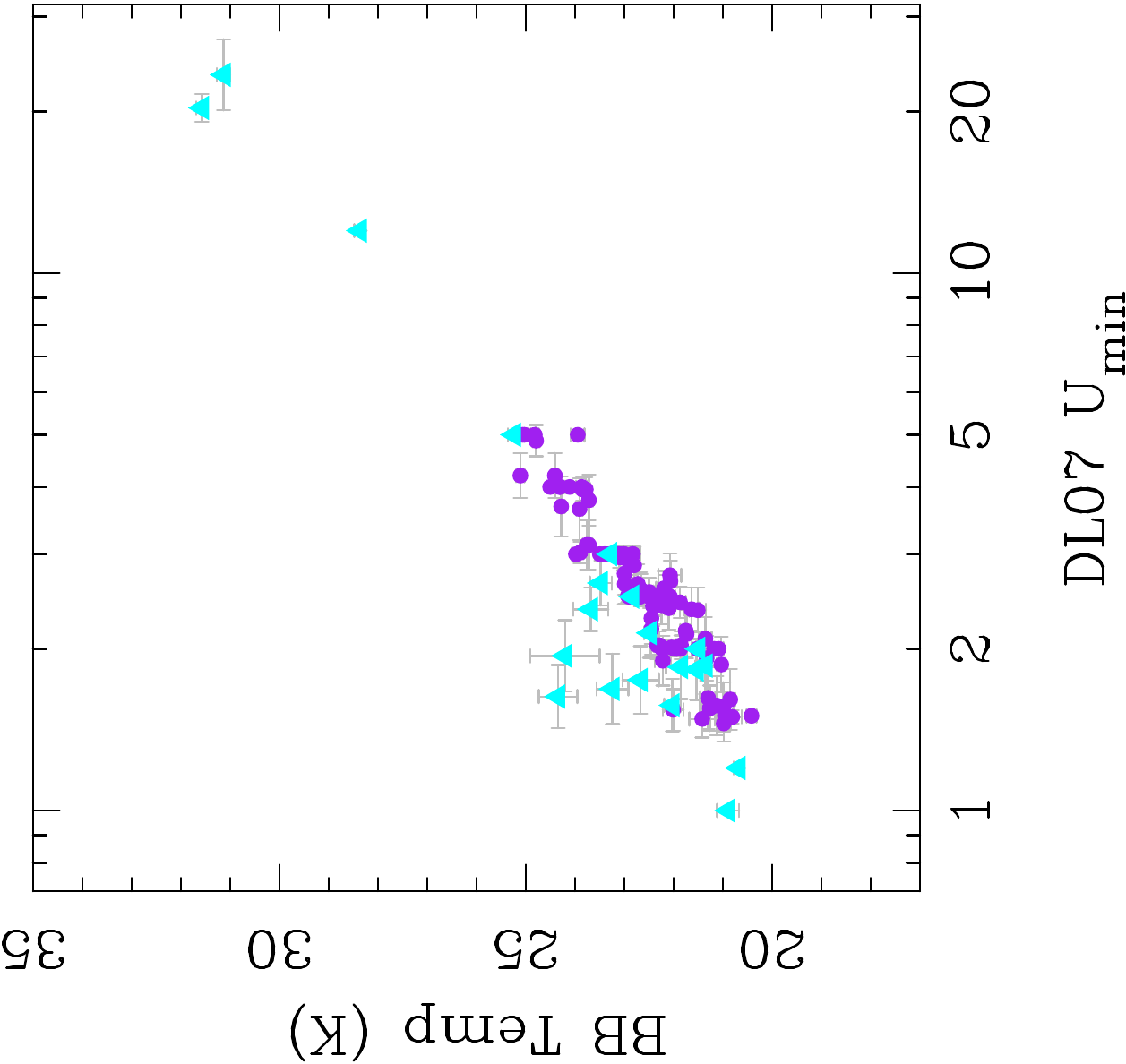}
\caption{\textit{Left:} DL07 dust masses relative to single dust temperature dust masses. \textit{Right:} Dust temperature from modified blackbody fits is compared to the DL07 ISRF parameter, U$_\mathrm{min}$. As expected, the two parameters are logarithmically related.}
\label{fig:dustcomp}
\end{center}
\end{figure}

The resulting dust mass distribution, seen in Figure~\ref{fig:BBdustmass}, is fairly similar to the one derived using the DL07 models from our Figure~\ref{fig:dustmass}. The temperature distribution also qualitatively matches the ISRF amplitude map from the DL07 models from our Figure~\ref{fig:dustparam}. Although the qualitative structure is similar, quantitative pixel-by-pixel comparison of the dust mass shows that systematically modified blackbody masses overestimate the dust mass relatively to the DL07 models at high ISRF values. In the left panel of Figure~\ref{fig:dustcomp}, the ratio between derived dust masses from the DL07 models and modified blackbody models are shown for each galaxy as a function of the amplitude of the ISRF. At higher ISRF, the light-to-mass ratio of the dust is higher. In the central region of NGC\,5195 where the ISRF peaks, the dust mass is overestimated by about a factor of 3. Without accounting for this variation in the light-to-mass ratio, the modified blackbody masses overestimate the dust mass relative to the DL07 dust masses. This is analogous to the optical regime in which young stellar populations exhibit much larger light-to-mass ratios (particularly in the blue) than evolved populations. We also show the relation between dust temperatures and the DL07 U$_\mathrm{min}$ from the different models in the right panel. As expected, the two are logarithmically related, although a few values diverge from the trend in NGC\,5195.
 
\section{Discussion}\label{s:discussion}

In the previous section, we presented the output parameter maps resulting from the average best-fit parameters from 100 Monte Carlo simulations of the stellar emission and dust emission fitting routine.  Here we compare some of the parameters and discuss their implications. First we provide mass comparisons and 1D radial profiles of the stellar, gas and dust mass surface density for both galaxies in \S\ref{s:massdisc}. We then turn our discussion to the relation of interstellar heating of the dust in the context of other properties of the galaxies in \S\ref{s:heating}. We focus on the peculiar properties of the post-starburst NGC\,5195 with its high amplitude ISRF, $\sim$30\,K central dust temperature and lack of star formation. In \S\ref{s:quench} we synthesize its properties and discuss why it is no longer forming stars despite containing molecular gas.


\subsection{Mass comparisons}\label{s:massdisc}

\subsubsection{Radial projections and total masses} 

\begin{figure}[t]
\begin{center}
\includegraphics[angle=-90,width=3.5in]{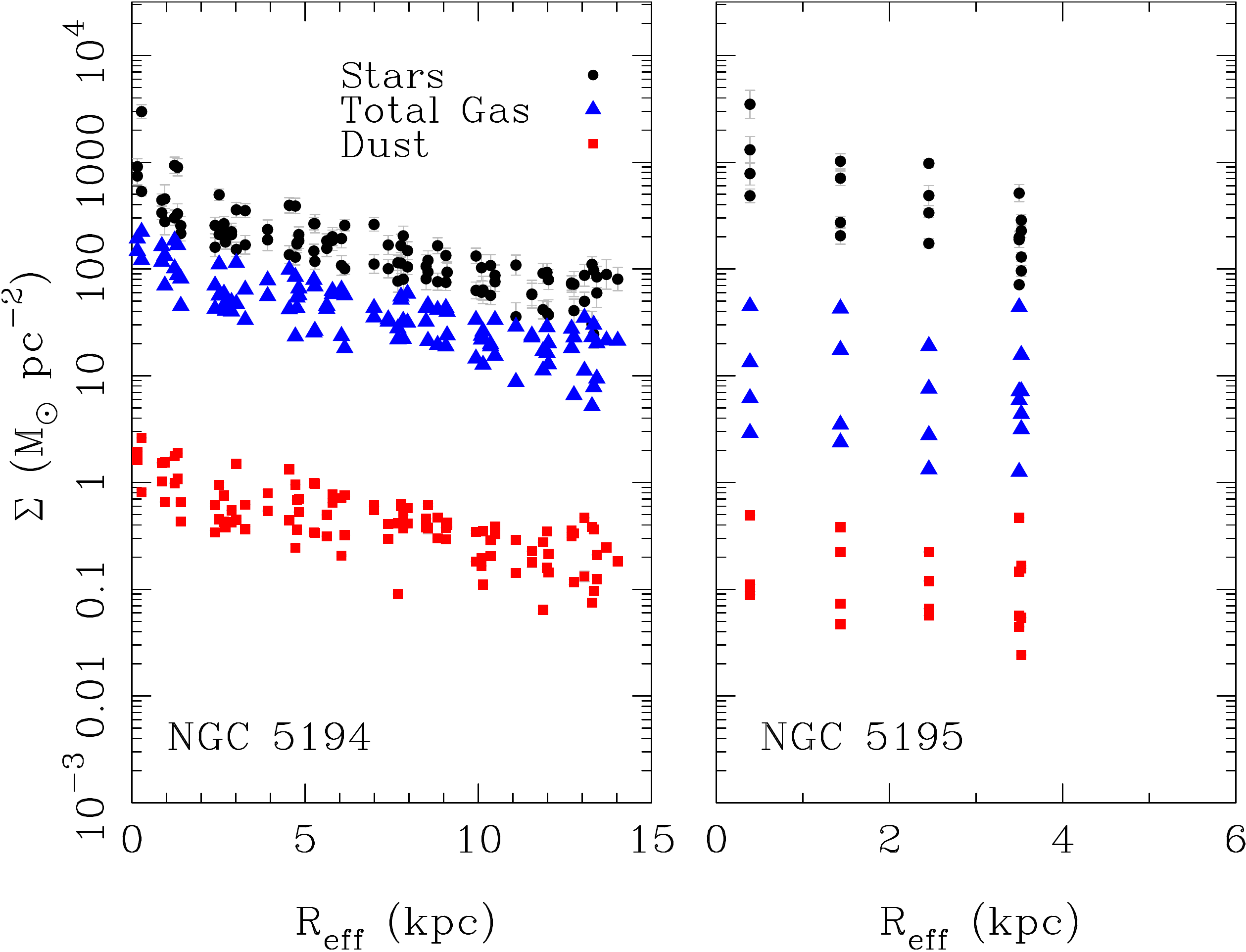}
\caption{Radial mass profile of the stars (black dots), gas (blue triangles), and dust (red squares) in NGC\,5194 (left) and NGC\,5195 (right). Each data point represents a 30\arcsec~wide pixel box encompassing one FWHM of the convolved beam size. Error bars are smaller than the symbol size if they are not visible.}
\label{fig:profile}
\end{center}
\end{figure}

In Figure~\ref{fig:profile}, we plot the radial mass surface density profiles of the stars, gas and dust in black points, blue triangles and red squares respectively for NGC\,5194 in the left panel and NGC\,5195 in the right panel. The effective radius, $R_\mathrm{eff}$, is determined using the structural parameters listed in Table\,\ref{tab:param}. The apertures of both galaxies were chosen at a constant stellar mass surface density of 50\,M$_\sun$pc$^{-2}$. This threshold ensured that the apertures were not overlapping.

In the spiral galaxy, NGC\,5194, the profiles of the stars, gas and dust all show similar slopes, but in NGC\,5195, the slope of the stars is slightly steeper and there is no significant slope in the gas and dust distribution. As can be seen in the gas and dust mass surface density plots in Figures~\ref{fig:gasimages} and \ref{fig:dustmass}, the gas and dust in NGC\,5195 appear to be connected to the spiral arm of NGC\,5194, rather than centrally concentrated and coupled to the stellar mass in NGC\,5195. Both galaxies reach stellar mass surface densities of over 2000\,M$_\sun$\,pc$^{-2}$ in the nuclear regions, then decrease radially to values of $\sim$50\,M$_\sun$\,pc$^{-2}$ at the edge of the apertures (by definition). The central surface density (in a 30\arcsec~square region) is only slightly higher in the NGC\,5195 at a value of $3490\pm900$\,M$_\sun$\,pc$^{-2}$
compared to $2980\pm450$\,M$_\sun$\,pc$^{-2}$ in NGC\,5194.

The dust mass density in the central region of NGC\,5194 peaks just above 2\,M$_\sun$\,pc$^{-2}$ in the central region, decreasing to 0.75\,M$_\sun$\,pc$^{-2}$ along the spiral arms and then a constant 0.25\,M$_\sun$\,pc$^{-2}$ throughout the rest of the system. The dust mass density is comparably lower in NGC\,5195 at values near 0.25\,M$_\sun$\,pc$^{-2}$, consistent with being an extension of the spiral arm of NGC\,5194.

\begin{table}[t]
\caption{Stellar, gas and dust mass in each aperture}
\begin{center}
\begin{tabular}{c|c|c|c}
	  		& M$_\star$ ($10^{10}$~M$_\sun$)		& M$_\mathrm{gas}$	($10^{10}$~M$_\sun$)	& M$_\mathrm{dust}$ ($10^{8}$~M$_\sun$)\\
\hline	
NGC\,5194  	& $4.7\pm0.1$  & $1.10\pm0.03$		& $1.19\pm0.01$\\	
NGC\,5195  	& $2.5\pm0.2$	& $0.052\pm0.005$	& $0.064\pm0.001$	\\
Total	\footnote{Total mass in combined system including the mass within both apertures, as well as outside the apertures in the image in which our S/N criteria (see S\ref{s:errors}) was met.}		& $7.9\pm0.2$	& $1.41\pm0.05$ 		& $1.45\pm0.01$\\
\end{tabular}
\end{center}
\label{tab:masses}
\end{table}%

We provide the total stellar, gas and dust masses in each aperture in Table~\ref{tab:masses}. We also measure the total mass in the combined system, summing up the mass in each pixel over the entire system. A small percentage ($<5$\%) of the galaxy is missing from these `total' values as we only added up pixels which satisfied our signal-to-noise criteria (see \S\ref{s:errors}) and low surface brightness regions of the galaxy are not accounted for.  About $\sim80$\% of the gas and dust mass is found in the spiral NGC\,5194, with the remaining found outside the aperture.  Less than 1\% of the gas and 3\% of the dust of the total system is located within the aperture defining NGC\,5195. The stellar mass is distributed more evenly between the two galaxies with NGC\,5194 being twice as massive as the spheroid NGC\,5195. The bulk of the stellar mass is found within our defined apertures due to relatively steeper mass profiles. Only about 10\% of the stellar mass is located outside the two apertures.

\subsubsection{Dust-to-stellar mass ratio} 

\begin{figure}[t]
\begin{center}
\includegraphics[angle=-90,width=3.3in]{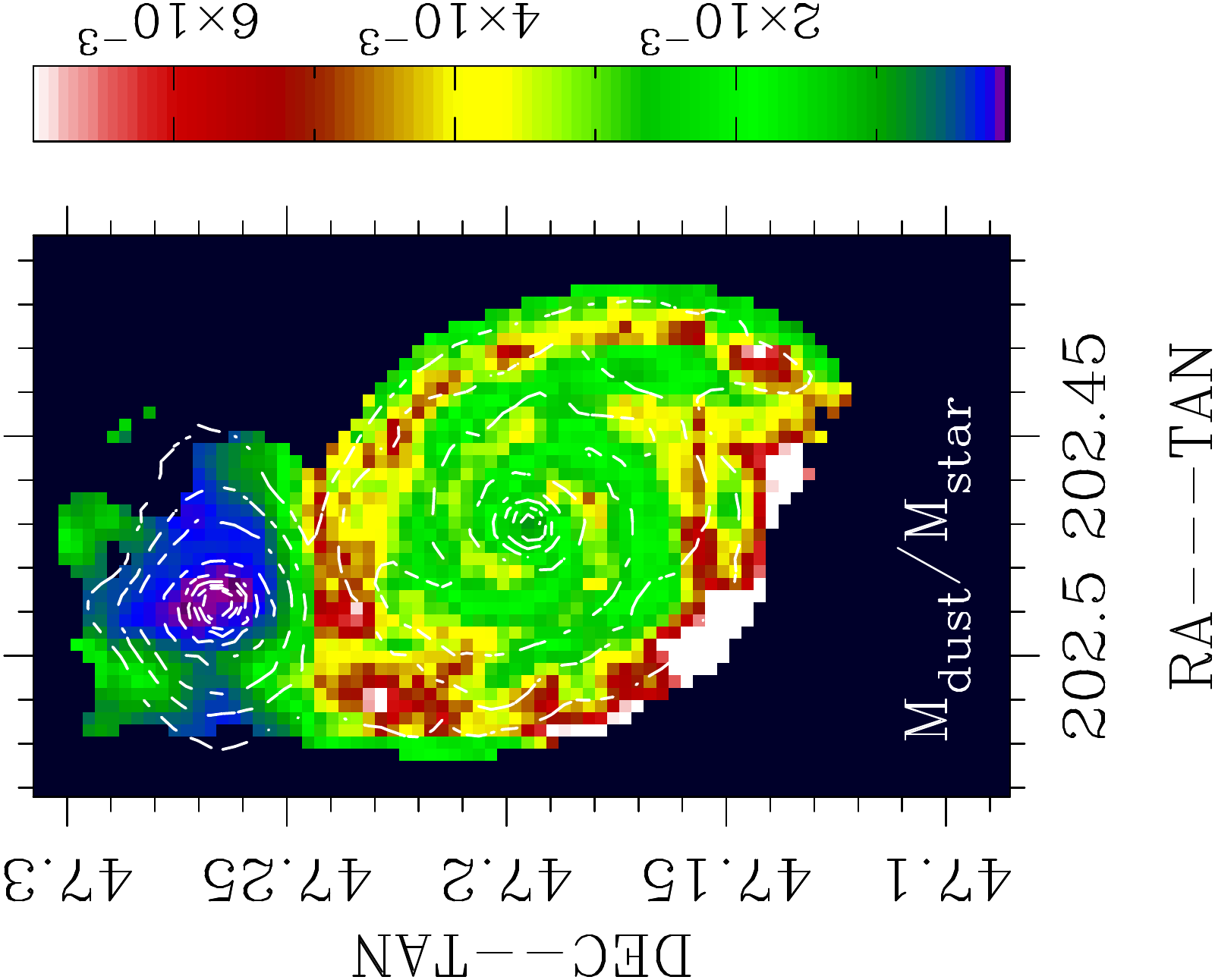}
\caption{A dust-to-stellar mass ratio map with the stellar mass density given in contours at the levels of 50, 100, 200, 500, 1000, 1500, 2000, 2500, 3000 M$_\sun$\,pc$^{-2}$ . NGC\,5195's dust-to-stellar mass ratio is an order of magnitude smaller than NGC\,5194 and increases with radius as the stellar mass decreases.}
\label{fig:dust-to-stellar}
\end{center}
\end{figure}

Recall from Figures~\ref{fig:stellarmass} and \ref{fig:dustmass} which show the stellar and dust mass surface density maps of the system, that the dust mass is slightly more structured than the stellar mass map. In the stellar mass map, there are some enhancements in stellar mass along the spiral arms of NGC\,5194, but they are not as pronounced as the spiral arms in the $B$-band image (which can be found in the appendix) for example or those found in the dust mass map. Both the $B$-band and dust mass surface density map are enhanced by the younger stellar populations. Because of this, the dust-to-stellar mass ratio, shown in Figure~\ref{fig:dust-to-stellar}, is a factor of two larger along the arms. 

\begin{figure}[t]
\begin{center}
\includegraphics[angle=-90,width=2.8in]{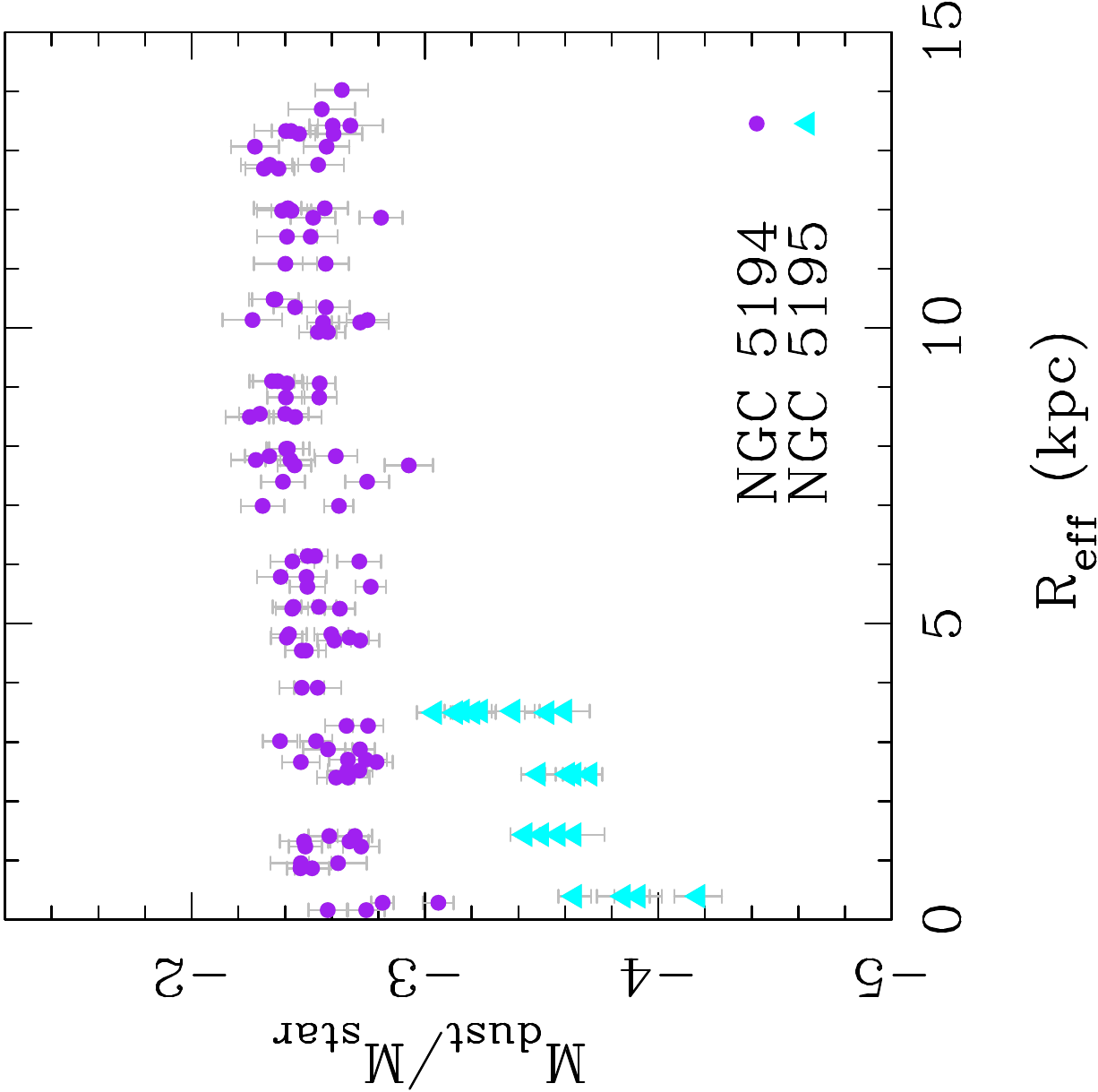}
\caption{Radial profile of the dust-to-stellar mass ratios in NGC\,5194 (purple) and NGC\,5195 (cyan). The ratio is relatively constant in NGC\,5194 but increases with radius in NGC\,5195 primarily due to the decreasing stellar mass density as demonstrated in Figure~\ref{fig:dust-to-stellar}. }
\label{fig:dust-to-stellar1D}
\end{center}
\end{figure}

We find an average dust-to-stellar mass ratio of $M_\mathrm{dust}/M_\star$ = ($3.1\pm1.0) \times10^{-3}$ (or log($M_\mathrm{dust}/M_\star)= -2.5\pm0.2$) in NGC\,5194. The dust-to-stellar mass ratio is an order of magnitude less in NGC\,5195 at $M_\mathrm{dust}/M_\star=(3.6\pm2.2)\times10^{-4}$ (or log($M_\mathrm{dust}/M_\star)= -3.5\pm0.3$). In Figure~\ref{fig:dust-to-stellar1D}, the radial profiles of the dust-to-stellar mass ratio is given for both galaxies.

The dust-to-stellar mass ratio of NGC\,5194 is consistent with results of other spiral galaxies in the \textit{Herschel Reference Survey} (HRS; \citealt{bos10}) which are found to have $\log(M_\mathrm{dust}/M_{\star})=-2.59\pm0.03$ \citep{cor12}. Both galaxies are consistent with the mass ratio found for a large range of galaxy types, $\log(M_\mathrm{dust}/M_{\star})=-2.95\pm0.68$ \citep{ski11}, from the KINGFISH survey \citep{ken11}, a \textit{Herschel} extension of the SINGS survey \citep{ken03}. Furthermore, the late-type NGC\,5194 is consistent with late-types in KINGFISH which \citet{ski11} found to be $\log(M_\mathrm{dust}/M_{\star})=-2.70\pm0.53$ and NGC\,5195 is in line with their findings for the early-type galaxies in the survey with $\log(M_\mathrm{dust}/M_{\star})=-3.77\pm0.83$. Early-type galaxies in the volume based HRS show the spheroid population has a lower dust-to-stellar mass ratio with $\log(M_\mathrm{dust}/M_{\star})=-4.1\pm0.1$ \citep{smi12} compared to NGC\,5195, although a couple of objects do have ratios as high as $\log(M_\mathrm{dust}/M_{\star})=-3.5$. For clarity, all the comparisons listed here have stellar and dust masses that were derived using a similar IMF (either a \citealt{cha03} or \citealt{kro01} IMF) and dust opacity and do not suffer from  systematic offsets in the derived masses.

In comparison to larger volume-based infrared surveys, recent results from the H-ATLAS survey \citep{bou12} reveal that for blue galaxies there is a trend for the dust-to-stellar mass ratio to decrease with stellar mass. They adopt a higher opacity value than ours with $\kappa_{(\lambda = 250\,\micron)} = 0.89~\mathrm{m^2\,kg^{-1}}$ compared to our choice of $\kappa_{(\lambda = 250\,\micron)} = 0.398~\mathrm{m^2\,kg^{-1}}$. For comparison, their dust masses and dust-to-stellar mass ratios need to be shifted up by 0.35 dex. Stellar masses should be roughly consistent as their chosen IMF \citep{cha03} leads to stellar masses only 0.03~dex lower than our chosen IMF from \citet{kro01}. Accounting for this, NGC\,5194's dust-to-stellar mass ratio is consistent with the blue galaxies from the H-ATLAS survey (with $\log(M_\mathrm{dust}/M_{\star})\sim-2.8$), while NGC\,5195 is slightly lower, consistent with the green galaxies in their sample, which it would belong to.

\subsubsection{Gas-to-dust mass ratio} 

\begin{figure*}[t]
\begin{center}
\includegraphics[angle=-90,width=2.1in]{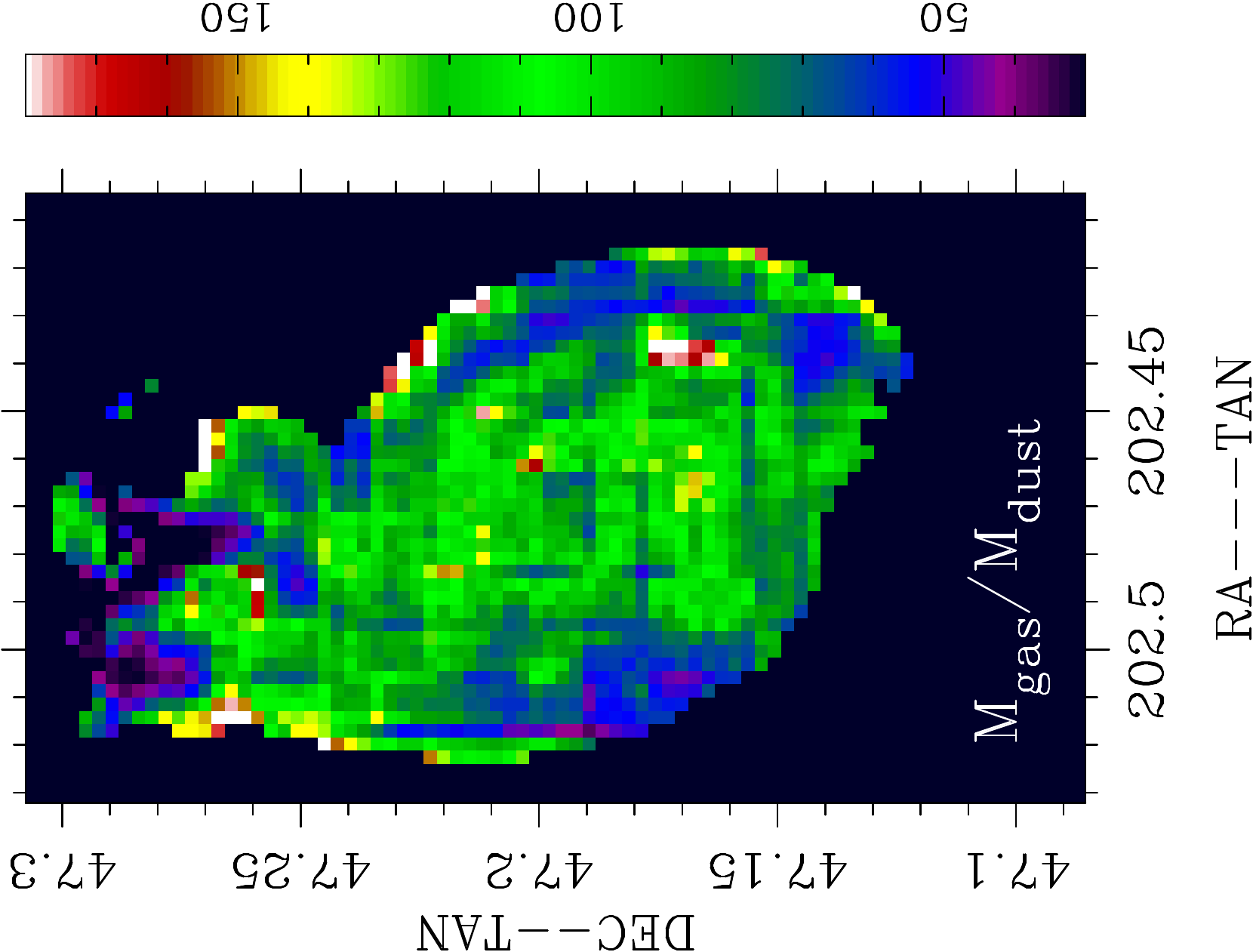}
\includegraphics[angle=-90,width=3.6in]{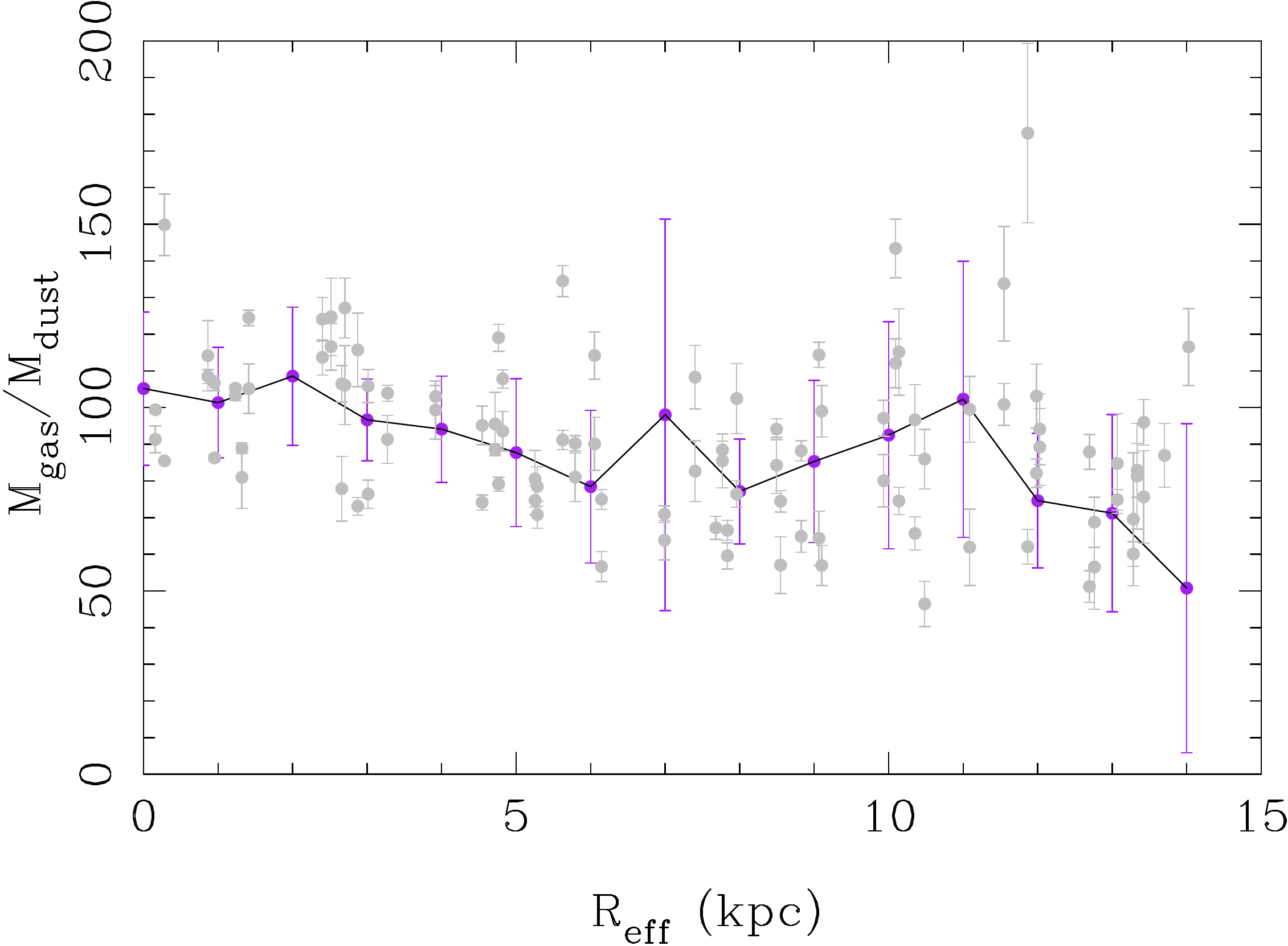}
\caption{Gas-to-dust mass ratio map is shown on the left. The right shows the 1D profiles of the ratio as a function of radial distance from the center of NGC\,5194. The grey points show the pixel values and the connected line shows the binned radial profile. }
\label{fig:gtd}
\end{center}
\end{figure*}

The map of the gas-to-dust ratio is shown in Figure~\ref{fig:gtd}, along with the 1D radial profile for NGC\,5194. The ratio is fairly constant across NGC\,5194 and appears to slightly decrease radially from a value of $105\pm21$ in the inner 60\arcsec~ to a value of $71\pm27$ at a distance of 13\,kpc~from the galaxy's center. Recalling the metallicity map of the system in Figure~\ref{fig:metals} and metallicity gradient results from \citet{mou10}, the metallicity drops by a factor of $\sim$2 from the central region to our aperture edge. Recent work by \citet{mag11} and \citet{foy12} show that a metallicity dependent $X_{CO}$ conversion factor is needed for regions dominated by molecular gas, which is the case for NGC\,5194. Because of the radially decreasing metallicity, such a conversion would cause the gas-to-dust ratio to increase at larger radii leading it to be even more constant across the galaxy as was seen for M83, also observed in the VNGS \citep{foy12}.  

In most cases, when a gradient is found in the gas-to-dust ratio, it often goes in the other direction, increasing with radial distance and is usually attributed to the galaxy containing a larger extended gas disk relative to its dust disk. This was seen in
 NGC\,2403 \citep{ben10b} and in the sample of SINGS galaxies \citep{mun09}. However, \citet{mun09} actually find a fairly constant gas-to-dust ratio for NGC\,5194 in agreement with our findings.

\subsection{Heating}\label{s:heating}

\begin{figure*}[tb]
\begin{center}
\includegraphics[angle=-90,width=3.in]{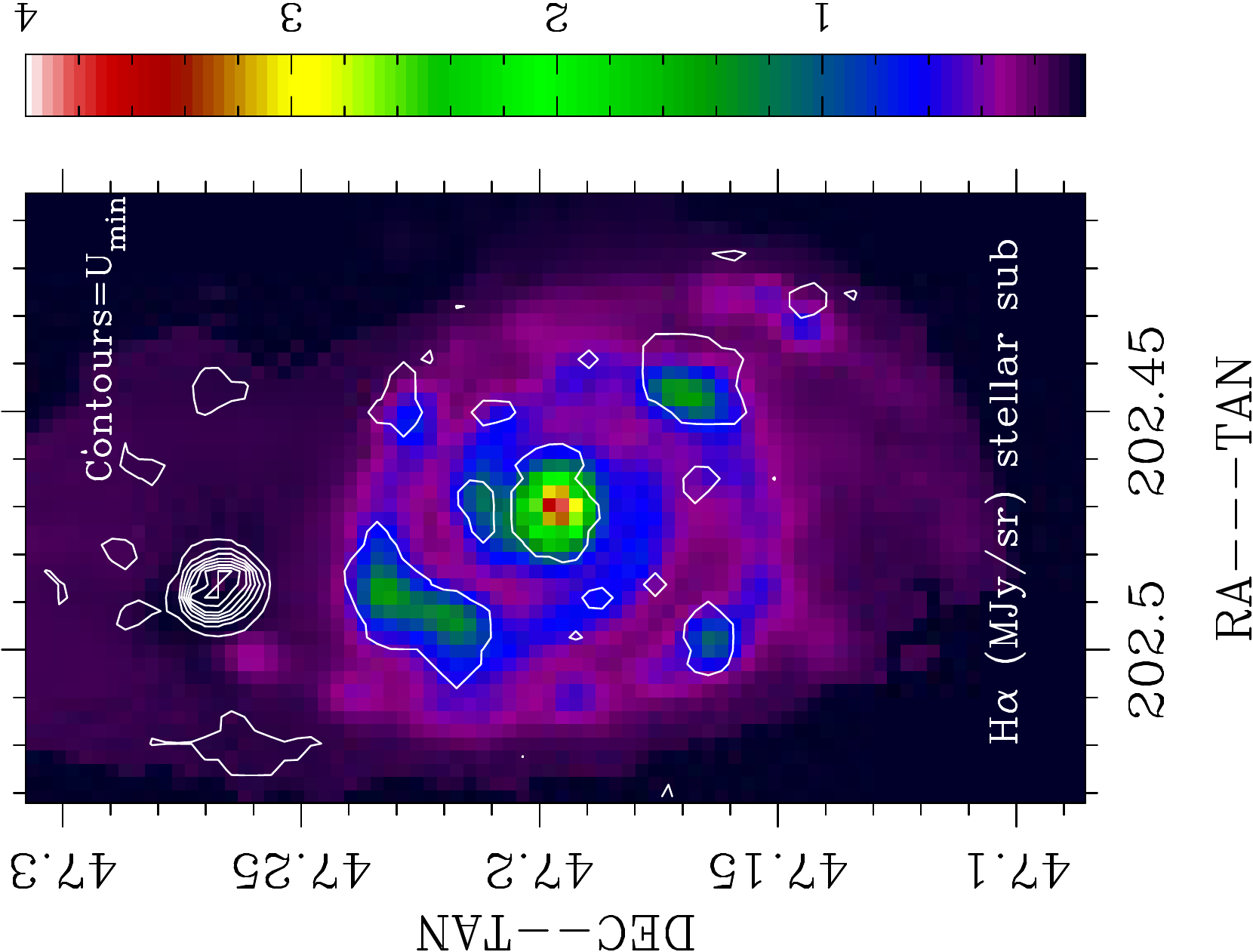}
\includegraphics[angle=-90,width=3.in]{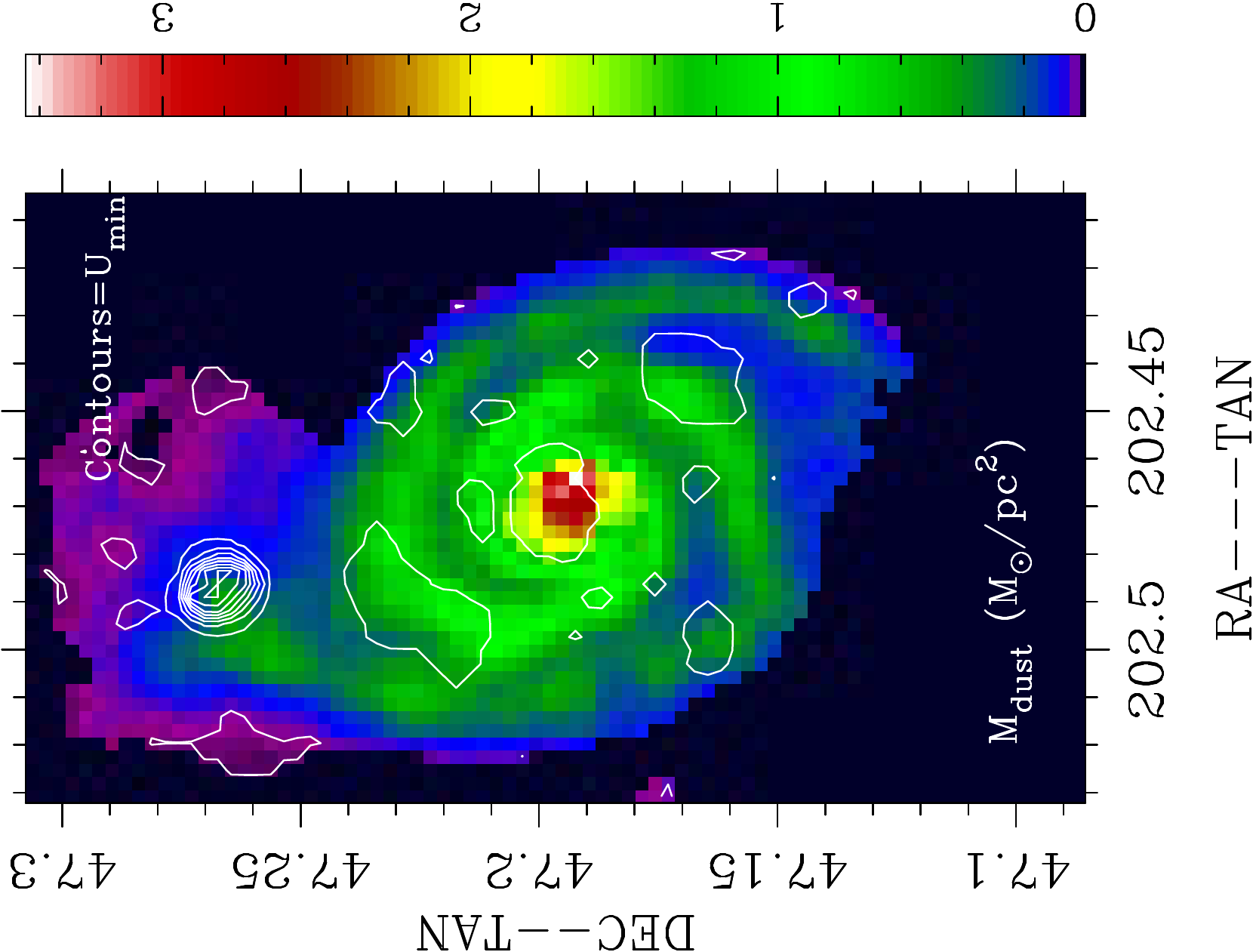}
\caption{\textit{Left:} Map of the stellar subtracted H$\alpha$ intensity with U$_\mathrm{min}$ shown in contours. The highest ISRF is found in the post-starburst galaxy, NGC\,5195, where a clear lack of H$\alpha$ emission is found in the nuclear region. The ISRF is enhanced along the spiral arms, as is H$\alpha$ emission but does not show an increase in the nuclear regions, in contrast to the H$\alpha$ emission which is strongly peaked in the bulge. \textit{Right:} Mass of the dust mass surface density in M$_\sun$\,pc$^{-2}$ with contours of the amplitude of the ISRF of the ISM, quantified by U$_\mathrm{min}$ from DL07 models. While both are enhanced along the spiral arms, there are some noticeable anti-correlations, specifically in the nuclear regions of both galaxies.  }
\label{fig:heating}
\end{center}
\end{figure*}

\subsubsection{Decoupling of dust mass and heating}

There is a subtle decoupling between the distribution of the radiation intensities across the system and the dust mass surface density. Although both show spiral structure, when we superimpose them together as shown in the rightmost plot of Figure~\ref{fig:heating}, it appears that there is a slight offset between the peaks of each quantity. This has been seen already in the nearby spiral M83 \citep{foy12}, also observed as part of the VNGS, as well as in NGC\,4501 and NGC\,4567/8 in the Virgo Cluster \citep{smi10}. As in \citet{foy12}, we find the peaks in the ISRF (or dust temperature values) lie downstream of the spiral arm distribution of the dust, assuming a trailing arm rotation of NGC\,5194 (counter-clockwise).

There are two more obvious discrepancies between the U$_\mathrm{min}$ map and the dust mass map in the central regions of both galaxies. In the center of NGC\,5194, there is an obvious increase in the dust mass density, relative to both the large scale mass density as well as the mass density in the spiral arms. On the other hand, the ISRF is enhanced but only at the same level as the spiral arms. Even though both the stellar, dust and gas mass densities are all increasing, the local ISRF value stays moderately low. This is unlike the central region of NGC\,5195 which has extremely high ISRF values (with peak dust temperatures of 30\,K), but no relative increase in the dust mass. Although there is less dust in the vicinity of NGC\,5195, the dust that is there is quite warm. This is in agreement with measurements of the gas temperature by \citet{koh02} who found that low HCN-CO ratios suggest that the molecular gas in NGC\,5195 is quite warm and unable to collapse and form stars.

\begin{figure}[tb]
\begin{center}
\includegraphics[angle=-90,width=2.5in]{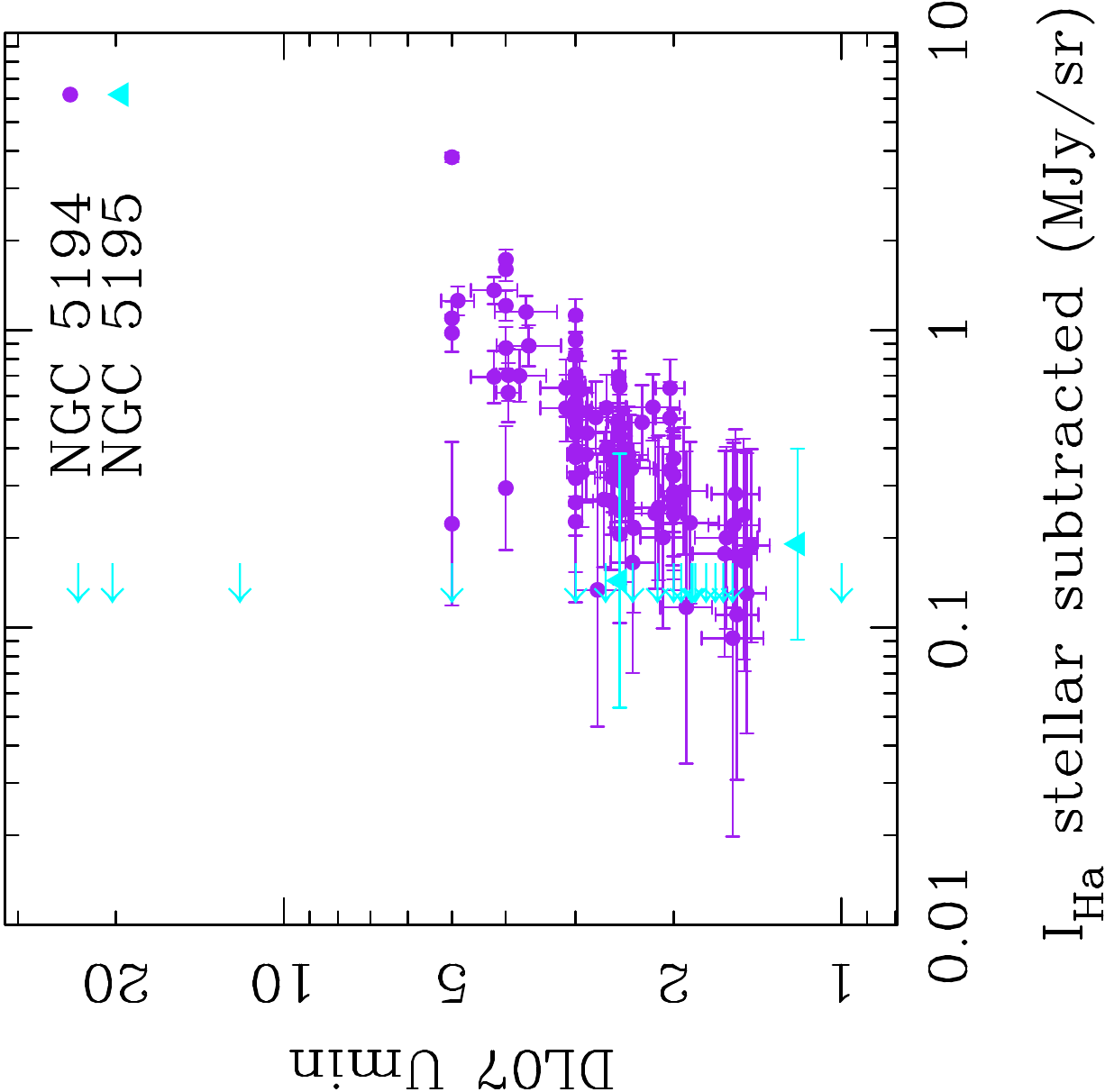}
\caption{The H$\alpha$ emission is moderately correlated with the ISFR in NGC\,5194 (purple points), while NGC\,5195 (cyan triangles and upper limits)  has minimal to zero H$\alpha$ emission despite having high ISRF values. }
\label{fig:UminvsHa}
\end{center}
\end{figure}

\subsubsection{Ionized gas and the ISRF}

In the left panel of Figure~\ref{fig:heating}, we show the narrowband H$\alpha$ image with stellar emission subtracted. This was done by calculating the stellar continuum transmitted through the H$\alpha$ narrowband filter curve using the best-fit SED model without nebular emission included. The strength of the ISRF, as parametrized by U$_\mathrm{min}$ is shown in contours (and also in Fig.~\ref{fig:dustparam}). Qualitatively, the H$\alpha$ emission and ISRF are spatially correlated in the disk of NGC\,5194. They are both enhanced along the spiral arms and in the central bulge region, although H$\alpha$ is more peaked than the ISRF in the nucleus perhaps due to AGN emission. In contrast, in NGC\,5195, there is no detected star formation and hence no H$\alpha$ emission. We note however, that in optical spectroscopy from \citet{mou10}, a modest amount of H$\alpha$ emission is detected in the nuclear region at a $\sim$2$\sigma$ level that appears to be blue shifted relative to H$\alpha$ absorption in stellar photospheres. Surprisingly, this is where the ISRF is the brightest in the entire map. In Figure~\ref{fig:UminvsHa} we show a 1D comparison between the ionized gas intensity and the strength of the ISRF determined from the dust emission model fitting. While the ionized gas intensity correlates with the ISRF in NGC\,5194, it clearly follows another trend in NGC\,5195 where high ISRF values correspond to no H$\alpha$ emission.

\subsubsection{Dust heating in NGC\,5195}
\begin{figure*}[t]
\centering
\includegraphics[angle=-90,width=2.3in]{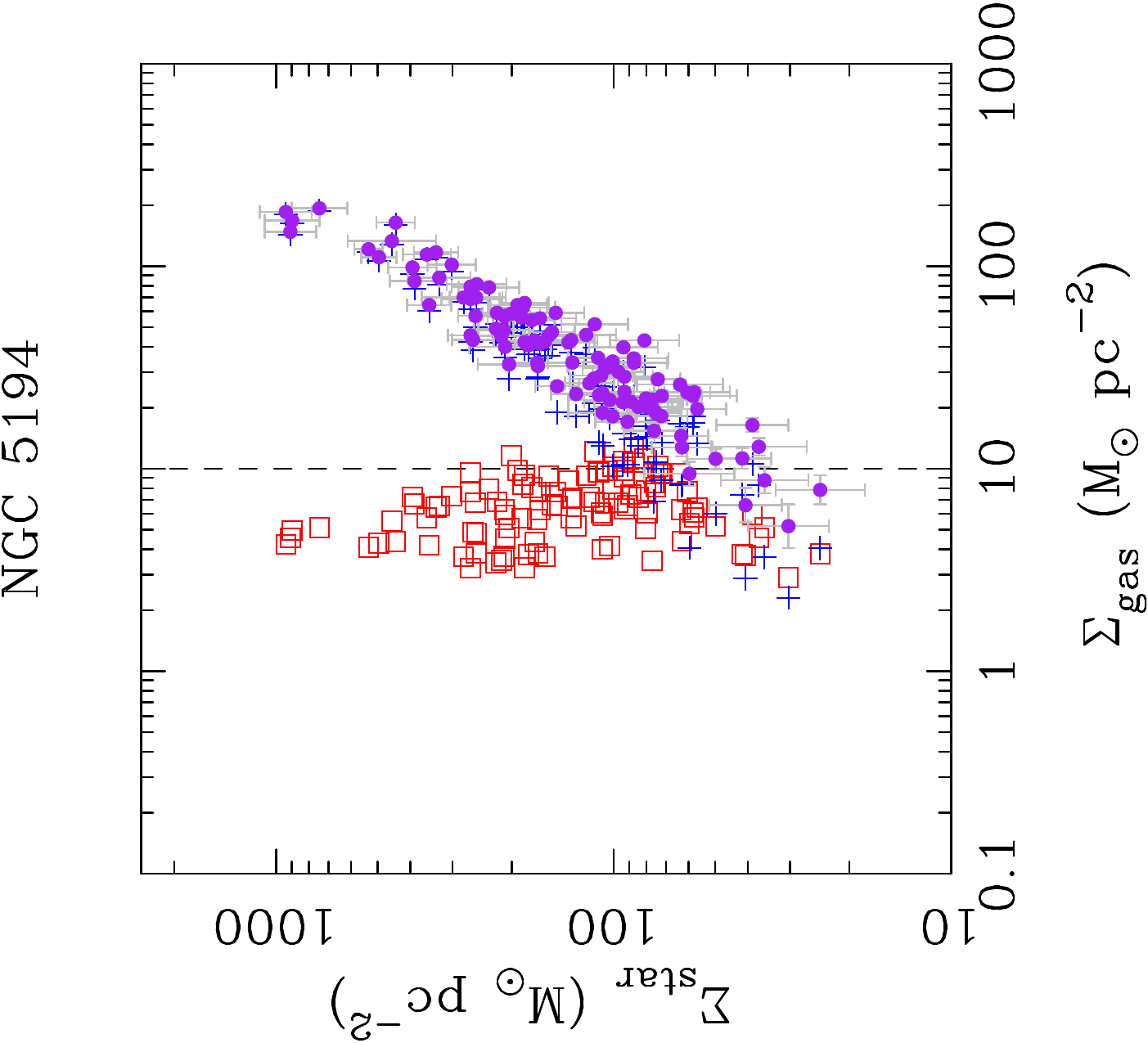}~~~~~
\includegraphics[angle=-90,width=2.3in]{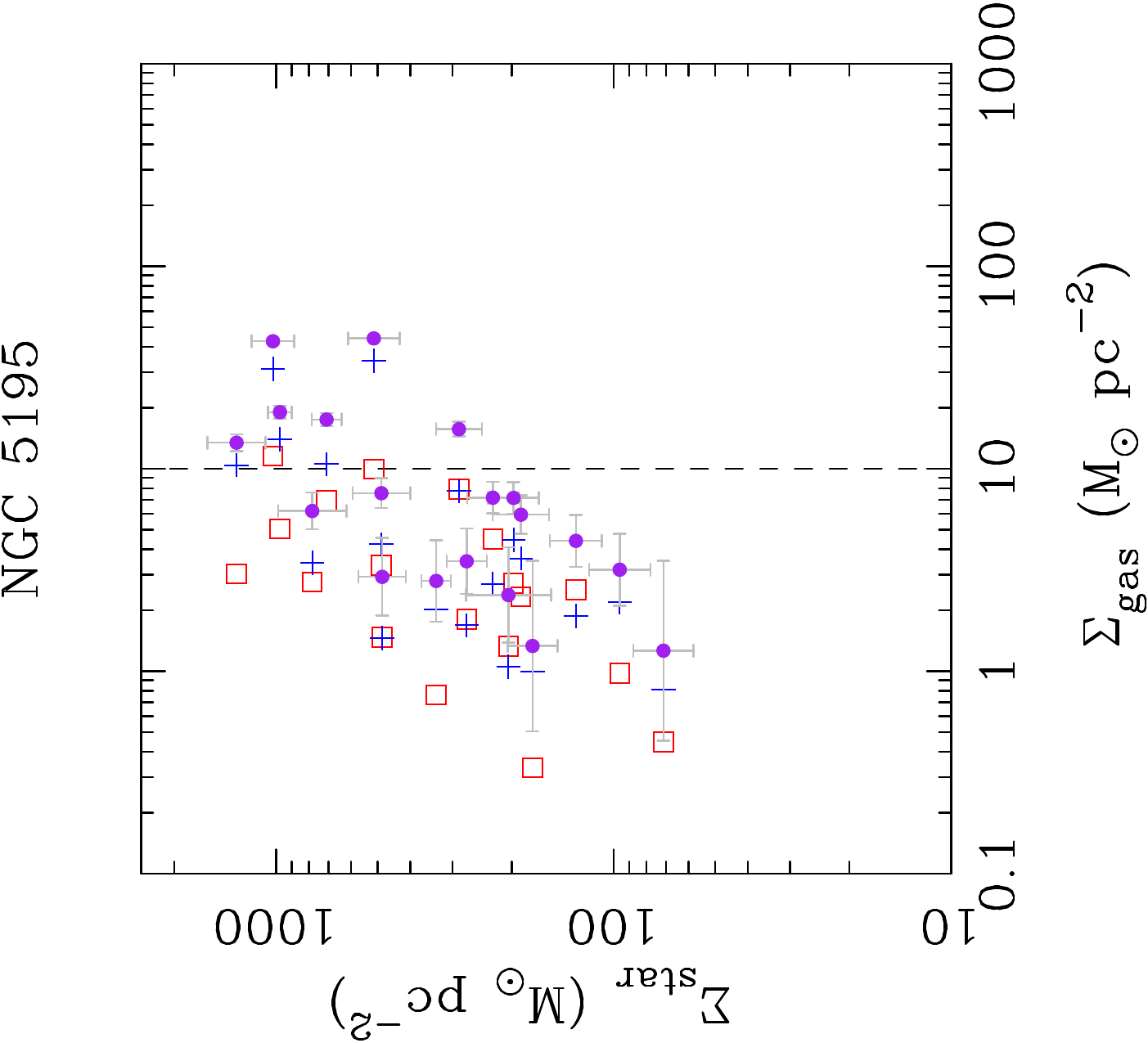}\\
\includegraphics[angle=-90,width=2.3in]{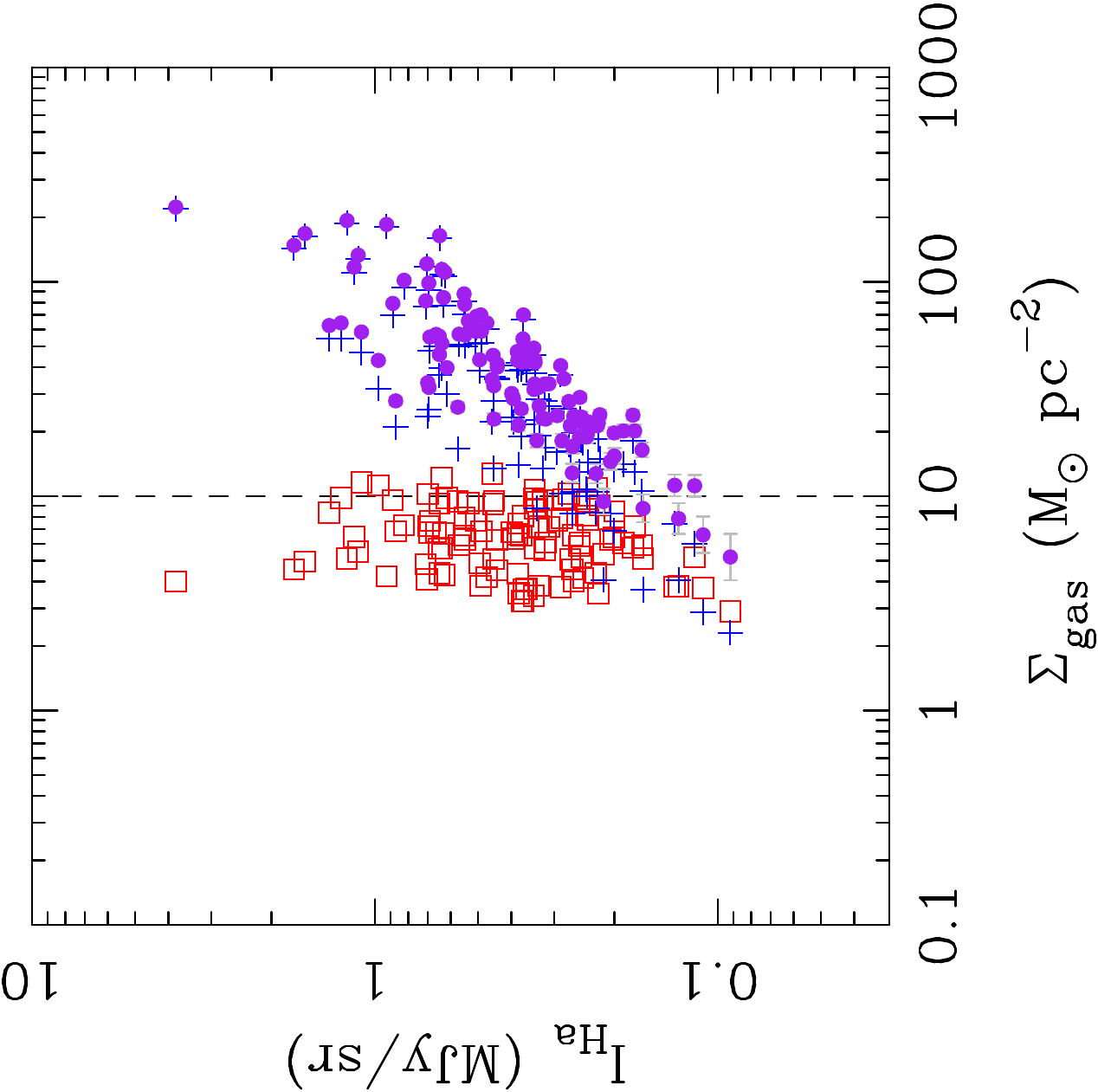}~~~~~
\includegraphics[angle=-90,width=2.3in]{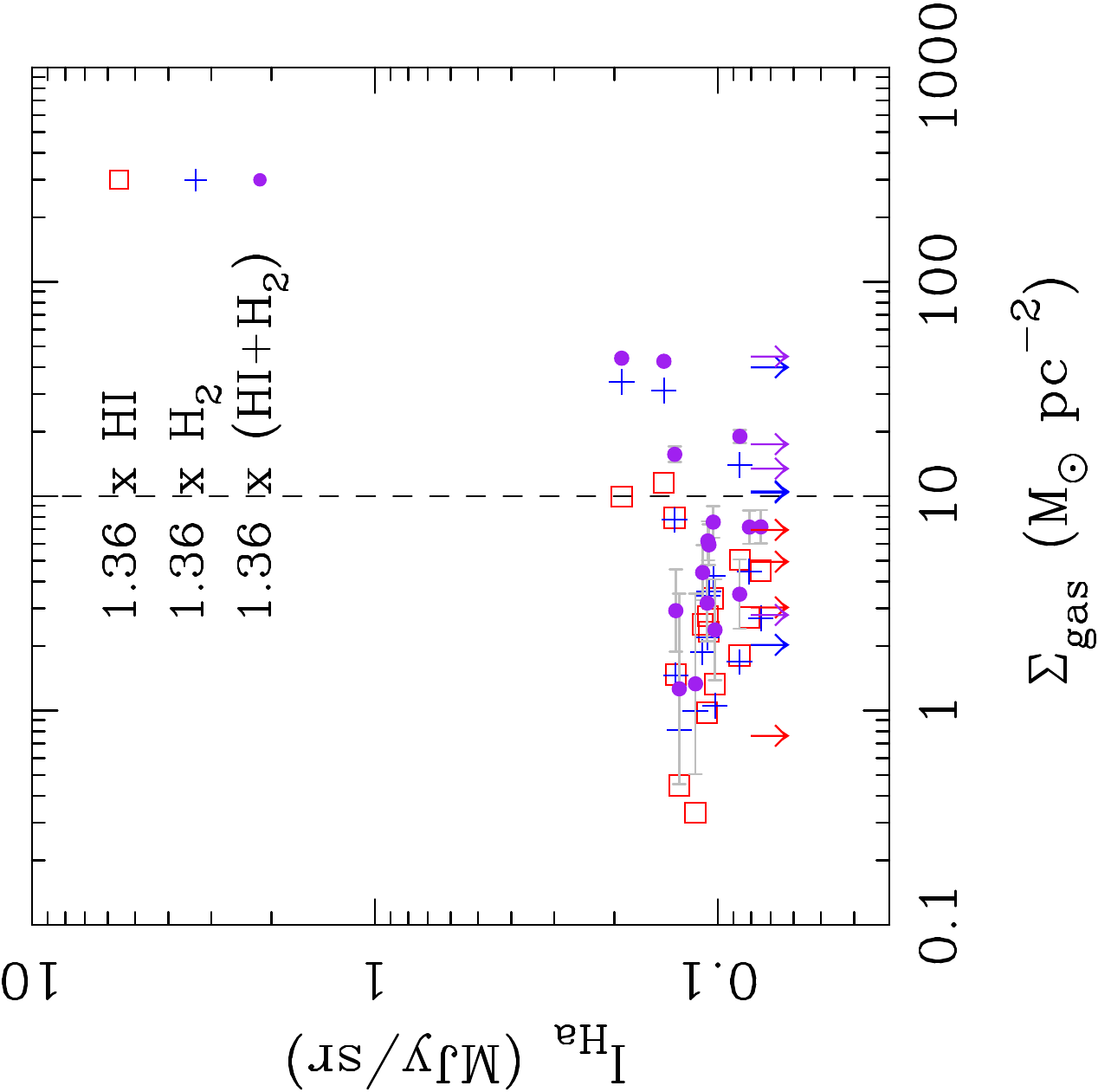}\\
\caption{Stellar mass density versus atomic (red squares), molecular (blue plus signs) and total (purple dots) gas mass surface density for NGC\,5194 in the top-left and for NGC\,5195 in the top-right. Stellar subtracted H$\alpha$ intensity as a function gas mass surface density is shown in the bottom panels. In each panel, we note the surface density threshold of 10\,M$_\sun$\,pc$^{-2}$ at which atomic gas converts to molecular gas. Note that some regions in NGC\,5195 contain molecular gas at densities lower than this threshold. }
\label{fig:gasdensity}
\end{figure*}

One of the most interesting findings of this analysis is that the ISRF of the post-starburst spheroid NGC\,5195 is up to 20--25 times the characteristic amplitude of the ISRF in the Milky Way despite its lack of star formation. Often high dust temperatures are associated with starbursting and actively star forming galaxies. This is not the case for NGC\,5195. The galaxy has a high ($29\pm3\,$K) characteristic dust temperature in the nuclear region and an average temperature of $23\pm3$\,K in the galaxy's aperture. To provide some context, recent results from the H-ATLAS survey \citep{bou12} indicate that, for blue galaxies in their sample, dust temperatures increase with stellar mass up to a stellar mass of  $\sim6\times10^{10}$\,M$_\sun$ reaching a peak dust temperature of $\sim25-30$\,K and then anti-correlate at higher stellar masses (see their Figure~14). NGC\,5194, a blue galaxy according to their criteria, is consistent with this trend as is NGC\,5195. However, NGC\,5195 is not a blue galaxy, but is about 0.4 mag redder than NGC\,5194 and would thus fall into their green galaxy color bin. In this colour bin, a few galaxies exhibit high temperatures similar to NGC\,5195, but their temperatures are poorly determined likely because the dust model used to estimate the temperature was a poor fit to the observations. The typical scenario suggests the redder the galaxy, the cooler the dust temperature, but there are hints, particularly from the large scatter and uncertainty in constraining the temperature, that the transition from blue-to-red galaxy is more complex in the context of dust heating.

As a post-starburst galaxy, NGC\,5195 is in a transition from starburst to quiescent. The lack of ionized gas \citep{thr91} indicates no recent star formation is ongoing but did peak during its close encounter with NGC\,5194 $\sim400$\,Myr ago. Other signposts that it is in a transition are evident by its low dust-to-stellar and gas-to-stellar mass ratios. Its gas fraction at $\sim$2\% is very low and we show, in Figure~\ref{fig:gasdensity}, the gas that is in the galaxy appears to mostly be below the empirical threshold observed in the THINGS galaxy survey \citep{big08} at which atomic gas transitions to molecular gas, a requirement for star formation. In this figure we show the stellar mass density as a function of gas mass density in the top panels, with NGC\,5194 in the top-left and NGC\,5195 in the top-right. The axes are kept on the same scale for each panel to show the different density parameter space in each galaxy. For a given stellar density, the gas density is considerably lower in NGC\,5195 and there is a large scatter between stellar mass density and the total gas mass density (purple points). On the other hand, NGC\,5194 shows a tighter relation between the total gas (purple) and stellar mass densities and there appears to be a threshold gas density of $\sim$10\,M$_\sun\,$pc$^{-2}$ (marked by vertical dashed lines) at which the atomic gas (open red squares) is converted from neutral to molecular (blue plus signs). This threshold was noticed in the Kennicutt-Schmidt (K-S) relations drawn from the THINGS survey in \citet{big08}. In their plots they used star formation density on the y-axis and found the same transition from atomic to molecular gas at a fixed gas density of 9\,M$_\sun$\,pc$^{-2}$. In the bottom panel, we provide plots analogous to these K-S plots by plotting the intensity of H$\alpha$ emission (a proxy for star formation rate density) in our narrowband images after subtracting stellar continuum based on our SED fitting. We do not correct for dust attenuation or convert the intensity to SFR density. The reader is referred to \citet{big08} who already performed this analysis for NGC\,5194. For NGC\,5195, no star formation is found in the galaxy so no Kennicutt-Schmidt relation can be observed in this galaxy.

So why is the dust temperature so high? We attempt to shed some light on the mechanism leading to the high temperature observed in this galaxy. Because this galaxy represents a nearby transitional galaxy, studying its resolved multi-wavelength properties can help isolate the source of heating. In the following few sections we will explore some of the possible heating mechanisms: (1) heating from star formation, (2) heating from the evolved stellar population and (3) heating by its active galactic nucleus. 

\medskip

\noindent \textit{Heating by ongoing star formation:} It is clear that the ionizing radiation of young massive stars is not heating the dust as evident by its lack of ionized gas shown in Figure~\ref{fig:UminvsHa}. The pixels in NGC\,5195 are all upper limits in H$\alpha$ so no ionized gas is detected except for a few regions near the edge of the galaxy, at which the emission from NGC\,5195 begins to blend with the spiral and the local stellar density is lower.

\medskip

\begin{figure*}[tb]
\begin{center}
\includegraphics[angle=-90,width=2.1in]{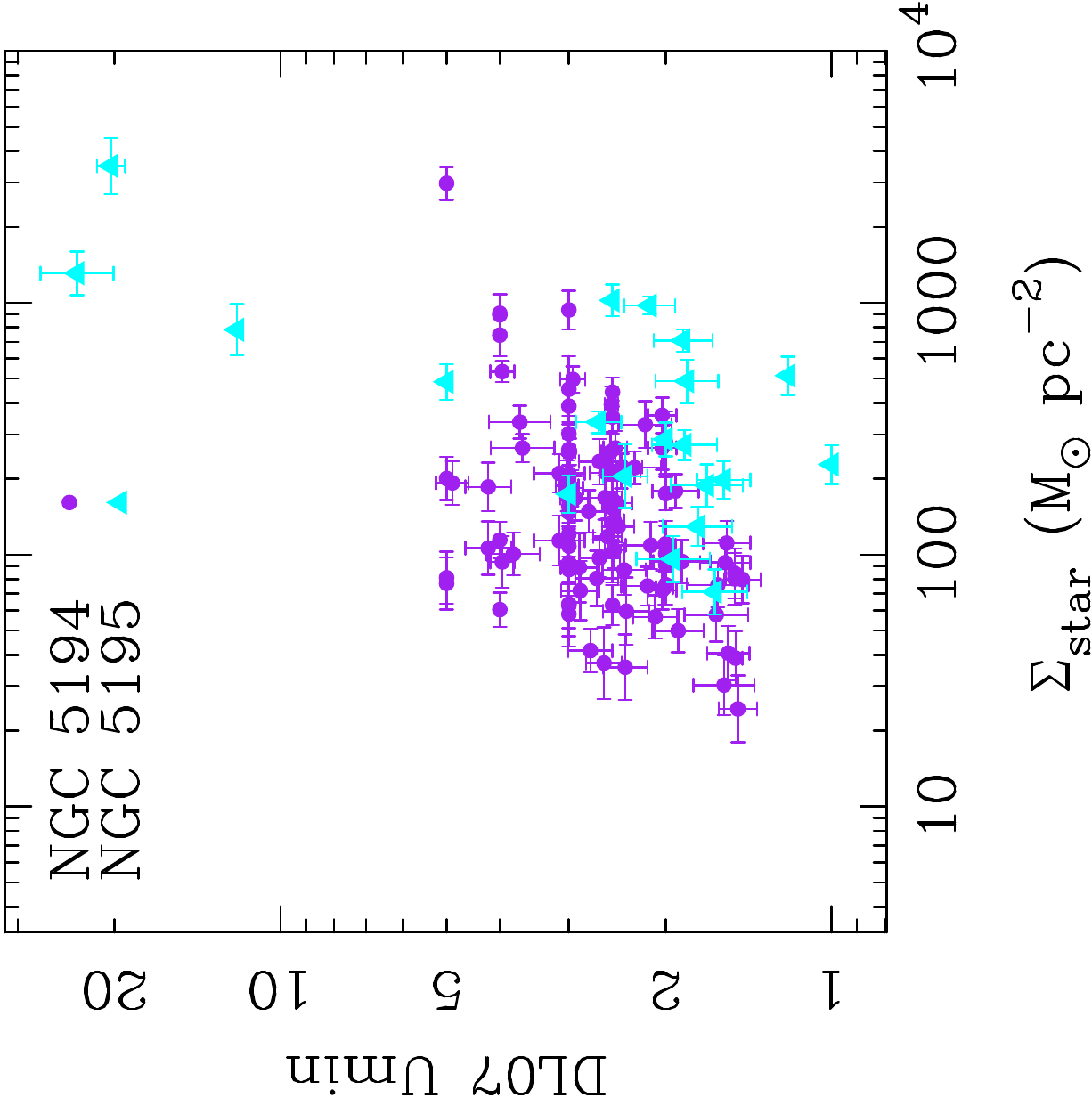}~~
\includegraphics[angle=-90,width=2.1in]{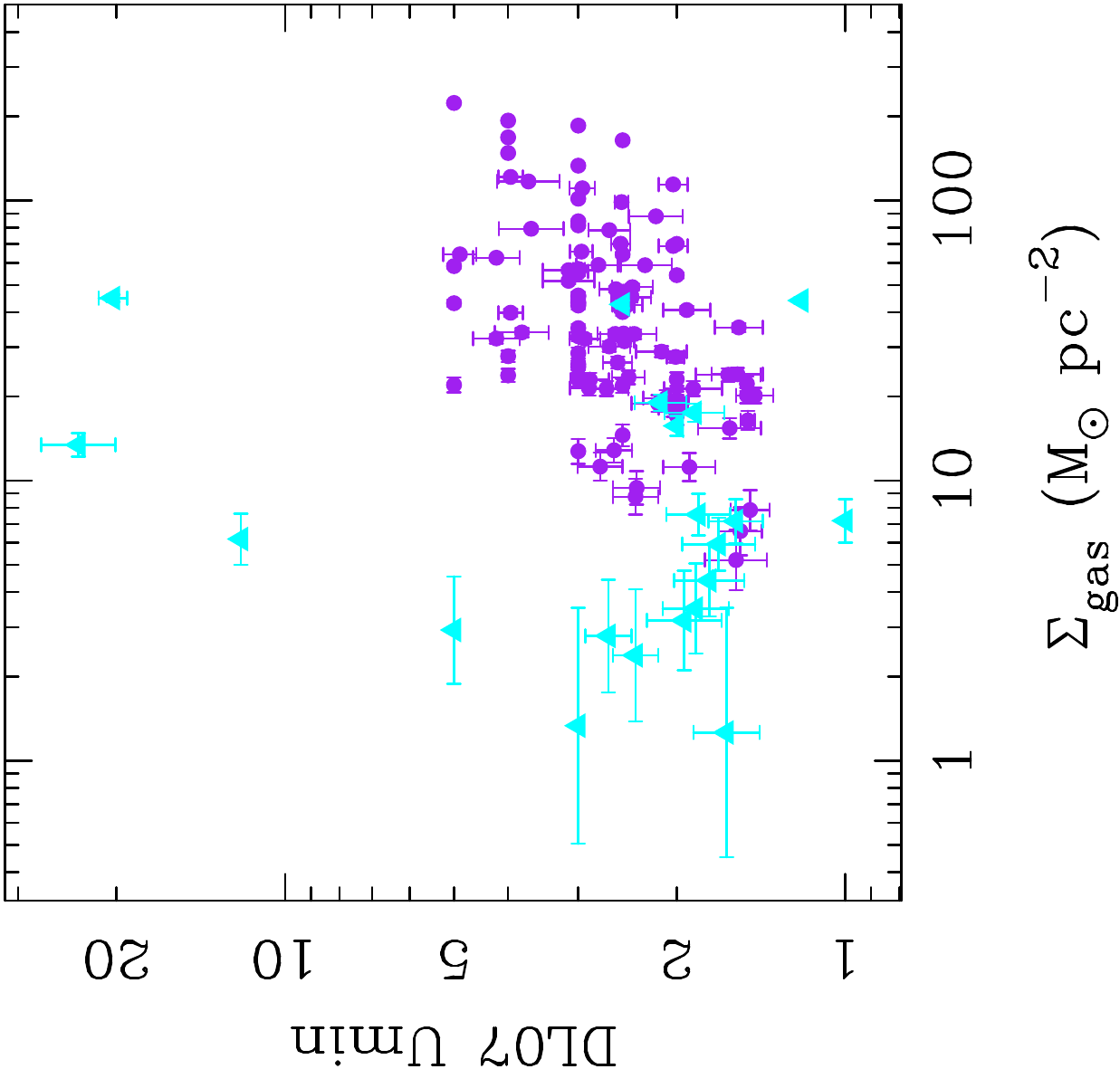}~~
\includegraphics[angle=-90,width=2.1in]{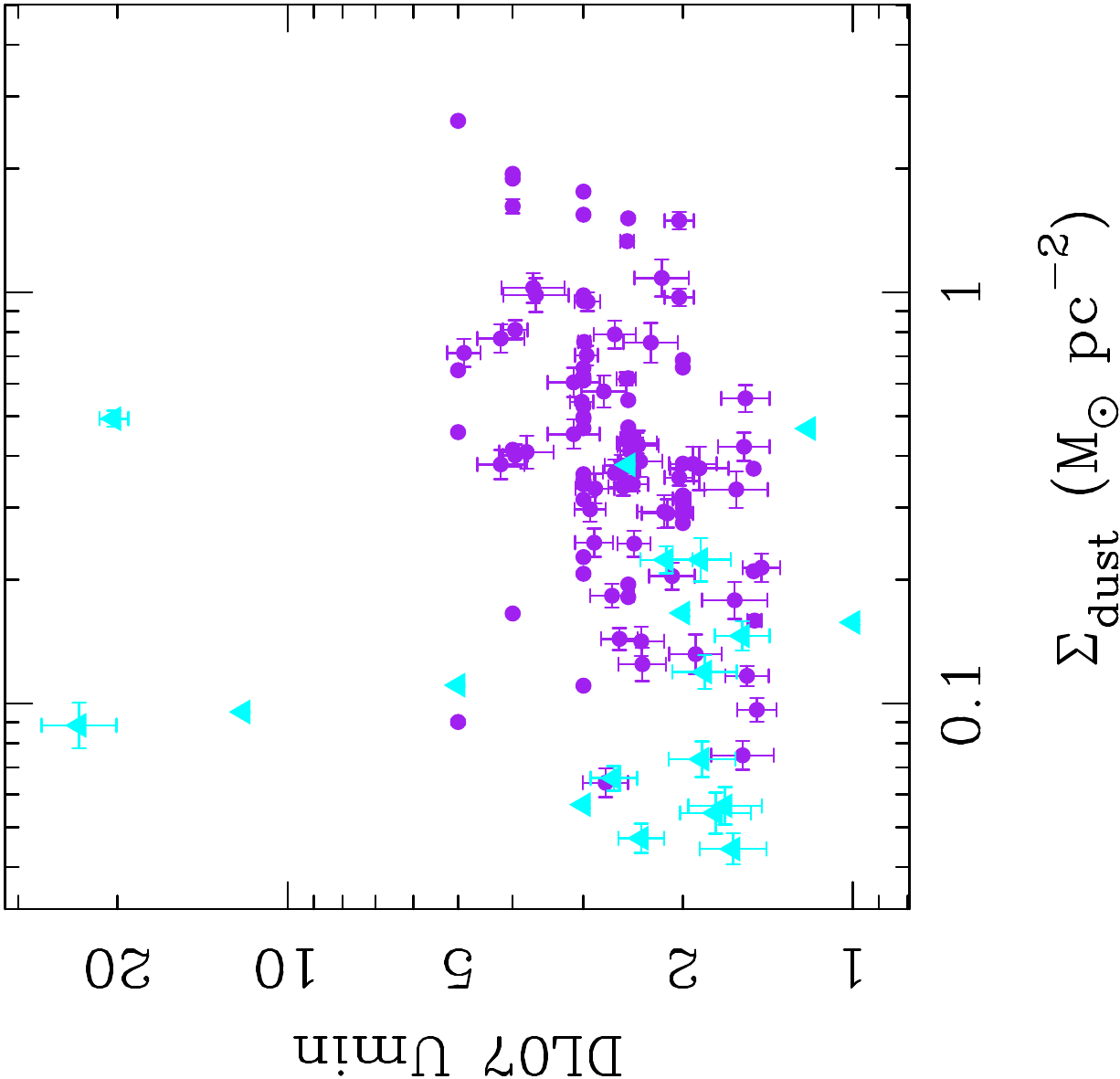}~~
\caption{The amplitude of the ISRF as a function of stellar mass (left), total gas mass (middle), and dust (right) for NGC\,5194 (purple dots) and NGC\,5195 (cyan triangles). There is a modest trend for higher ISRF values to be found at higher stellar densities.}
\label{fig:Uminvsmass}
\end{center}
\end{figure*}

\noindent \textit{Heating by the evolved population:}  Recent studies by \citet{ben12} and \citet{bos12} indicate that evolved stellar population heating plays a strong role in dust heating, although this is most important for lower temperature dust heating at longer infrared wavelength beyond 160\,\micron. In Figure~\ref{fig:Uminvsmass}, we plot the amplitude of the ISRF as a function of stellar, total gas and dust mass density. The ISRF does not correlate with either the stellar, dust or gas mass density.  In addition, the bulge in the spiral reaches similarly high stellar mass densities ($\sim3\times10^4$\,M$_\sun\,$pc$^{-2}$) but has a much lower ISRF value compared to NGC\,5195, so the enhanced stellar mass density cannot be the primary explanation for the enhanced ISRF observed in NGC\,5195.

\medskip

\noindent \textit{Heating by its active galactic nucleus:} NGC\,5195 has a compact radio source in the nucleus \citep{van88} and has spectral indications of harboring an AGN \citep{mou10}. In a spectrum presented in covering the nuclear region, a small amount of H$\alpha$ emission is seen blueshifted relative to H$\alpha$ seen in absorption, indicative of an outflow of ionized gas \citep{mou10}. In our observations, the ISRF is strongly peaked in the nuclear region and is suggestive that the AGN plays a role in heating the dust. 

However, it is not completely clear that this is the dominant mechanism heating the dust and/or shutting down star formation. Mid-infrared observations with ISOCAM show a lack of AGN emission lines \citep{bou96} and rule out LINER activity as the dominant source of mid-infrared emission. Furthermore, NGC\,5194 also harbors an AGN and based on equivalent widths and line intensities presented in \cite{mou10}, the ionized emission associated with the AGN is 5-10 times brighter in NGC\,5194. We caution that this comparison is complicated by unknown dust attenuation due to strong Balmer absorption in NGC\,5195 \citep{ho95}. Unlike the nuclear region of NGC\,5195, the ISRF is only slightly enhanced in the nuclear region in NGC\,5194 (see left panel in Figure~\ref{fig:heating}), and at the same intensity as is seen in the spiral arms. 

Unfortunately none of these mechanisms stand out as the clear heating source. Although evolved population heating appears to be the most likely, with the other two almost ruled out. The stellar mass density is similar in both galaxies so have to conclude that the relatively lower gas and dust fractions in NGC\,5195 result in less efficient gas and dust cooling, leading to a relatively higher ISRF.



\subsubsection{PAH destruction}

Looking back on Figure~\ref{fig:dustparam}, the PAH fraction decreases slightly to $\sim4$\% in the nuclear region of NGC\,5194 due to a low-luminosity active galactic nucleus \citep{mou10}. This is not too surprising. Many studies report \citep{gen98,dal06} the destruction PAH molecules around AGN and within hard radiation fields with PDRs \citep{bos04,cal05,ben08}. On the other hand, the decrease in PAH fraction in NGC\,5195 is interesting. The PAH fraction traces the stellar mass surface density and decreases as a gradient across NGC\,5195. One can see in the bottom-right SED of NGC\,5195 in Figure~\ref{fig:SED} that PAH emission is easily detected and is many times the stellar emission at 5.8 and 8.0\,$\mu$m. So the decrease in the mass fraction of PAHs relative to the total dust mass is not a detection issue. NGC\,5195 also harbours an AGN \citep{mou10} leading to a lower PAH fraction. But unlike NGC\,5194, where the depression is concentrated in the nuclear region, the decrease in PAH fraction across NGC\,5195 is a gradient. This may suggest NGC\,5195 is accreting gas, dust and PAH material from the nearby spiral arm, but as the PAHs approach the AGN, the fraction decreases. 

%

%

\subsection{An example of morphological quenching}\label{s:quench}

Unlike NGC\,5194, the molecular gas in NGC\,5195 is found at densities lower than $\sim$10\,M$_\sun\,$pc$^{-2}$. If this threshold marks the density at which atomic gas becomes molecular, then it is likely that the molecular gas has come from NGC\,5194 via accretion rather than being formed in situ. In addition to this, the molecular gas that does exist is unable to form stars because of the high stellar density (and thus its gravitational potential). The critical gas density required for the gas to cool and collapse is not satisfied. Modelled as a simple single-fluid Toomre stability model \citep{too64}, \citet{ken89} show that at higher stellar densities, the molecular gas must reach a higher critical density in order to cool and collapse. This is because localized shear motions from the gravitational potential are greater than the self-gravitational potential of the molecular cloud thus making the gas stable. In regions where the average surface density of gas is less than 5 times the critical density, star formation is completely suppressed \citep{ken89} because the molecular gas does not become dense enough to collapse into cold, dense molecular clouds required for star formation. 

This scenario is supported by observations of the molecular gas in NGC\,5195 by \citet{koh02}. They measured both CO(1-0) and HCN(1-0) in the galaxy and found an HCN-to-CO intensity ratio 5-15 times smaller than that seen in starburst regions \citep{hel93,koh02}  and 5 times smaller than that usually found in spiral disks \citep{hel95,koh99}, indicative of a lack of dense molecular cores.  

Half a billion years ago, when the two galaxies were closer together, tidal forces would have disrupted such high shear velocities and likely involved higher rates of gas and dust accretion so both the density required for clumps to form would have been lower and the gas supply available would have been more dense, allowing star formation to occur. But as the galaxies drew apart, the tidal forces were reduced and the dense molecular gas was exhausted as it was converted into stars during the galaxy's starburst episode.

This process, recently termed ``morphological quenching'' by \citet{mar09}, occurs naturally within hydrodynamical simulations of galaxy evolution and can be extremely efficient under the right circumstances. \citet{mar09} showed that gaseous molecular disks, even those up to several 10$^9$\,M$_\odot$, are unable to form stars if they are embedded within a sufficiently large gravitational potential such as a bulge dominated galaxy. Star formation will cease even if gas accretion continues. The authors claim that this mode of star formation quenching is important for galaxies in dark matter halos less massive than $\sim10^{12}$\,M$_\sun$ and thus an important mechanism for converting L$_\star$ and lower mass galaxies from the `blue cloud' to the `red-sequence' and regulating galaxy growth, particularly because other quenching mechanisms such as AGN feedback or gas stripping are only effective in more massive halos.  NGC\,5195 provides a great nearby laboratory to further understand and validate this star formation quenching mechanism.


\section{Conclusions}

We have fit optical, near-, mid- and far-infrared images of the Whirlpool galaxy system to a large library of stellar and dust spectral energy distribution models. By performing 100 Monte Carlo simulations of the photometry fitted to the models in each pixel, we have obtained maps of output parameters from stellar models generated by the SPS code PEGASE.2 \citep{fio97} and dust model parameters generated by the dust emission model of \citet{dra07}. Our images were matched to a common plate scale and spatial resolution defined by a gaussian PSF with a FWHM=28\arcsec~resolving the galaxy on physical scales of $\sim1$\,kpc. Here we summarize the main conclusions of this work:

\begin{enumerate}

\item The SPS model fits reveal a burst of star formation occurred in both galaxies roughly 340--500\,Myr ago, consistent with dynamical models of the interaction history of the galaxies \citep{sal00,dob10} as well as colour-magnitude diagrams of individual stars \citep{tik09}. There is little spatial dependence of the burst age and other SFH output parameters, suggesting a resolved SED approach is not necessary at this spatial resolution, corresponding to physical scales of $\sim$1\,kpc. Metallicity and dust attenuation show some radial dependence but were not very well constrained by our method.

\item  The dust-to-stellar mass ratio in NGC\,5194 is $\log(M_\mathrm{dust}/M_\star)= -2.5\pm0.2$, while NGC\,5195 has a magnitude lower mass ratio at NGC\,5195 at $\log(M_\mathrm{dust}/M_\star)= -3.5\pm0.3$. Both galaxies are consistent, according to galaxy type, with other galaxies observed with \textit{Herschel} \citep{cor12,smi12,bou12}.

\item High radiation field values concentrated in the nuclear region of NGC\,5195 provide evidence that the dust is located within the galaxy, not in its foreground. Despite a high ISRF (often an indication of star formation), there is no current star formation in the central region of NGC\,5195 as indicated by a lack of H$\alpha$ emission and lack of dense molecular clouds \citep{koh02}. 

\item Morphological quenching \citep{mar09} is a plausible scenario for the suppression of star formation in NGC\,5195. With its higher gravitational potential, the gas density threshold required for star formation is higher than the available low-density gas and is thus unable to form stars.

\end{enumerate}

\bigskip

{\noindent \textit{Acknowledgements:}} EMC would like to acknowledge the continued support for this research provided by the Natural Sciences and Engineering Research Council (NSERC) of Canada through her PDF fellowship. This publication would not be possible without the legacy left by the great efforts of those involved and in the Spitzer Infrared Nearby Galaxies Survey (SINGS), the Two Micron All Sky Large Galaxy Atlas (2MASS LGA) and the BIMA SONG survey. This research was supported by grants from the Canadian Space Agency and the Natural Sciences and Engineering Research Council of Canada (PI: C. Wilson). PACS has been developed by a consortium of institutes led by MPE (Germany) and including UVIE (Austria); KU Leuven, CSL, IMEC (Belgium); CEA, LAM (France); MPIA (Germany); INAF-IFSI/OAA/OAP/OAT, LENS, SISSA (Italy); IAC (Spain). This development has been supported by the funding agencies BMVIT (Austria), ESA-PRODEX (Belgium), CEA/CNES (France), DLR (Germany), ASI/INAF (Italy), and CICYT/MCYT (Spain). SPIRE has been developed by a consortium of institutes led by Cardiff University (UK) and including Univ. Lethbridge (Canada); NAOC (China); CEA, LAM (France); IFSI, Univ. Padua (Italy); IAC (Spain); Stockholm Observatory (Sweden); STFC and UKSA (UK); and Cal- tech, JPL, NHSC, Univ. Colorado (USA). This development has been supported by national funding agencies: CSA (Canada); NAOC (China); CEA, CNES, CNRS (France); ASI (Italy); MCINN (Spain); SNSB (Sweden); STFC (UK); and NASA (USA). HIPE is a joint development by the Herschel Science Ground Segment Consortium, consisting of ESA, the NASA Herschel Science Center, and the HIFI, PACS and SPIRE consortia. This research has made use of the NASA/IPAC Extragalactic Database (NED) which is operated by the Jet Propulsion Laboratory, California Institute of Technology, under contract with the National Aeronautics and Space Administration.

\newpage

\appendix 

\begin{figure}[h]
\centering
\includegraphics[angle=-90,width=4in]{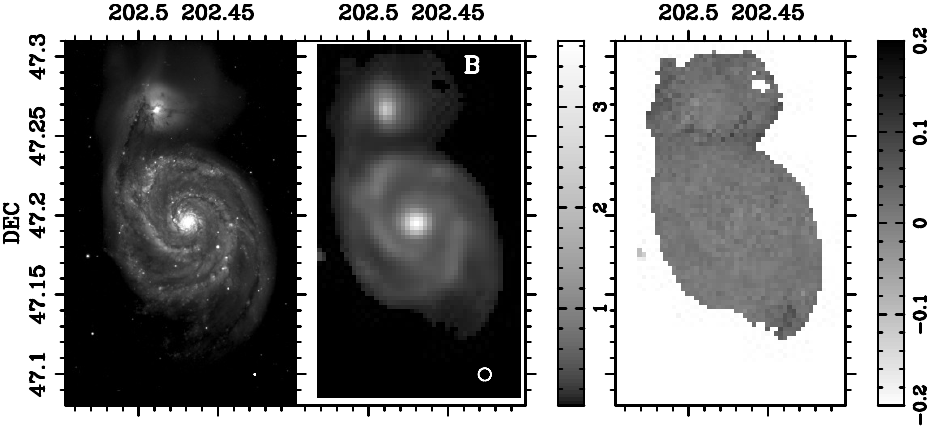}\\ 
\includegraphics[angle=-90,width=4in]{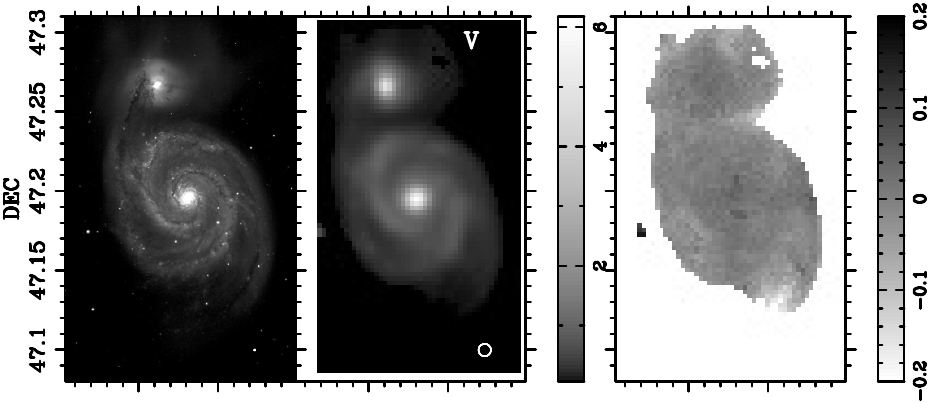}\\
\includegraphics[angle=-90,width=4in]{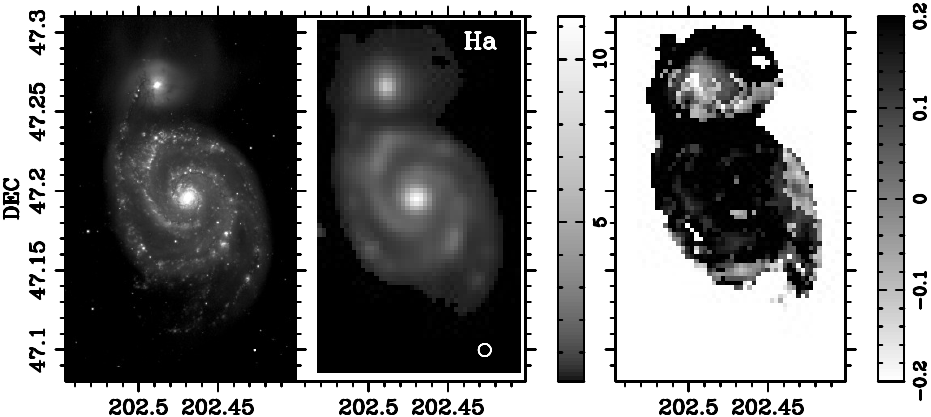}\\
\caption{Optical images of NGC\,5194/5195 from the SINGS survey \citep{ken03} from top to bottom: $B$,$V$ and $H\alpha$. On the left we show the original image at its native resolution in each band and then the processed images after convolution with the beam size of each filter in the bottom right of each panel. Both left and middle panels are in intensity units of MJy~sr$^{-1}$, whose range is indicated by the gradient on the right of the middle panel. The right column shows the residual image, defined in Eq.~\ref{eq:res}, between the observed image after it is PSF matched (middle) and a synthetic galaxy image (not shown) generated by the best fitting SED in our analysis. In the bottom row, the observed and processed H$\alpha$ narrow band image is shown (and includes stellar continuum emission in this band as there was no need to stellar continuum subtract in our analysis). The image was not include in our SED fitting process, but we do show that overall the models -- which include nebular emission-- underestimate the amount of nebular emission by more than 20\%, except in NGC\,5195, where the models accurately predict the amount of emission (mostly stellar continuum emission in this case) in the H$\alpha$ bandpass.} 
\label{fig:montageSINGS}
\end{figure}

\begin{figure}[h]
\centering
\includegraphics[angle=-90,width=4in]{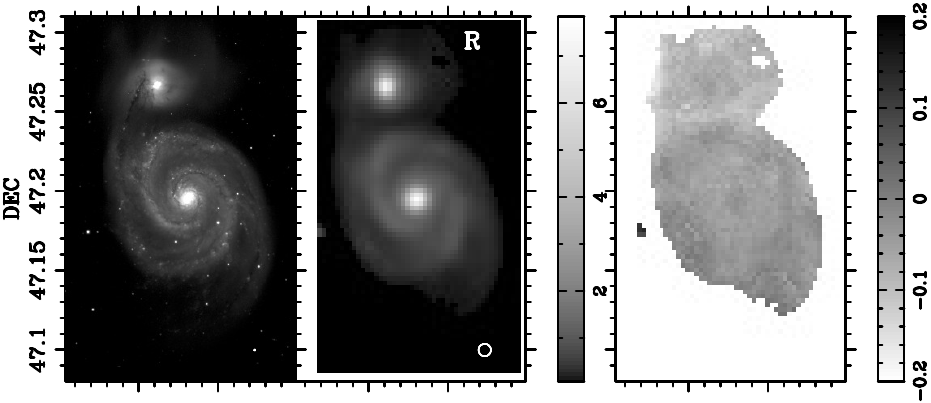}\\
\includegraphics[angle=-90,width=4in]{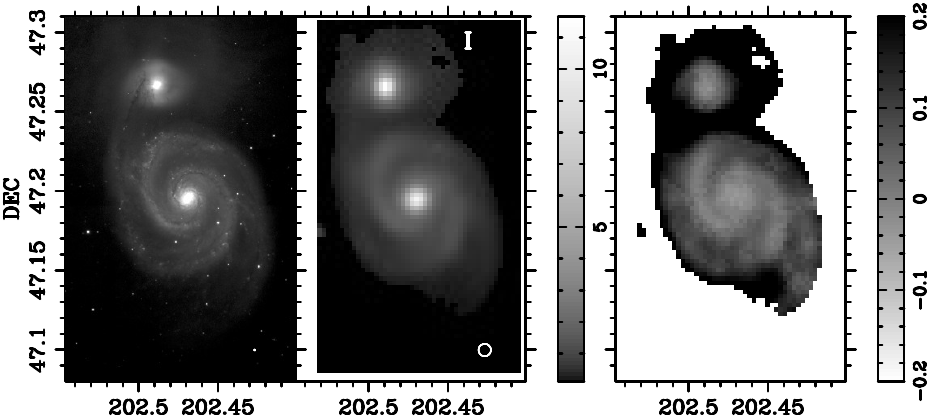}\\
\caption{Optical images of NGC\,5194/5195 from the SINGS survey \citep{ken03} from top to bottom: $R$ and $I$. As in Figure~\ref{fig:montageSINGS}, the left panel is the original image from the archive and the processed image in the middle, both in MJy~sr$^{-1}$. The right shows the residual image between the models and observations..} 
\label{fig:montageSINGS2}
\end{figure}

For reference, we show the image processing of the archival data used for our analysis. As described in \S\ref{s:data}, all images were matched to a common spatial resolution (FWHM=28\arcsec) and platescale (10\arcsec\,pix$^{-1}$). Figures~\ref{fig:montageSINGS} \& \ref{fig:montageSINGS2} show the optical images taken from the ancillary catalogue of the SINGS survey \citep{ken03}.  The left image is the original image of the galaxy at native resolution prior to any post-processing. The middle panel shows the galaxy after foreground stellar emission is replaced by emission in surrounding pixels. Following this, we are able to match the spatial resolution to the far-infrared images using convolution kernels as described in \S\ref{s:data}. The far right image shows a normalized residual map which we quantify as in Equation~\ref{eq:res}.

Most of the optical images are well fit by the modes with the residual images revealing that the models fit the data to within 10\%, and thus within their uncertainties. The only image that is not well fit is the H$\alpha$ image. This image was not used in the fitting routine but since the SEDs include nebular emission, ideally, it should be predicted by the SED models. The predicted nebular emission is reasonable in NGC\,5195, which has no star formation, but in NGC\,5194, the amount of H$\alpha$ emission is underestimated. This is primarily because our chosen SFH is too simplistic and only modelled the very old population and the burst population from $\sim400$\,Myr ago. A more complex SFH, such as a declining or constant star forming burst, is needed to properly describe the recent star formation occurring in NGC\,5194.

\begin{figure}[h]
\centering
\includegraphics[angle=-90,width=4in]{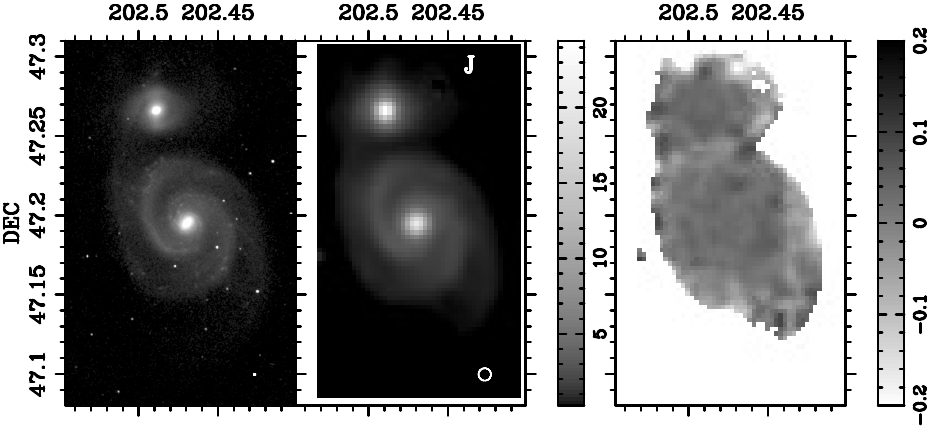}\\
\includegraphics[angle=-90,width=4in]{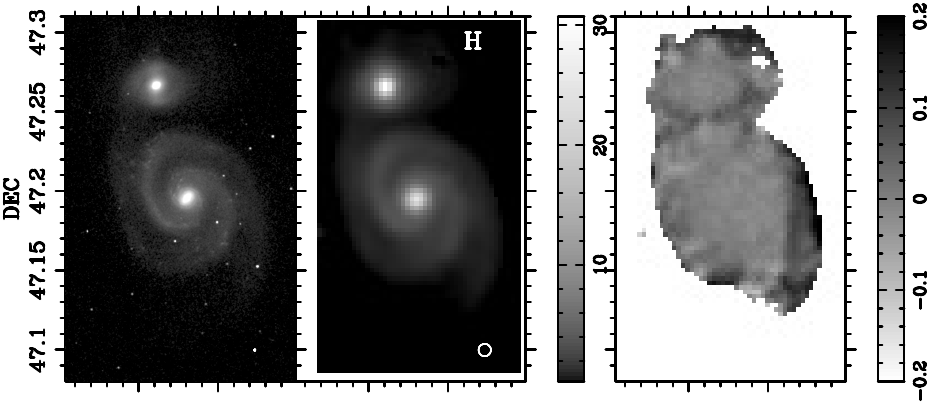}\\
\includegraphics[angle=-90,width=4in]{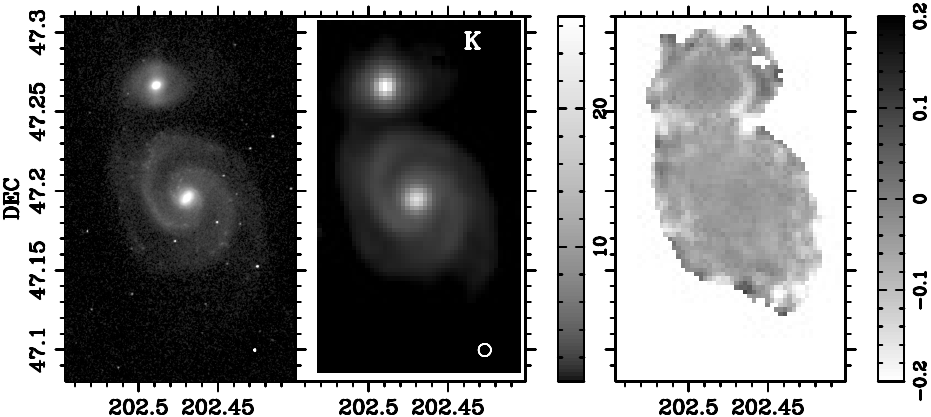}\\
\caption{Near-infrared $J$$H$$K$ images of NGC\,5194/5195 from the 2MASS Large Galaxy Atlas \citep{jar03}. As in Figure~\ref{fig:montageSINGS}, the left panel is the original image from the archive and the processed image in the middle, both in MJy~sr$^{-1}$. The right shows the residual image between the models and observations.}
\label{fig:montage2MASS}
\end{figure}

\begin{figure}[h]
\centering
\includegraphics[angle=-90,width=4in]{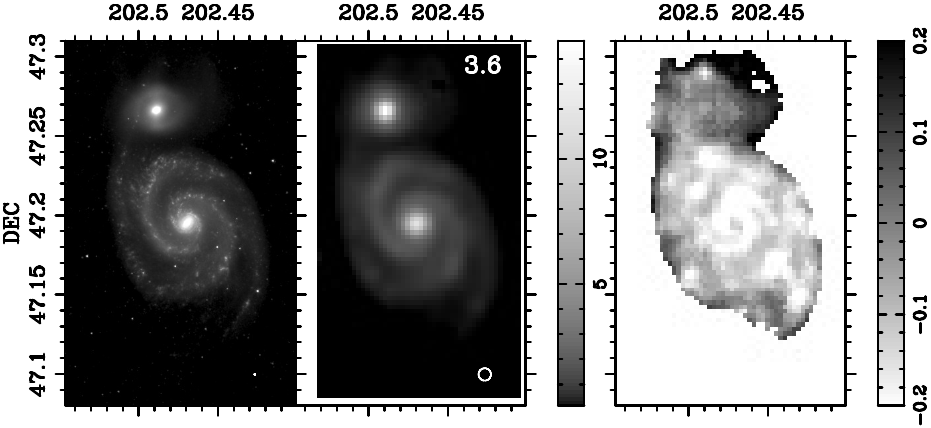}\\
\includegraphics[angle=-90,width=4in]{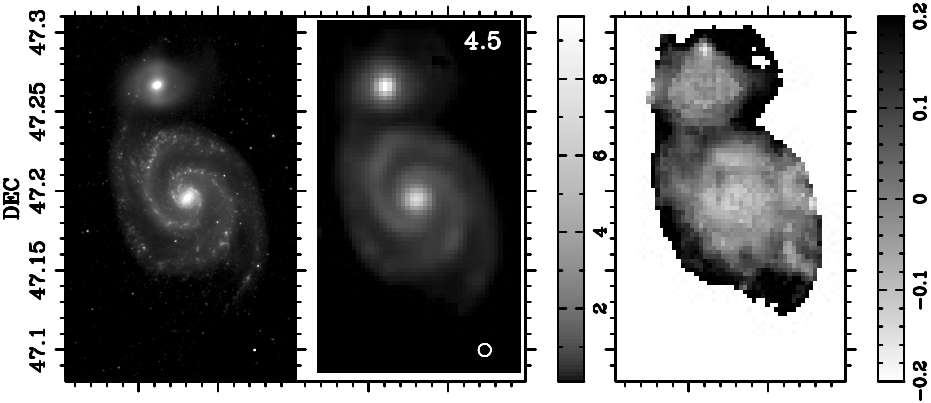}\\
\includegraphics[angle=-90,width=4in]{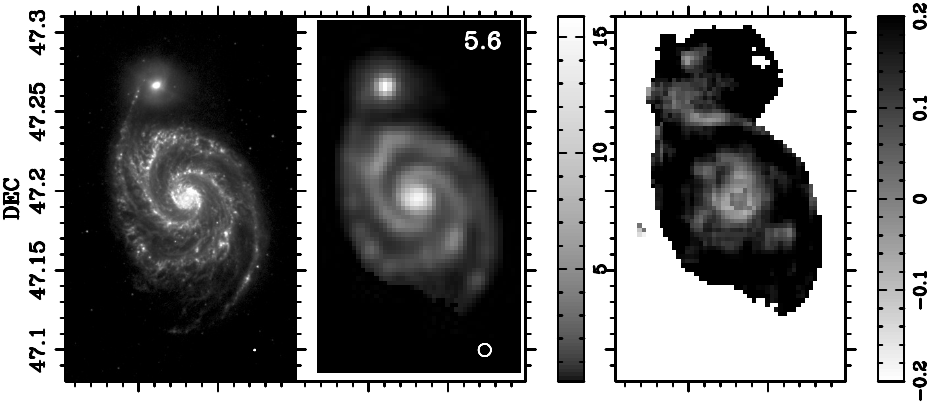}\\
\includegraphics[angle=-90,width=4in]{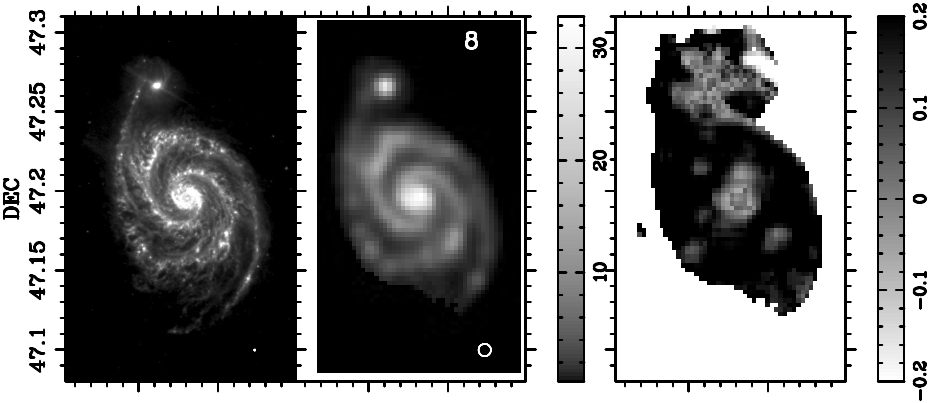}\\
\includegraphics[angle=-90,width=4in]{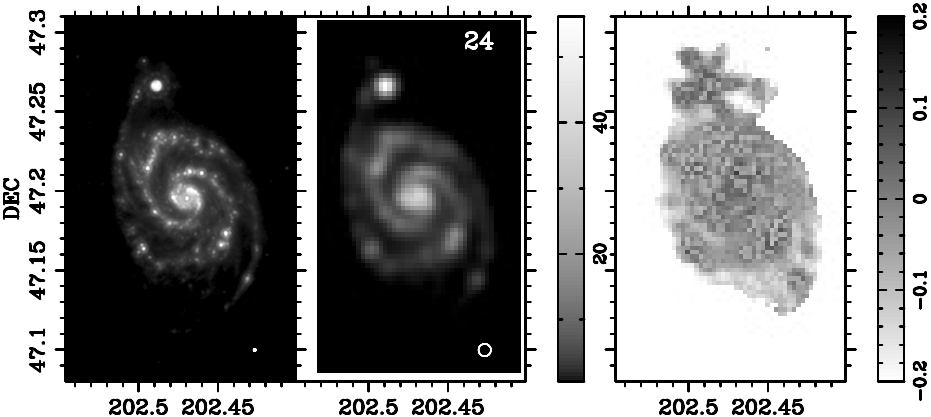}\\
\caption{Mid-infrared images of NGC\,5194/5195 from \textit{Spitzer} as part of the SINGS survey \citep{ken03}.  As in Figure~\ref{fig:montageSINGS}, the left panel is the original image from the archive and the processed image in the middle, both in MJy~sr$^{-1}$. The right shows the residual image between the models and observations. The top two images were not used in either fit.}
\label{fig:montageSpitzer}
\end{figure}

The near-infrared images from the 2MASS Large Galaxy Atlass \citep{jar03} are shown in Figure~\ref{fig:montage2MASS}. The models lead to small residual images, indicating a good fit to the NIR images. On the other hand, mid-infrared images are not as well fit as shown in Figure~\ref{fig:montageSpitzer}. This is primarily because the mid-infrared images at 3.6 and 4.5\,\micron~are not used to fit the SEDs and the emission is being extrapolated from models that have fit on one side of the spectrum the stars and on the other side the dust and PAH emission. Both models combined tend to overestimate the emission at 3.6 and 4.5\,\micron. More analysis is needed to determine whether the stellar or PAH emission is being overestimated or both. The 5.6 and 8.0\,\micron~images are used to fit the DL07 dust emission SED model. In the nuclear regions of both galaxies, the models are a good representation of the emission, but outside of these regions, the models underestimate the emission. Recalling from the PAH mass fraction map on the right in Figure~\ref{fig:dustparam} that in the regions outside of the nuclear region or edges of the galaxies, the PAH mass fraction reaches the highest value permitted by the DL07 models. This is likely the reason why the 5.6 and 8.0\,\micron~are underestimated by the models. A higher PAH fraction is needed to properly describe the observations. At longer wavelengths, just like our \textit{Herschel} observations shown in Figure~\ref{fig:montageHerschel}, the 24\,\micron~image is accurately described by the DL07 models.

\end{document}